\pdfoutput=1
\documentclass[11pt]{article}
\usepackage[margin=2.5cm]{geometry}
\usepackage[T1]{fontenc}
\usepackage{lmodern}
\usepackage{amsmath,amssymb}
\usepackage{booktabs}
\usepackage{graphicx}
\usepackage[numbers,sort&compress]{natbib}
\usepackage[hidelinks]{hyperref}
\usepackage{setspace}
\newcommand{\vN}{v(N)}
\newcommand{\phii}{\phi_i}

\title{Organizational Learning in Materials Discovery Campaigns: Which Databases, Calculations, and Agent Teams Pay Off}
\author{Kinston Ack{\"o}lf\textsuperscript{1} and Taylor D. Sparks\textsuperscript{1,*}\\[4pt]
\small \textsuperscript{1}Department of Materials Science and Engineering, The University of Utah, Salt Lake City, UT 84112, USA\\
\small \textsuperscript{*}Corresponding author: taylor.sparks@utah.edu\\
\small ORCiD: K.~Ack{\"o}lf 0000-0002-5579-4878; T.~D.~Sparks 0000-0001-8020-7711}
\date{}

\begin{document}
\maketitle

\begin{abstract}
\noindent More databases can make a materials campaign worse. Through an organizational lens, we trace each source's contribution. One band-gap campaign ranks seven anion families from archived measured gaps, one vote per source. In the first half of its budget, all seven pooled rank worse (rank agreement 0.737; 1 = measured order) than the best pool of three (0.851). As a stability check, choosing the pool on half the runs and testing on the rest kept the gap. Splitting Materials Project afterwards points to its transition-metal oxides and fluorides. Two databases from one pipeline add nothing over one, but pooling gives their shared ordering two votes. New calculations repay their cost early only when two or three campaigns reuse them. Flat language-model agent teams trailed a random-pick baseline (mean 0.876, sd 0.096, 20 searches). An adaptive Al--Mg--Si team, mostly interatomic potentials, was scored on unseen HSE06 calculations. In its first quarter of compute it found 49\% of the ten lowest-formation-energy compounds, against at most 40\% for fixed teams and 73\% for MACE-MP-0 alone. Merging members whose orderings agreed carried most of its gain. This suggests pooling a few sources that fit the measurements, counting each pipeline once and charging for computation.
\end{abstract}


\section{Introduction}
\label{sec:intro}

A group screening wide-gap semiconductors can consult several computed band-gap databases before it measures anything. The Materials Project \citep{Jain2013MaterialsProject}, JARVIS \citep{Choudhary2020JARVIS}, and SNUMAT \citep{Kim2020SNUMAT} hold gaps computed with the generalized gradient approximation (GGA) \citep{Perdew1996PBE}, meta-GGA ($r^{2}$SCAN) \citep{Kingsbury2022r2SCAN}, modified Becke--Johnson (TB-mBJ) \citep{Tran2009mBJ}, and hybrid functionals (HSE06) \citep{Heyd2003HSE,Heyd2006HSEErratum,Krukau2006HSE06}, and each level departs from measurement by different amounts in different chemistries \citep{Borlido2019BandGapBenchmark}. Multi-fidelity learning fuses such levels into one predictor \citep{Pilania_2017,Hutchinson2017transfer,Chen_2021}, and multi-source Bayesian optimization draws on several sources within the search \citep{khatamsaz2021mis,zanjanifoumani2023mitigating}, in one variant with a cost-aware stopping rule \citep{zanjanifoumani2025costaware}; here each level is kept as a separate source, so that what it adds to a campaign can be read. The materials genome approach \citep{dePablo2019MGI,Agrawal2016MaterialsInformatics,Ramprasad2017MaterialsInformatics} invites a group to use all of them. Databases are pooled as wholes, however, and whether each added database helps, repeats another, or misleads the campaign is rarely known before the campaign is run.

Organization science has long asked whether reused knowledge helps. Organizations learn from their own and others' experience \citep{Levitt1988OrganizationalLearning,ArgoteMironSpektor2011OrgLearning}; transferred knowledge can raise performance or, when it does not fit the receiving context, lower it \citep{Argote2000KnowledgeTransfer}, and transfer is hard to achieve \citep{Szulanski1996Stickiness,Cohen1990Absorptive}; redundant ties carry little new information \citep{Granovetter1973WeakTies,Burt1992StructuralHoles}; and effort is divided between exploiting what is known and exploring what is not \citep{March1991Exploration}. A materials campaign can be read, by analogy, as such an organization: its members are databases, calculations, interatomic potentials, acquisition rules, and language-model agents; its organizational memory \citep{Walsh1991OrganizationalMemory} is the archive; and its structure is who shares results with whom. Self-driving laboratories coordinate such members against shared budgets \citep{Abolhasani2023SelfDrivingLabs,Kusne2023MultiAgentLab,Lo2024FrugalTwin}, and teams of language-model agents now work on research and software tasks together \citep{Park2023GenerativeAgents,Ghafarollahi2025SciAgents,Hong2023MetaGPT,Chen2023AgentVerse,Guo2024LLMMultiAgentSurvey,Zhan2026Agensh}.

Benchmarks of these systems report the acceleration of the whole loop \citep{Rohr2020BenchmarkingSL,Liang2021BenchmarkingBO,Kavalsky2023ClosedLoop}. A whole-loop number cannot say which member carried the campaign's knowledge forward and which held it back, and a leaderboard of members run one at a time cannot either, because members interact: two databases computed with one pipeline repeat each other, and a misaligned ordering can pull a pooled prior away from a good one. Recent multi-agent language-model systems make some organization-science constructs executable as team configurations \citep{Chen2026OrgScienceMAS}, but not as values read from what a campaign found. Here every subset of the knowledge sources is run as a campaign of its own, replayed from a version-locked archive or from calculations made once, so the outcome of ``the campaign without source $X$'' is the result of a run rather than a counterfactual. Cooperative-game methods \citep{Shapley1953Value,Harsanyi1963Bargaining,Myerson1977Graphs,Myerson1980Conference} trace from those outcomes how far each source advances the campaign, and how that changes when members meet, are coordinated, or learn whom to rely on. Shapley values already divide the gains of federated and collaborative learning among data holders that train one model \citep{Song_2019,Wang_2020,Sim2020incentive}; here the players are the sources of a campaign that buys measurements.

Two campaigns are studied: ranking wide-gap semiconductor candidates by measured band gap, using a version-locked archive of band-gap records from eight sources and HSE06 gaps computed on demand, and finding the Al--Mg--Si compounds with the lowest formation energy per atom, where every query triggers a density functional theory (DFT) or molecular-dynamics (MD) calculation with a measured cost. Four questions of a materials group are each read through an organization-science construct that the campaigns illustrate rather than test, and each is answered as a rule: pool a few well-aligned databases rather than all of them (knowledge integration, negative transfer); count databases from one pipeline once (redundancy); compute an ordering no database holds only when the calculations will be reused (exploration); and, when the best member is not known in advance, learn from the record whom to rely on, which beats fixed structures but does not replace the best member (communication structure, trust).

\section{Results}
\label{sec:results}

The organizational terms used below (learning, trust, coordination, teams) are an interpretive lens on the campaign records; the organizational constructs themselves are not tested.

\subsection{How a campaign draws on each knowledge source}
\label{sec:measure}

A worked example fixes the terms. The band-gap campaign puts seven anion families (borides, oxides, halides, nitrides, chalcogenides other than oxides, pnictides other than nitrides, and others) in order of band gap. Each step buys the values of a tray of four candidates from a receiver source (the source whose values the campaign buys), one archived record per candidate, none drawn twice. The orderings of the donor sources form a prior with a trust weight, fixed or learned from the receiver values acquired so far, and each purchase corrects it. Quality is the Kendall $\tau$ \citep{Kendall1938Tau} between the fitted ranking and the receiver's own fixed order of family median gaps ($\tau=1$ when the two orders agree, $0$ when they are unrelated); its uncertainty is not propagated into the intervals. For a set $S$ of donors, $v(S)$ is the mean $\tau$ over the first half of the budget (the early window) of a campaign whose prior uses only the donors in $S$, averaged over matched seeds; $v(\emptyset)$ is the campaign without donors. All $2^{n}$ subsets are run.

\begin{figure}[t]
\centering
\includegraphics[width=\textwidth]{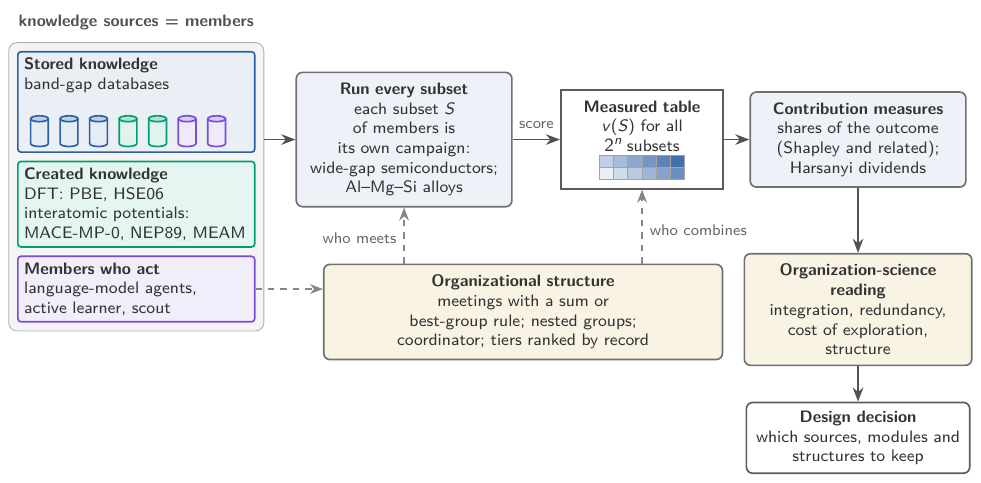}
\caption{How a campaign draws on each knowledge source. The members of a campaign are its knowledge sources: stored knowledge (band-gap databases), created knowledge (Perdew--Burke--Ernzerhof (PBE) and HSE06 calculations, interatomic potentials), and members who act (language-model agents, an active learner, a scout). Every subset $S$ of members is run as a campaign of its own in the wide-gap semiconductor or Al--Mg--Si alloy system, giving the measured table $v(S)$ over all $2^{n}$ subsets, which contribution measures divide among the members. Meetings (dashed) restrict which members combine. The measures also use Owen levels, nested groups (a share is divided first between groups, then within each), and Beta(16,1) weights, which count small pools most.}
\label{fig:method}
\end{figure}

From the measured table $v$, the Shapley value $\phi_i$ \citep{Shapley1953Value} is the average change that source $i$ makes when it joins the others in every order, and the values sum to $v(N)-v(\emptyset)$; leave-one-out, $v(N)-v(N\setminus\{i\})$, is what removing one source at a time from the full pool shows; and the Harsanyi dividend $d(S)$ \citep{Harsanyi1963Bargaining} is what a set of sources creates beyond all its subsets (negative for overlap, positive for complementarity). Communication structure restricts which sources can combine (Methods; Figure~\ref{fig:method}).

\subsection{Pooling band-gap databases: more stored knowledge does not make a better campaign}
\label{sec:databases}

A group that combines public band-gap databases to advise a measured-gap campaign has to decide which databases to keep. Seven databases advised the campaign on the measured gaps of the compilation of \citet{Zhuo2018BandGapML}: the Materials Project (MP) GGA records \citep{Jain2013MaterialsProject}, the MP GGA records of experimentally synthesized compounds identified by ICSD provenance \citep{Belsky2002ICSD}, MP $r^{2}$SCAN, JARVIS OptB88vdW and TB-mBJ \citep{Choudhary2020JARVIS,Klimes2010optB88}, and SNUMAT GGA and HSE06 \citep{Kim2020SNUMAT}. All 128 subsets of this seven-source game were run with 10 matched seeds (Supporting Information Section~S2).

\textbf{Early in the budget, integrating every database held the campaign back.} The campaign without donors reached $\tau = 0.695$ and the full pool of seven $0.737$, but the best pool of three (MP $r^{2}$SCAN and both SNUMAT sources) reached $0.851$, and from three to seven donors the best attainable value fell with each added source. Chosen on five seeds and scored on the other five, the best pool of three was the same in both halves, so this check shows that the choice is stable rather than testing it on independent data: the held-out pool exceeded the full pool by $0.115$ (95\% interval from $0.037$ to $0.193$; the ten-seed interval, not adjusted for selection); in the final window the halves chose different pools and the difference was not resolved ($+0.057$, $[-0.023,+0.137]$; Supporting Information Section~S2). The two SNUMAT databases contributed most (Shapley value $+0.054$ each) and MP GGA least ($-0.029$); the intervals of this game are read as descriptive (Methods). With seven families one exchange of two neighboring families moves $\tau$ by $0.095$; differences below about $0.03$ are reported but not ranked.

\begin{figure}[t]
\centering
\includegraphics[width=\textwidth]{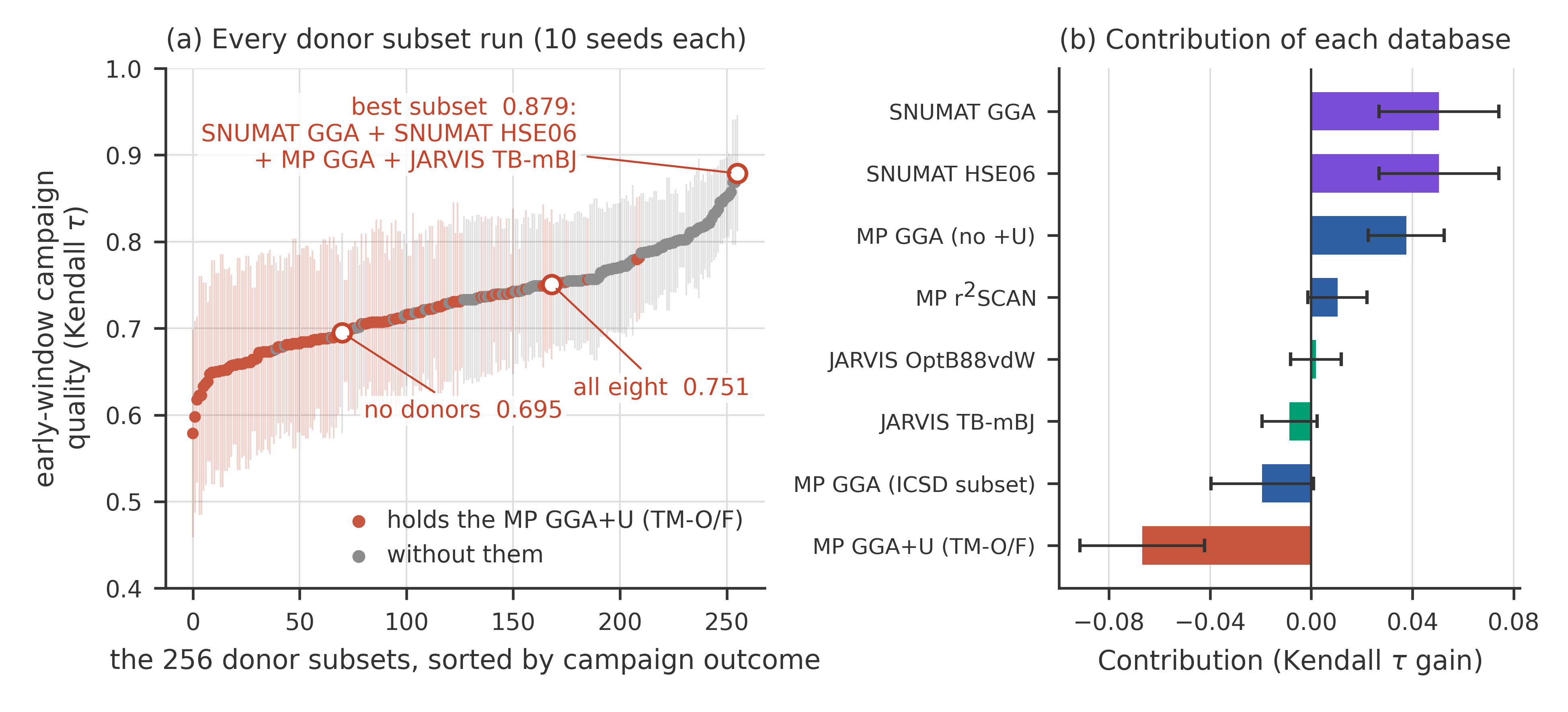}
\caption{In the post hoc eight-donor game, integrating all eight band-gap sources reaches a lower early ranking quality than the best pool of four. (a) All 256 donor subsets, each a campaign run with 10 matched seeds: dot, seed mean; line, 95\% $t$ interval over seeds; orange, subsets that hold the MP GGA+U (TM-O/F) subset. Full pool $\tau = 0.751$, best pool of four $0.879$, no donors $0.695$. (b) Shapley contribution of each source with the 95\% $t$ interval over seeds; colors mark the database family (Materials Project blue, its TM-O/F subset orange, JARVIS green, SNUMAT purple). Quality: Kendall $\tau$ of the anion-family ranking against the measured gaps, mean over the first five of ten budget checkpoints.}
\label{fig:databases}
\end{figure}

\textbf{A post-hoc eight-donor game repeats the pattern.} After the seven-source game had been run, the MP GGA records were split at the Materials Project rule that applies GGA+U \citep{Jain_2011} into the transition-metal oxides and fluorides, MP GGA+U (TM-O/F), and MP GGA (no +U), and all 256 donor subsets were run with 10 matched seeds (Figure~\ref{fig:databases}; Methods). This game is exploratory and serves only to locate the loss; the headline values are those of the seven-source game.

The full pool reached $0.751$, but the best pool of four (JARVIS TB-mBJ, MP GGA (no +U), and both SNUMAT sources) reached $0.879$ (the in-sample best of the 256 subsets, not adjusted for selection), and from four to seven donors the best attainable value fell with each added source (all eight, $0.751$; Supporting Information Section~S3); chosen on five seeds and scored on the other five, the best pool still exceeded the full pool by $0.128$ (95\% interval from $0.061$ to $0.195$; Supporting Information Section~S2). The two SNUMAT databases carried the largest contributions ($+0.051$ each). In this post-hoc split, which is confounded with chemistry and generates a hypothesis rather than testing one, the TM-O/F subset held the campaign back most ($-0.067$, interval $[-0.091,-0.042]$) while the rest of MP GGA advanced it: a block of records that does not fit the receiving context \citep{Argote2000KnowledgeTransfer,Zhang2023NegativeTransferSurvey}, where the control does not resolve whether the loss comes from the +U records or from the chemistries they cover (Supporting Information Section~S2).

\begin{figure}[t]
\centering
\includegraphics[width=\textwidth]{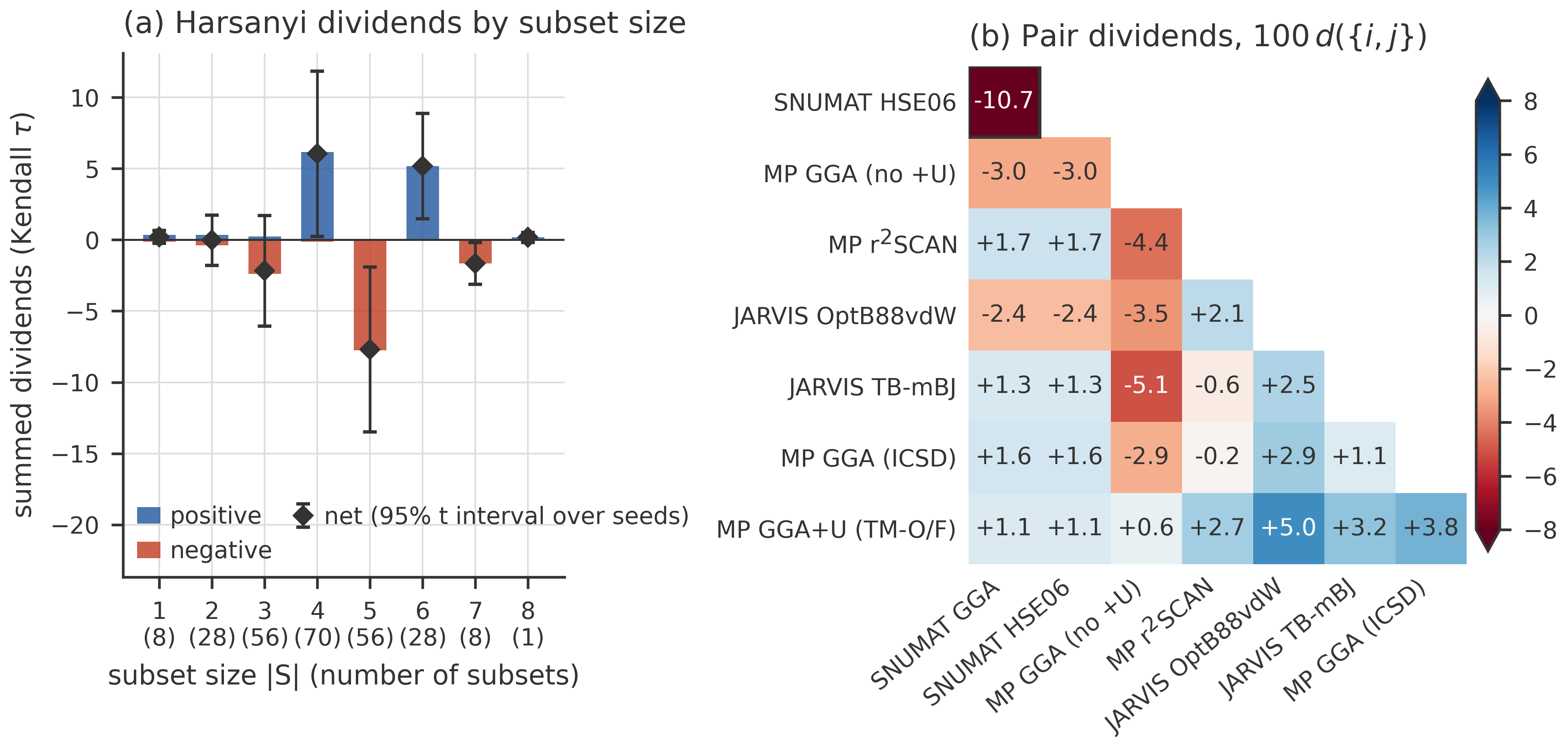}
\caption{In point estimate, the value of integrating databases is lost in large pools rather than between pairs (post hoc eight-donor game); the ten-seed intervals do not separate the two. (a) Harsanyi dividends of the eight-donor campaign summed by subset size: positive and negative mass, and the net with its 95\% $t$ interval over 10 seeds; order sums can exceed the range of $\tau$ because terms of opposite sign cancel. A dividend is what a set of sources adds beyond all its smaller subsets; orders three to eight carry large dividends of alternating sign. (b) Dividend of every pair ($\times 100$, Kendall $\tau$; color scale clipped at $\pm 8$, the outlined cell lies beyond it): 17 of 28 pairs have a positive point estimate, none with an interval above zero; the only pair whose interval lies below zero is the two SNUMAT databases.}
\label{fig:dividends}
\end{figure}

\textbf{The loss appears when many sources are integrated.} The pooled prior acts as a vote whose weight each added source dilutes. The dividends (Figure~\ref{fig:dividends}) give two readings. First, pairs mostly help: 17 of the 28 pairs have a positive dividend in point estimate, though none has an interval above zero, and the only pair whose interval lies below zero is the two SNUMAT databases ($-0.107$, interval $[-0.200,-0.014]$), exactly minus one of them, because either one alone produces the same campaign as both. Second, in the post hoc eight-donor game the loss sits, in point estimate, in interactions of three or more sources: the eight sources alone add $+0.219$ over the campaign without donors and the pairs $-0.031$, but higher-order dividends of alternating sign net $-0.132$, so the full pool keeps only $+0.056$. Over ten seeds the interval of the pair sum includes zero, and the higher orders do not settle the location either: orders four to seven each have an interval that excludes zero, but they alternate in sign, and their net is a point value by subtraction with no interval of its own. This is a reading of the point estimates, not a resolved location. These high-order terms are what a saturating, mean-of-orderings prior produces and are the noisiest entries of the table, so they name no mechanism among particular databases. Past a few well-aligned sources, more stored knowledge, pooled with one vote per source, lowers what the campaign knows early.

\subsection{Databases from one pipeline: redundant knowledge and what a removal screen misses}
\label{sec:redundancy}

\begin{figure}[t]
\centering
\includegraphics[width=\textwidth]{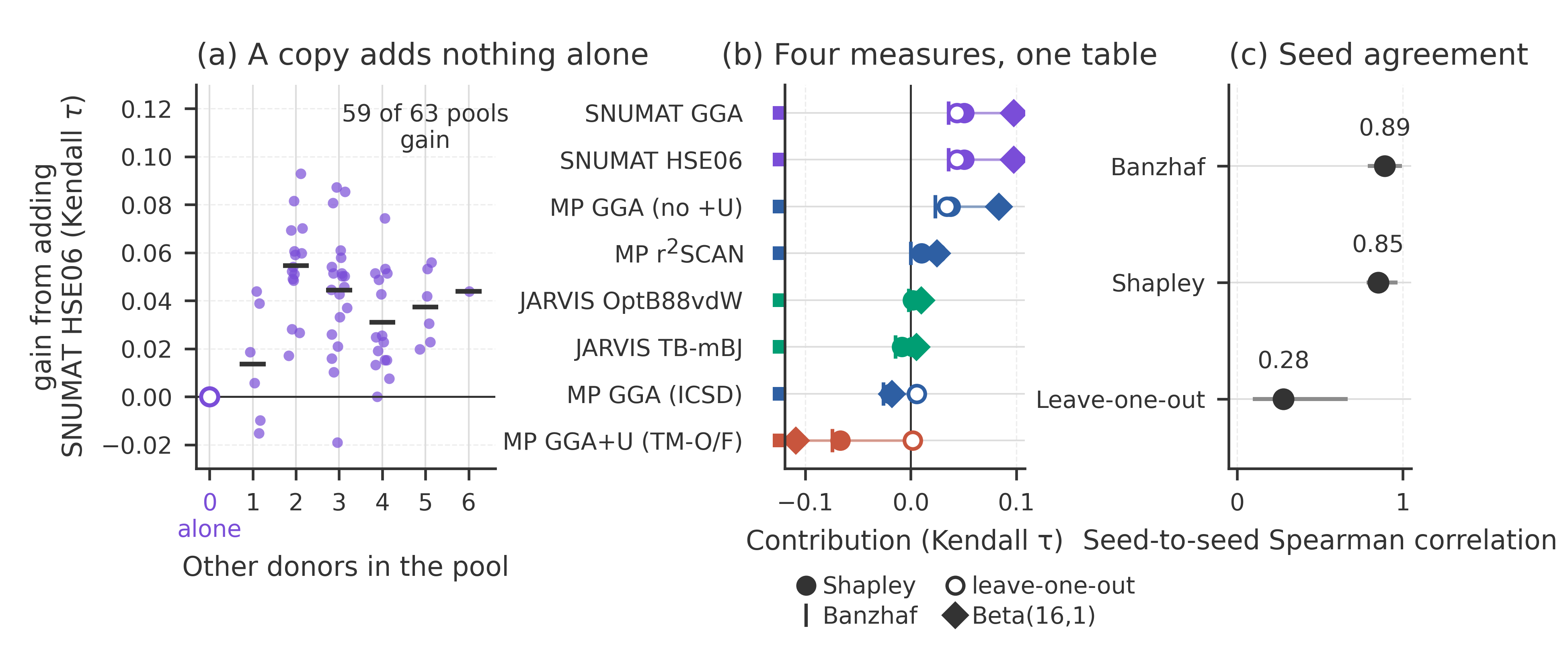}
\caption{Redundant knowledge counts once alone and twice in a pool, and the measures disagree on which source is worst. (a) Gain in early ranking quality from adding SNUMAT HSE06 to each pool that holds SNUMAT GGA, against the number of other donors in the pool: zero when SNUMAT GGA is alone, positive in 59 of the 63 pools with another donor (dots, pools; lines, means). (b) Contribution of each source under four measures on the same 256 measured campaigns; sources ordered by Shapley value, colors by family (MP GGA+U (TM-O/F) subset orange). Only the Shapley values sum to $v(N)-v(\emptyset)$. (c) Agreement of each measure's ranking between two seeds: dot, mean Spearman correlation over the 45 seed pairs; whisker, 95\% bootstrap interval over seeds.}
\label{fig:redundancy}
\end{figure}

\textbf{Two copies of one pipeline count once alone and twice in a pool.} The two SNUMAT databases carry identical family orderings, the database analogue of the structurally equivalent contacts of \citet{Burt1992StructuralHoles}, so the second adds nothing to a pool that holds only the first. In 59 of the 63 pools of the eight-donor game that also held another donor, however, it added to the outcome (mean $+0.040$, up to $+0.093$; Figure~\ref{fig:redundancy}a), because it doubles the weight of the SNUMAT ordering in the vote. With one vote per distinct ordering instead, the second copy adds exactly nothing to every pool that holds the first, the full pool falls to $0.707$, and the contribution of each copy falls to $+0.032$ (Supporting Information Section~S3). What redundancy does therefore depends on how the campaign pools its sources, not on the redundancy alone. The doubling belongs to per-source pooling, and the Shapley value, which pays both copies by symmetry, passes it on; replication-robust valuations of data sources were proposed for this reason \citep{Ohrimenko2019Collaborative,Han2023replication,Xu2021VolumeValuation}.

\textbf{A removal screen misses what the subset runs show.} A group that screens its subscriptions by removing one database at a time from the full pool computes leave-one-out values (Figure~\ref{fig:redundancy}b), and every mean was positive, so the screen would keep all eight donors: at the full pool the TM-O/F subset and the ICSD subset, which still holds GGA records of the same chemistries, mask each other, as redundant ties do \citep{Granovetter1973WeakTies,Burt1992StructuralHoles}. Over the 10 seeds, leave-one-out singled out the source that Shapley ranked lowest in only two. Read from the measured values instead, dropping in turn the source whose removal raises the value most or lowers it least removes the TM-O/F subset, the ICSD subset, JARVIS OptB88vdW and MP $r^{2}$SCAN and reaches the best of all 256 pools after four steps (Supporting Information Section~S3). \textbf{The useful pool size decides which sources to keep.} Weighting small pools, Beta$(16,1)$ gave JARVIS TB-mBJ a positive contribution where Shapley gave a negative one ($+0.005$ against $-0.009$) and the TM-O/F subset its most negative value ($-0.109$), so the measure should match the pool size a group intends to use. Across seeds the Banzhaf ranking, which averages each source's gain uniformly over pools, was the most consistent (mean Spearman correlation between two seeds $0.89$, against $0.85$ for Shapley and $0.28$ for leave-one-out; Figure~\ref{fig:redundancy}c), although its margin over Shapley lies within the bootstrap interval of the difference (Supporting Information Section~S3).

\subsection{Computing what no database holds: progress once the compute is charged}
\label{sec:exploration}

\begin{figure}[t]
\centering
\includegraphics[width=\textwidth]{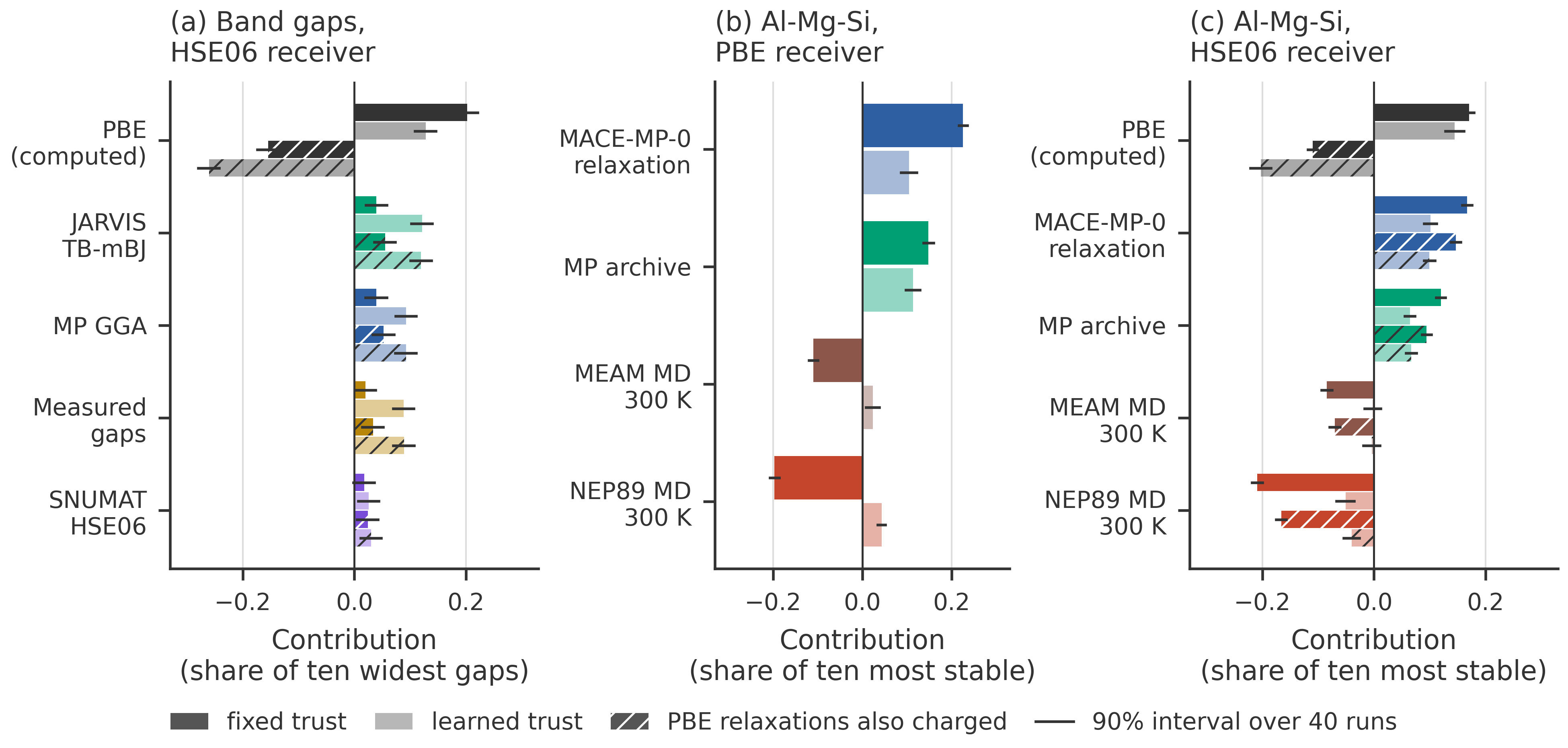}
\caption{Created knowledge advances a campaign until its cost is charged. Charging the PBE relaxations of the whole pool turns the PBE ordering from the largest contribution to a negative one in both systems (hatched bars), the most negative for band gaps under both trust settings and for Al--Mg--Si under learned trust. Value: early share of the ten targets found; all subsets run (32, 16, and 32), each over 20 seeds and batches of one and four. Solid, fixed trust; light, learned trust; hatched, relaxations also charged. (a) Band gaps, HSE06 computed on demand. (b) Al--Mg--Si, PBE receiver, four low-cost sources. (c) Al--Mg--Si, HSE06 receiver, five sources. Whiskers: 90\% percentile-bootstrap intervals over the 40 runs, conditional on the one fixed pool; in (b) and (c) every MD value is drawn from its measured sampling error.}
\label{fig:exploration}
\end{figure}

The games above reuse stored knowledge. Exploration creates it, at a cost \citep{March1991Exploration}. In the first exploring campaign, 171 semiconductors and ordered semiconductor alloys were relaxed with the Perdew--Burke--Ernzerhof (PBE) functional \citep{Perdew1996PBE} in Quantum ESPRESSO \citep{Giannozzi2017QE} and their HSE06 gaps computed at the PBE geometry; 24 of the 32 alloys are held by no database. The goal was the ten widest HSE06 gaps among the 142 candidates without halogens, three of which are such alloys. Five sources advised: the computed PBE gaps and four archives (MP GGA, JARVIS TB-mBJ, SNUMAT HSE06, and the measured compilation), with $v(S)$ the early share of the ten found, over 20 seeds and batches of one and four, all 32 subsets run (Figure~\ref{fig:exploration}a).

\textbf{The source that covered the unexplored candidates advanced the campaign most.} Charging HSE06 compute only, the computed PBE ordering contributed $+0.203$ under fixed trust, against $+0.018$ to $+0.040$ for the archives, because it was the only source that held the alloys; three of the ten targets are such hypothetical ordered alloys. \textbf{Charged its own cost, exploration held the campaign back.} A campaign that borrows the PBE ordering must first relax all 142 candidates, 9.9\% of that campaign's combined PBE and HSE06 compute; charging that cost, the computed source contributed the most negative value ($-0.15$ fixed, $-0.26$ learned trust), and the best subsets held archived sources only. A group that reuses the relaxations in several campaigns charges each an equal share \citep{Littlechild1973Airport}. Under this accounting the PBE ordering contributed a positive value when charged less than $0.57$ of the relaxations (fixed trust; 95\% bootstrap interval $[0.50,0.63]$) or $0.34$ (learned trust; $[0.28,0.39]$), that is, once the relaxations served about two or three campaigns, judged at a quarter of the receiver compute. Over the whole budget the relaxations pay after $1.2$ to $2.2$ campaigns, and with the alloys removed from the pool, or kept as candidates but barred as targets, the fixed-trust multiple becomes $2.1$ or $2.4$, so the break-even is a statement about the early window of this band-gap pool (Supporting Information Section~S11).

The same reversal held for the Al--Mg--Si alloy, where every query was a calculation. The targets were the ten of 168 Al--Mg--Si compounds with the lowest formation energy per atom, 141 of them ordered fcc decorations that no archive holds, with PBE as the receiver and four low-cost sources: the MACE-MP-0 foundation potential \citep{Batatia2024MACEMP} relaxed at 0~K, MD at 300~K with NEP89 \citep{Liang2025NEP89} and with the modified embedded-atom method (MEAM) potential of \citet{Jelinek2012MEAM}, and the Materials Project archive. Under fixed trust MACE-MP-0 contributed $+0.23$ and the archive $+0.15$, while NEP89 MD and MEAM MD held the campaign back ($-0.20$, $-0.11$); the full pool ($0.59$) stayed below MACE-MP-0 alone ($0.88$), and learned trust made all four contributions non-negative. With HSE06 as receiver and PBE as a fifth source, PBE and MACE-MP-0 led ($+0.17$ each) until the PBE relaxations of the pool (9.3\% of the Al--Mg--Si campaign's combined compute) were charged; PBE then fell to $-0.11$ (fixed) and $-0.20$ (learned), and MACE-MP-0 alone was the best subset. The most accurate created knowledge is therefore not the most valuable once the exploration that produced it is paid for, and a statement of what a source contributed has to name the compute it charges.

\subsection{Data providers, meetings, and coordinators: communication structure}
\label{sec:communication}

\begin{figure}[t]
\centering
\includegraphics[width=\textwidth]{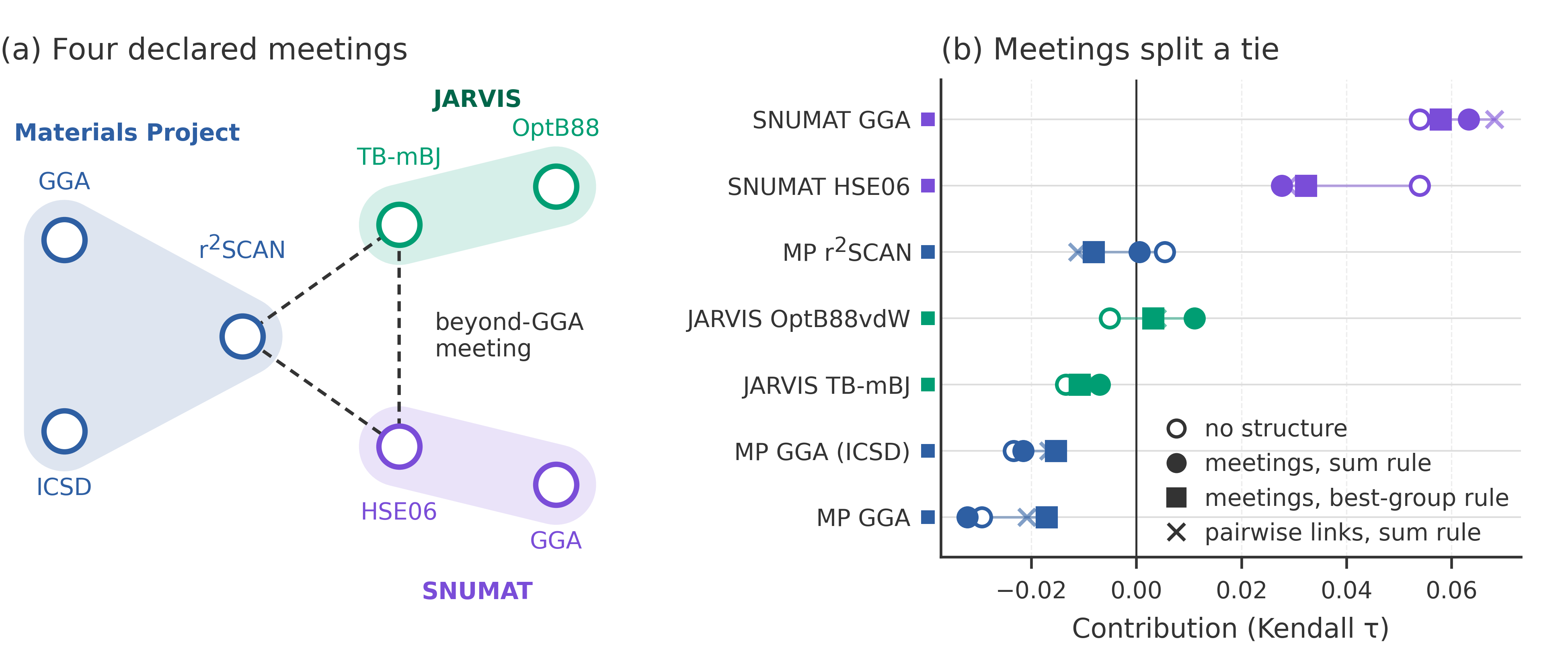}
\caption{Meetings reorder the contributions of the databases. (a) The declared meetings: one per data provider (shaded) and a cross-organization beyond-GGA meeting (dashed) of JARVIS TB-mBJ, MP $r^{2}$SCAN, and SNUMAT HSE06; members combine only through a meeting they share. (b) Contribution of each database to the early ranking quality without structure (Shapley) and under the meetings, valued by the sum rule (a separated team is worth the sum of its groups) and by the best-group rule (worth its best group), with the equivalent pairwise links for comparison. In point estimate (one run), the meetings break the tie between the SNUMAT databases and move JARVIS OptB88vdW from fourth to third under both rules.}
\label{fig:meetings}
\end{figure}

Knowledge moves more readily between units that are tied to each other \citep{Hansen1999SearchTransfer}; here the databases of one organization share a workflow and a release schedule, and this provenance declares meetings. A meeting is a set of members who share results, convened only when all of them are present (Methods). \textbf{Meeting structure reordered the contributions} (Figure~\ref{fig:meetings}). With organization meetings joined by a cross-organization beyond-GGA meeting, the tie between the SNUMAT databases broke in favor of SNUMAT GGA ($+0.063$ against $+0.028$), and JARVIS OptB88vdW rose from rank 4 to rank 3, though its sign is not resolved (positive in 7 of 10 seeds; Supporting Information Section~S4). A screen of a group's sources should therefore be read under the structure the group actually uses.

\begin{figure}[!t]
\centering
\includegraphics[width=0.85\textwidth]{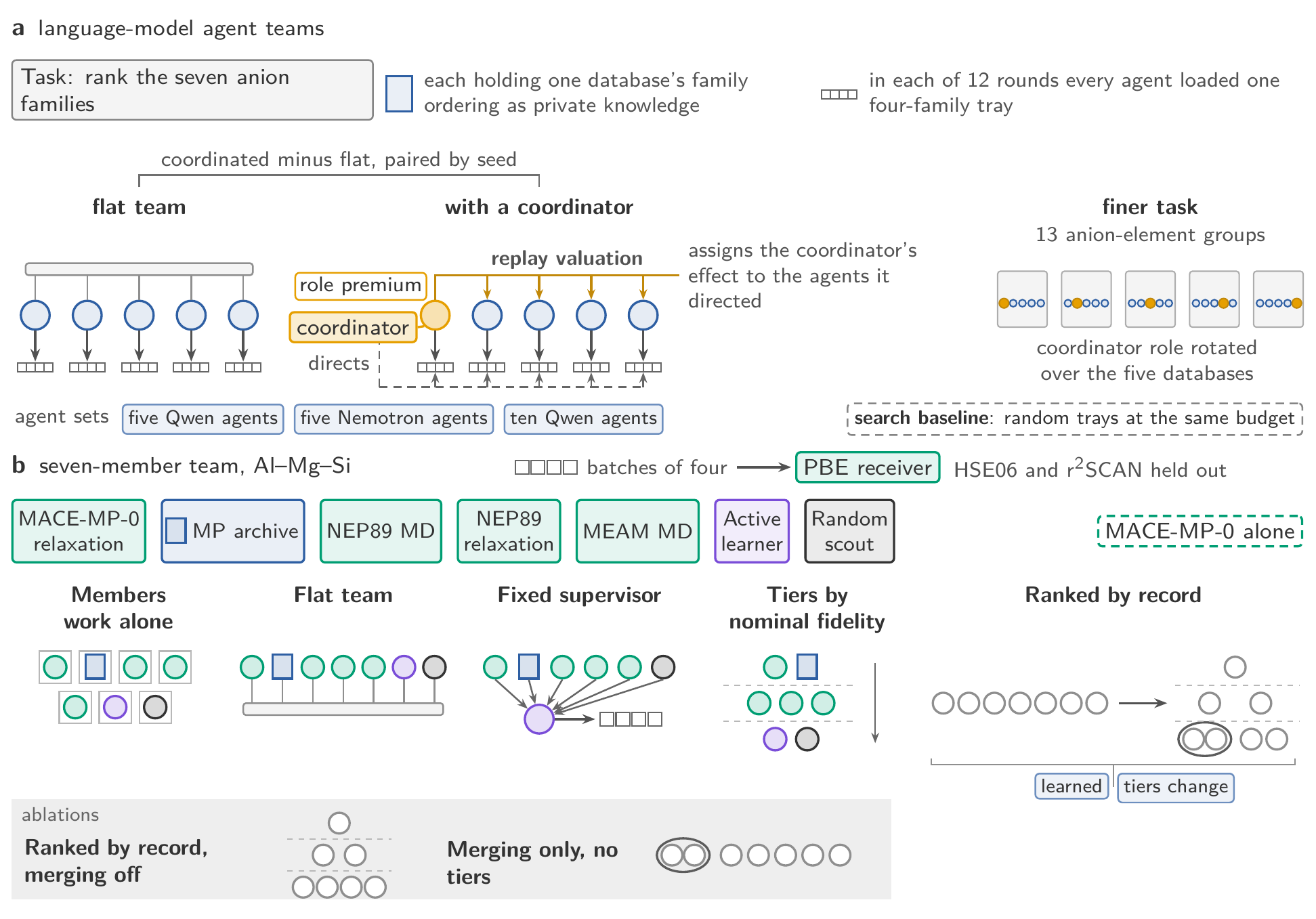}
\caption{The team structures compared. (a) Language-model agent teams: flat, with all agents acting at once on a shared notebook, or led by a coordinator (agent 1) that receives the agents' messages and assigns the trays. The coordinator team is compared with the flat team of the same seed in three agent sets and on a finer task with the coordinator role rotated over the five databases. Random trays at the same budget are the search baseline; replay valuation assigns the coordinator's effect to the agents it directed. (b) The seven-member Al--Mg--Si team in the structures of Figure~\ref{fig:organization}: members working alone, a flat team, a fixed supervisor, tiers by nominal fidelity, and the team ranked by record, which re-ranks members into tiers after every batch and merges agreeing members (ablations: Ranked by record, merging off; Merging only, no tiers; Supplementary Section~S6). MACE-MP-0 alone is the reference.}
\label{fig:teamstructures}
\end{figure}

Figure~\ref{fig:teamstructures} shows the team structures compared: the language-model agent teams (a), discussed here, and the seven-member Al--Mg--Si team (b). Five language-model agents (Qwen3-4B \citep{Yang2025Qwen3}) ranked the seven anion families of the measured compilation, each holding one database's family ordering as private knowledge, a setting closer to opinion aggregation than to a hidden profile, in which the facts that decide the answer are split among members \citep{Stasser1985HiddenProfile}; in each of 12 rounds every agent loaded one four-family tray. \textbf{Flat teams loaded few distinct trays and trailed the random-tray search baseline}: only 5--10\% of their trays were distinct, and the five-agent flat team reached $0.756$ against $0.876$ for random trays at the same budget, a search baseline (sd $0.096$ over 20 random searches of 240 evaluations each), a coordination loss against this baseline rather than against a nominal group (the same members working alone, results pooled), akin to process loss \citep{Steiner1972GroupProcess,Malone1994Coordination,Cemri2025MASfail}.

\begin{figure}[t]
\centering
\includegraphics[width=\textwidth]{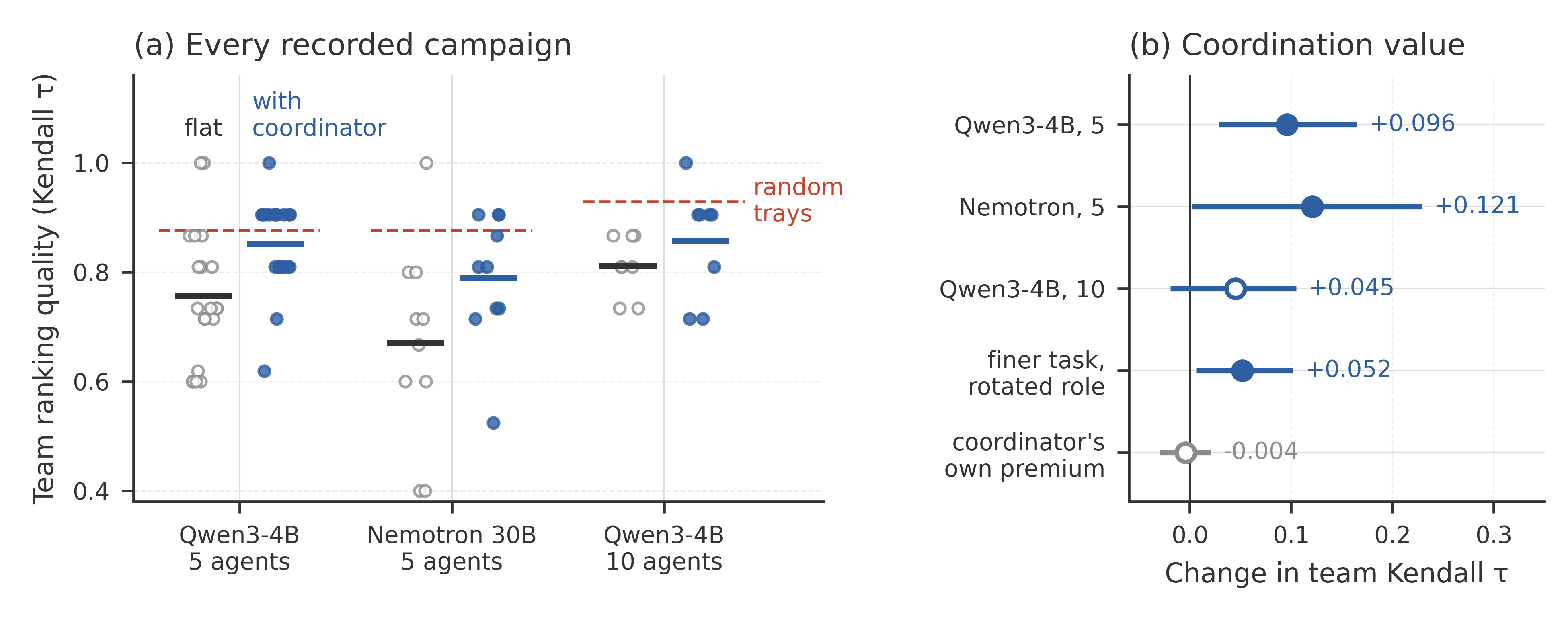}
\caption{A coordinator raises the team; replay valuation assigns the gain to the agents it directs. (a) Team ranking quality of every recorded campaign of language-model agents, flat (open) and with a coordinator (filled), in three agent sets; bars, means; dashed, random trays at the same budget as a search baseline (mean of 20 random searches; sd $0.096$ for five agents, $0.073$ for ten), above every coordinated mean. (b) Interventional coordinator effect on team $\tau$ (coordinated minus flat, paired by seed) in each set and on the finer task with the coordinator role rotated over the five databases, and the coordinator's own share (role premium) under replay valuation; lines, 95\% bootstrap intervals (for the finer task resampled by seed, since each flat run serves all five rotations); filled markers, intervals excluding zero; open markers, intervals including zero (not resolved).}
\label{fig:teams}
\end{figure}

\textbf{A coordinator raised the team, but no coordinated team beat the random-tray search baseline.} Coordinated minus flat team $\tau$, paired by seed, was $+0.096$ $[+0.029,+0.165]$ for five Qwen agents, $+0.121$ $[+0.002,+0.229]$ for five Nemotron agents, a lower bound just above zero, and $+0.052$ $[+0.006,+0.102]$ on a finer task of 13 anion-element groups, a lower bound narrowly clear of zero, with the effect uneven across seeds (10 of 20 seeds positive); that interval resamples whole seeds rather than treating the 100 coordinated-minus-flat differences as independent, because each seed's flat run serves all five coordinator rotations; for ten Qwen agents the interval includes zero ($+0.045$ $[-0.019,+0.105]$), and in no agent set did a coordinated team beat the random-tray search baseline ($0.876$, sd $0.096$, five agents, a spread rather than the coordinator effect; $0.929$, sd $0.073$, ten agents; Figure~\ref{fig:teams}). As a process count, coordinated minus flat, paired by seed, the share of distinct trays rose in every set, by $+1.75$ $[+0.42,+3.25]$ percentage points for five Qwen agents, $+11.67$ $[+8.17,+15.33]$ for five Nemotron agents and $+2.81$ $[+1.56,+3.85]$ for ten Qwen agents, and coordinated teams still repeated most of their trays. The agents were small models, decision order and the database the coordinator held remain confounded with its effect, and six of the twelve predictions made before the runs failed (Supporting Information Section~S5). Replay valuation \citep{Janzing2020FeatureRelevance,Heskes2020CausalShapley} assigns the coordinator's effect to the agents it directed.


\subsection{An adaptive team in the Al--Mg--Si campaign: learned trust}
\label{sec:organization}

\begin{figure}[t]
\centering
\includegraphics[width=\textwidth]{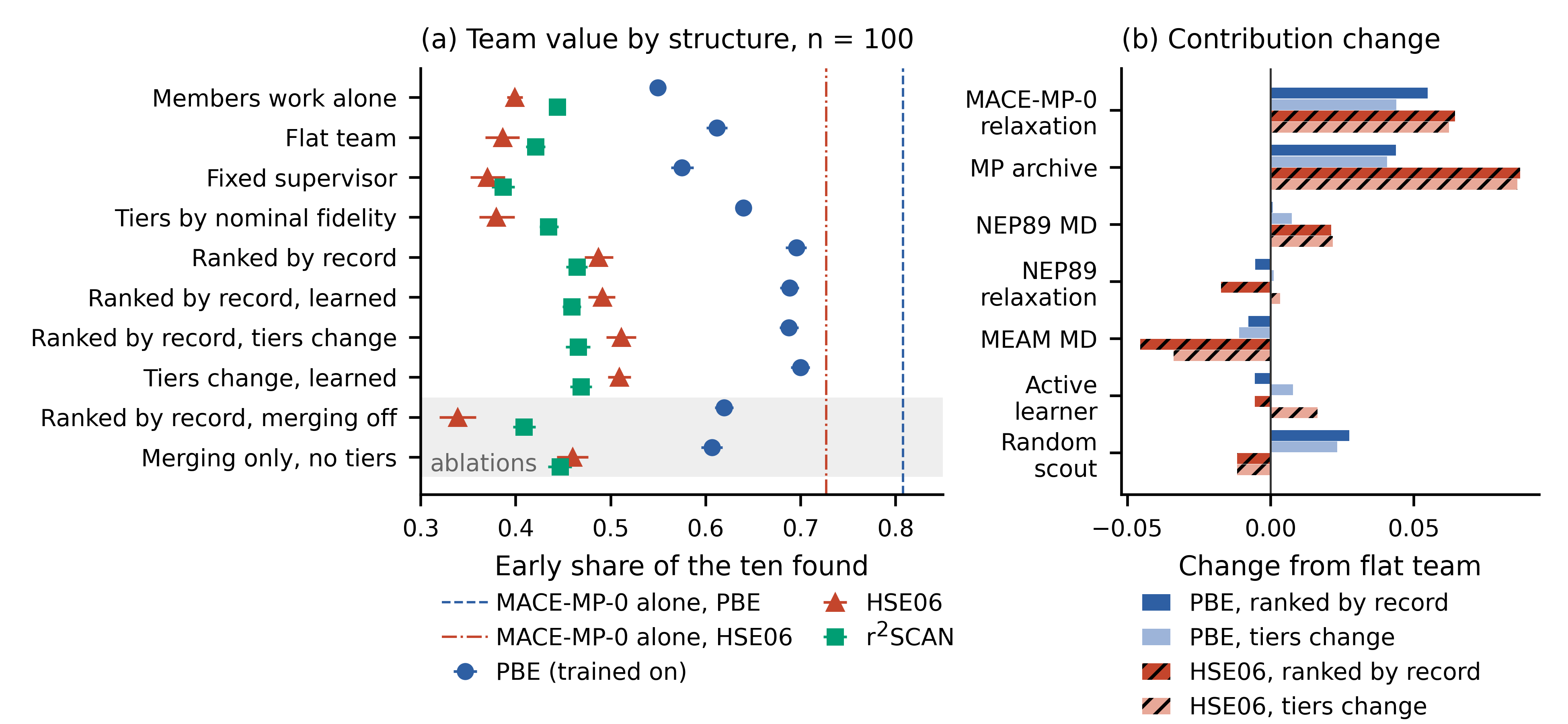}
\caption{A seven-member team ranked by its own record found more of the ten Al--Mg--Si compounds with the lowest formation energy per atom early than any fixed structure, on the receiver its settings were tuned on and on the held-out ones, but fewer than MACE-MP-0 alone. (a) Early share of the ten found (first quarter of the receiver compute), mean over 100 campaigns per structure with the 90\% bootstrap interval, for the PBE receiver (on which learned settings were tuned) and the held-out HSE06 and $r^{2}$SCAN receivers; dashed lines, MACE-MP-0 alone ($0.808$ on PBE in this seven-member setting, against $0.88$ in the four-source game of Section~2.4); shaded rows, ablations of the record-ranked team. (b) Shapley value of each member in the record-ranked team minus its value in the flat team; all 128 sub-teams run per structure. The differences are descriptive, not normalized, since the two structures have different totals.}
\label{fig:organization}
\end{figure}

Fixed structures are declared before a campaign. A group running many advisers at once can instead let its structure adapt while the alloy campaign runs. Seven members ran the Al--Mg--Si campaign together (168 compounds, batches of four, PBE receiver, with HSE06 and $r^{2}$SCAN held out): five low-cost orderings (MACE-MP-0 relaxation, NEP89 MD and relaxation, MEAM MD, and the archive), an active learner with no prior, and a random scout. Every sub-team of every organization was run in 100 campaigns, with seeds that differ between organizations, so intervals that compare organizations are unpaired (Figure~\ref{fig:organization}).

\textbf{A team that learns from its record whom to rely on finds low-formation-energy compounds earlier than any fixed structure, though not as early as its best member alone.} After every batch each member was re-ranked into tiers by how well its scores had anticipated the results it could see, and members whose orderings agreed were merged for the rest of the campaign into one member that proposed once per batch. Scored on the held-out HSE06 receiver, this team, with its default settings, found $0.49$ of the ten compounds with the lowest formation energy per atom in the first quarter of the compute, against at most $0.40$ for fixed teams (a gap of $0.09$, 90\% bootstrap interval $0.07$--$0.11$ over 100 campaigns per structure), but MACE-MP-0 alone found $0.73$. On the PBE receiver, on which the learned settings were tuned, the order was the same ($0.70$ with default settings, against $0.64$ for tiers assigned by nominal fidelity and $0.61$ for a flat team). Merging carried most of the gain: once the potentials had lost trust in their own orderings they proposed from the same composition surrogate, and merging them returned their duplicated batch slots to the leader. This is a simple analogue of the credibility part of transactive memory \citep{Wegner1987TransactiveMemory,Lewis_2003}; it is not strong synergy, since the team stayed below its best single member \citep{Hill1982GroupIndividual,Larson2010Synergy}.

\textbf{In these campaigns, combining groups as records accumulated never lowered the outcome; separating them lowered it in most settings.} When a coordinator could fuse, nest, split, or detach groups, combining raised the outcome above the fixed structure with the same start in 10 of 12 settings and separating lowered the mean in all 12; members allowed to move to the group whose ordering they agreed with raised the early value by $+0.026$ to $+0.071$ across the three receivers, and these sharing rules kept the results of NEP89 MD away from the groups. The settings share members and starts, so these counts are descriptive, exploratory, and hypothesis-generating (Methods; Supporting Information Sections~S6--S9).

\section{Discussion}
\label{sec:discussion}

The measured campaigns suggest four working rules for a materials group, each consistent with an organization-science construct that the campaigns illustrate rather than test. Each rule is a hypothesis from the receivers and budget windows measured here.

\textbf{Pool a few well-aligned databases} (knowledge integration has an optimum \citep{Grant_1996}; in data integration, adding sources can lower quality \citep{Dong_2012}, and an ensemble of a few models can beat one of all \citep{Zhou_2002}). In both band-gap games the full pool scored below smaller pools, and the loss arose when many sources were integrated. A campaign should integrate the few sources whose orderings fit the receiver, chosen on the receiver's own results, rather than every source it can access. Because each pool is run as a campaign, its value is an outcome rather than a score of agreement, and a post-hoc split can locate a block of records that holds a loss, a hypothesis for the next campaign.

\textbf{Treat databases from one pipeline as one source} (redundancy depends on integration, as in groups pooling information \citep{Hinsz_1997}, and copying between sources inflates their votes in truth discovery \citep{Dong_2009}). Two databases from one pipeline repeat each other alone and double each other's weight in a pool, and a removal screen sees only the full pool, where one copy masks the other. A group should integrate them as one source unless it deliberately weights them: one vote per distinct ordering removes the double weight, whereas valuing them as one player or as one Owen union does not, because the second vote sits in the pooled prior itself.

\textbf{Charge the compute of created knowledge} (exploration must be paid for). Early in the budget, created knowledge that covers candidates no database holds advanced the campaign most, and the same knowledge held it back once the calculations that produced it were charged. A group deciding whether to compute its own low-level ordering should therefore do so when the ordering covers candidates the databases lack and the calculations will be reused across campaigns, because charged to one campaign they cost more early budget than they return; exploration that pays only over a horizon longer than one campaign is the trade-off \citet{March1991Exploration} describes.

\textbf{When the best member is not known in advance, learn from the group's own campaigns which members and sources to draw on} (trust learned from the record). Meetings reordered the contributions of the databases, a coordinator raised agent teams, and a team that learned from its record whom to rely on found low-formation-energy compounds earlier than every team whose weights were set in advance. Neither baseline was beaten: no language-model team reached the random-tray search baseline ($0.876$, sd $0.096$ over 20 random searches of 240 evaluations each), and MACE-MP-0 alone found more targets early than the adaptive team. Learning from the record does not replace finding the best member.

\textbf{Cost of the method.} Running every subset costs $2^{n}$ campaigns: a few core-hours for the 256-subset band-gap game, which replays archived records, and about 270 core-hours for the Al--Mg--Si PBE pool, which replays calculations made once (Supporting Information Section~S11). A larger group can remove sources one at a time and keep the best pool seen, as in Section~\ref{sec:redundancy}, in about $n^2/2$ campaigns; estimate Shapley values from sampled orderings (Supporting Information Section~S1); or reweight its members by their record during the campaign.

Data valuation assigns Shapley-type values to training points and data sources \citep{Ghorbani2019DataShapley,Jia2019EfficientDataValuation,Ghorbani2020DistributionalDataValuation,Rozemberczki2022ShapleyML}, and Beta Shapley and Data Banzhaf were proposed because Shapley rankings are noisy and weight large subsets heavily \citep{Kwon2022BetaShapley,Wang2023DataBanzhaf}; the higher seed consistency of the Banzhaf ranking and the change of sign of JARVIS TB-mBJ under Beta$(16,1)$ agree with that motivation, here on run campaigns rather than retrained models. Negative transfer is widely reported \citep{Zhang2023NegativeTransferSurvey}; here it appeared in large pools rather than in pairs. Benchmarks of self-driving laboratories report the acceleration of a whole loop \citep{Rohr2020BenchmarkingSL,Liang2021BenchmarkingBO}; subset campaigns decompose it into members, at the cost of $2^{n}$ runs. Computational organization theory studies these constructs in stylized agent-based models \citep{March1991Exploration,LazerFriedman2007Network}; run materials campaigns add real knowledge sources whose every subset's outcome is run rather than modeled.

\textbf{Limitations.} The band-gap campaigns replay archived records, and the structural results rest on one fully measured game (one measured receiver, an equal-vote prior, the early window); whether the optimum pool size and the TM-O/F negative transfer hold for another receiver, such as SNUMAT HSE06, is untested. The language-model teams were compared with a search baseline, not with the control organization science would use, a nominal group of the same members working alone with their results pooled afterwards. The eight-donor game is post hoc, since the GGA+U split and the lowered entry threshold came after the seven-source game had been run, and its split is confounded with chemistry. The computed source's lead rests partly on hypothetical ordered alloys; the computed band-gap game has one HSE06 receiver without spin--orbit coupling, the Al--Mg--Si receivers are calculations, and formation energies at temperature omit entropy. The agents were small models (Qwen3-4B, and Nemotron 30B-A3B in one set), so duplication and the gap to the random-tray search baseline (sd $0.096$ over 20 random searches) may reflect model capability as well as team structure, and the coordinator effect is not separated from turn order or from the database the coordinator held. The adaptive team had seven computational members and DFT receivers; it is a fixed re-ranking rule with set thresholds, and nothing in it closes the loop to an experiment. The constructs are read by analogy: the members are knowledge sources and agents without goals of their own, and people were not modeled.

\section{Conclusions}
\label{sec:conclusions}

When every subset of a materials campaign's knowledge sources is run, what carries the campaign's knowledge forward can be read from the campaign record rather than assumed. Early in one band-gap campaign, with one measured receiver and an equal-vote prior, integrating more databases helped up to a few well-aligned sources and then held the campaign back (Kendall $\tau$ $0.737$ for all seven databases against $0.851$ for the best three). Databases from one pipeline counted once alone and twice in a pool. Created knowledge advanced the campaign until its cost was charged. A team that learned from its record whom to rely on did better than structures fixed in advance, but not better than its best member. Because the values are read from runs on the receiver's own measurements, what carries to a new campaign is the rule (which kinds of source to pool, merge, or charge for), not the numbers.

\section{Methods}
\label{sec:methods}

\paragraph{Band-gap campaigns.} The archive held 261{,}053 records from eight sources, the measured gaps of \citet{Zhuo2018BandGapML} among them, version-locked by a manifest hash. In the eight-donor game the MP GGA records were split by the Materials Project GGA+U rule (Co, Cr, Fe, Mn, Mo, Ni, V, or W with O or F) into the TM-O/F subset (31{,}828) and MP GGA (no +U); a donor entered the prior when it held at least two anion families of at least 30 records each (four in the seven-source game). Each archived record was drawn at most once per campaign. Rankings were fitted with the Plackett--Luce model \citep{Plackett1975Permutations,Luce1959IndividualChoice}; the donor prior, the mean of the coalition's orderings with one vote per source, entered with trust weight $\lambda$, fixed or chosen by its prequential log-score \citep{Dawid1984Prequential} (Supporting Information Section~S1).

\paragraph{Computed campaigns.} The band-gap pool held 171 compounds, relaxed with PBE in Quantum ESPRESSO 7.4 \citep{Giannozzi2017QE} with HSE06 single points at the PBE geometry; its MgCaO and ZnBeO alloys are hypothetical. The Al--Mg--Si pool held 168 compounds. Each step computed four compounds chosen by blending the pooled donor ordering with a composition surrogate fitted to the receiver values so far. MD formation energies (NEP89 in GPUMD \citep{Fan2022GPUMD}, MEAM in LAMMPS \citep{Thompson2022LAMMPS}) were time averages; in each of 20 runs every MD value was drawn from its measured standard error. Every subset paid the PBE relaxation of each compound it queried, and a subset that borrowed the PBE ordering paid that of the whole pool, so no relaxation was charged twice. $v(S)$ was the mean share of the receiver's ten best candidates found over the first quarter of its compute (Supporting Information Section~S11).

\paragraph{Contribution measures.} The players have no preferences or strategies, so no bargaining claim is made. Beside the Shapley value \citep{Shapley1953Value}, leave-one-out \citep{Cook1977Influential} and Harsanyi dividends \citep{Harsanyi1963Bargaining} defined in the Results, the Banzhaf value averages the same marginal changes uniformly over subsets \citep{Dubey1979Banzhaf} and Beta$(\alpha,\beta)$ semivalues reweight subset sizes \citep{Kwon2022BetaShapley}. A communication structure restricts the game: with meetings $H$, only meetings contained in $S$ convene, so a meeting of three is not three pairwise links, and the value of $S$ is the sum (sum rule \citep{Myerson1977Graphs,Myerson1980Conference}) or the maximum (best-group rule) of the zero-normalized values $v(C)-v(\emptyset)$ of the groups $C$ the convened meetings link, so the empty team is worth zero and, under the best-group rule, the efficient values sum to $\max_C[v(C)-v(\emptyset)]$. Teams split into fixed groups use the Owen value \citep{Owen1977APrioriUnions}. All values were computed exactly by enumerating every subset (definitions and axioms in Supporting Information Section~S1).

\paragraph{Language-model teams.} Agents ran Qwen3-4B-Instruct-2507 or Nemotron 3.5 Lightning 30B-A3B, served with vLLM 0.29.0 \citep{Kwon2023vLLM} at temperature 0.7; each held one database's family ordering, and meetings were read from messages and notebook citations. Twelve predictions were recorded in time-stamped commits before the campaigns, and six held (Supporting Information Sections~S5 and S12).

\paragraph{Statistics.} In the eight-donor game intervals are $t$ intervals over 10 seeds, with Holm-adjusted tests of a zero Shapley value \citep{Holm1979Sequential}; elsewhere they are percentile bootstrap intervals \citep{Efron1979Bootstrap} over seeds or campaigns (paired by seed where the text says so, resampled by seed where one flat run serves several coordinated runs, and unpaired between organizations), 95\% for the seven-source and agent-team games and 90\% for the computed and organization games. The organization games are uncorrected for multiplicity and exploratory; the seven-source intervals and those of agent sets S2 and S3 (8 to 10 seeds) are descriptive (sample sizes in Supporting Information Section~S1). All intervals are conditional on the fixed pool of sources, candidates and receivers: they describe variation over seeds or campaigns, not over other pools.

\section*{Data availability statement}
Code, campaign logs, subset values, the version-locked archive manifest, and the scripts that draw every figure are openly available on Zenodo as release v1.3.0 (\url{https://doi.org/10.5281/zenodo.23413476}; this DOI resolves to the latest version), with code under the MIT License and data under CC BY 4.0, following the FAIR principles \citep{Wilkinson2016FAIR}. The band-gap archive is identified by its build fingerprint: it is rebuilt with the kit in this release, and the rebuilt manifest has SHA-256 \texttt{f2158ecdc78e022926f6b05a40433152}\allowbreak\texttt{f069d706ec4d2cb706977e99c7a531f3}; no separate record of the archive is deposited.

\section*{Funding}
Taylor D. Sparks and Kinston Ack{\"o}lf acknowledge support from the Army Research Office Materials Design, under Contract No. W911NF-23-1-0333. Taylor D. Sparks acknowledges support from the National Science Foundation through the AURORA Programmable Cloud Lab (NSF Award No. 2607535).

\section*{Conflict of interest}
The authors declare no conflict of interest.

\section*{Use of generative AI}
The language-model agents studied in the agent-team campaigns are part of the research and are described in Methods.

\section*{Author contributions}
Kinston Ack{\"o}lf: Conceptualization, Data curation, Formal analysis, Investigation, Methodology, Software, Validation, Visualization, Writing--original draft, Writing--review \& editing. Taylor D. Sparks: Funding acquisition, Resources, Supervision, Writing--review \& editing.
\bibliographystyle{unsrtnat}
\bibliography{references}

\clearpage
\singlespacing
\setcounter{section}{0}
\setcounter{figure}{0}
\setcounter{table}{0}
\setcounter{equation}{0}
\renewcommand{\thesection}{S\arabic{section}}
\renewcommand{\thetable}{S\arabic{table}}
\renewcommand{\thefigure}{S\arabic{figure}}
\renewcommand{\theequation}{S\arabic{equation}}
\renewcommand{\theHsection}{S\arabic{section}}
\renewcommand{\theHtable}{S\arabic{table}}
\renewcommand{\theHfigure}{S\arabic{figure}}
\renewcommand{\theHequation}{S\arabic{equation}}
\begin{center}
{\LARGE Supporting Information}
\end{center}
\addcontentsline{toc}{section}{Supporting Information}

\section*{How this Supporting Information is organized}
Sections S1--S3 cover the band-gap campaigns: the ranking model and estimators (S1), the seven-source and eight-donor games with their selection checks and controls (S2), and the contribution measures compared (S3). Sections S4 and S5 cover communication structure and the language-model agent teams. Sections S6--S10 cover the adaptive Al--Mg--Si teams and their variants. Section S11 gives the extended Al--Mg--Si games, the cost-share sweep and the composition fits, and S12 lists code, data and result files.

\section*{Organization-science constructs and the quantities that stand for them}
The main text reads each campaign question through one organization-science construct, by analogy. Table~\ref{tab:constructs} lists, for each construct, the campaign quantity that stands for it and where it is reported.

\begin{table}[h]
\centering
\small
\caption{Constructs and campaign quantities. Sections refer to the main text (2.x) and to this Supporting Information (S).}
\label{tab:constructs}
\begin{tabular}{@{}p{0.27\textwidth}p{0.5\textwidth}p{0.15\textwidth}@{}}
\toprule
Construct & Quantity in the campaigns & Where \\
\midrule
Knowledge integration; negative transfer & best value by pool size; full pool against the best pool; Shapley value of each database and of the MP GGA+U (TM-O/F) & 2.2; S2, S3 \\
Redundancy; structural equivalence & increment of the second SNUMAT copy; leave-one-out against Shapley; one vote per ordering & 2.3; S3 \\
Exploration and its cost & contribution of the computed PBE ordering under HSE06-only and charged cost; cost-share sweep & 2.4; S11 \\
Communication structure (meetings) & restricted-game values under declared and recorded meetings (sum and best-group rules) & 2.5; S4 \\
Coordination; process loss & coordinator effect paired by seed; share of distinct trays; random-tray search baseline with its spread & 2.5; S5 \\
Organizational learning (trust learned from the record) & early share of the record-ranked team against fixed tiers and a flat team; ablations; held-out receivers & 2.6; S6 \\
Boundary spanning & members that move between groups or sit in two & 2.6; S9 \\
\bottomrule
\end{tabular}
\end{table}

\section{Ranking model, trust, and sampling}
\label{si:model}
\subsection{The Plackett--Luce fit and the anchored update}
\label{app:math-pl}

An observed event $e$ that ranks its $k$ participants $r_1 \succ r_2 \succ \dots \succ r_k$ contributes the sequential-choice likelihood
\begin{equation}
P(e) \;=\; \prod_{s=1}^{k-1} \frac{w_{r_s}}{\sum_{l=s}^{k} w_{r_l}}, \qquad w_i = \exp(\theta_i) > 0,
\end{equation}
fitted by Hunter's MM iteration \citep{Hunter2004MM}, which collects over all stages of all events the statistics
\begin{equation}
W_i = \text{total weight of stages won by } i, \qquad
D_i = \sum_{\substack{\text{stages with remaining set } R \ni i}} \frac{\mathrm{weight}_e}{\sum_{j \in R} w_j},
\end{equation}
and updates $w_i \leftarrow (W_i + \lambda_0)/(D_i + \lambda_0/w_{\mathrm{ref}})$ with a light damping prior $\lambda_0 = 0.01$ anchored at $w_{\mathrm{ref}} = 1$. For $k = 2$ the fixed point of the undamped update satisfies $w_a/w_b = \mathrm{wins}_a/\mathrm{wins}_b$, exactly Bradley--Terry \citep{BradleyTerry1952RankAnalysis}. The anchored MAP fit used when prior and data are both present replaces the damping anchor by the mean-centered prior $\pi_i = \exp(\theta^{\mathrm{prior}}_i)$ with trust weight $\lambda \ge 0$,
\begin{equation}
w_i \;\leftarrow\; \frac{W_i + \lambda}{D_i + \lambda / \pi_i},
\end{equation}
The limiting regimes are as follows. With no data, the fixed point is $w = \pi$. With a flat prior, the update reduces exactly to the unanchored model. As events accumulate, the prior's influence decays as $\lambda / n_{\mathrm{events}}$. When the trust weight is learned, a grid of candidate $\lambda$ values is maintained. Each incoming event is scored under the model fitted to all previous events at each $\lambda$ (prequential scoring, \citealp{Dawid1984Prequential}). The cumulative log-score selects $\lambda$, with ties broken toward smaller $\lambda$.

\subsection{Monte-Carlo estimators and their observed convergence}
\label{app:math-mc}

The sampled-permutation Shapley estimator draws $m$ uniform permutations $\sigma^{(1)}, \dots, \sigma^{(m)}$ of $N$ and averages each player's marginal contribution at its arrival position,
\begin{equation}
\hat{\phi}_i \;=\; \frac{1}{m} \sum_{t=1}^{m} \Bigl( v\bigl(\mathrm{pre}_i(\sigma^{(t)}) \cup \{i\}\bigr) - v\bigl(\mathrm{pre}_i(\sigma^{(t)})\bigr) \Bigr),
\end{equation}
where $\mathrm{pre}_i(\sigma)$ is the set of players preceding $i$ in $\sigma$; the Banzhaf estimator averages marginal contributions over uniformly sampled subsets instead. On the fully measured seven-source game (MP GGA unsplit, Section~\ref{si:databases}), whose 128 coalition values are all cached, the median maximum absolute value error of the sampled Shapley estimate is $8.57\times 10^{-4}$ at $m = 3{,}000$ permutations. By the symmetry axiom, two players whose exact values are equal (the two SNUMAT donors, whose class orderings over the archive are identical; the HSE06 field is a separate calculation, not a shift of the GGA value: over the 10{,}481 paired records it lies $1.08$~eV above GGA on average with a standard deviation of $0.50$~eV and no equal pair, record-level Spearman $\rho = 0.94$, yet the order of the anion-family medians is unchanged) cannot be strictly ordered by a sampled estimate at any $m$, because the difference of their estimands is exactly zero. The truncated variant of \citet{Ghorbani2019DataShapley} sets the marginal contribution of every later player in a permutation to zero once the value of the prefix is within a tolerance of $\vN$; its error on the same game is reported in Section~\ref{si:measures}.

\subsection{Extended methods}

Each step buys the gaps of a tray of four candidates; before the first purchase the databases' own orderings of the families serve as a starting guess, and each purchase corrects that guess.

The reference ranking is the receiver's own fixed order of family median gaps (oxides 3.33, nitrides 3.20, and halides 3.02~eV at the top); its uncertainty is not propagated into the intervals. A purchased evaluation returns the receiver's order for a tray of four candidates, one archived record per candidate, and no record is drawn twice within a campaign: a campaign that can afford 240 evaluations draws 240 different archived measurements, none repeated.

Or it can estimate Shapley values from sampled orderings of its sources, running only the subsets those orderings visit; on the seven-source game, $1{,}000$ sampled orderings gave a median maximum error of $0.0014$ and no sign reversal in 50 repeats (Section~S1, which also reports $3{,}000$ orderings); with seven sources these orderings still visit nearly every one of the 128 subsets, so this checks the accuracy of the estimator, and the saving in campaigns appears only for larger groups.

\paragraph{Band-gap archive and campaigns.} The archive held 261{,}053 records from eight sources, version-locked by a manifest hash: MP GGA/GGA+U (95{,}920 records), MP GGA restricted to ICSD-provenance compounds (27{,}347), MP $r^{2}$SCAN (14{,}672), JARVIS OptB88vdW (75{,}993) and TB-mBJ (19{,}805), SNUMAT GGA and HSE06 (10{,}481 each), and the measured compilation of \citet{Zhuo2018BandGapML} (6{,}354). In the eight-donor game the MP GGA records were split by the Materials Project GGA+U rule, a formula holding Co, Cr, Fe, Mn, Mo, Ni, V, or W together with O or F, into the MP GGA+U (TM-O/F) subset (31{,}828) and MP GGA (no +U; 64{,}092); the ICSD subset was not split, and 5{,}433 of its records fall under the rule.

Candidate identity was the reduced formula with space group; the measured compilation lists no space group, so its records were matched by reduced formula alone (973 of its 4{,}921 formulas have more than one computed space group in the archive), which leaves the anion family of every record unchanged; no source was designated ground truth.

Rankings were fitted with the Plackett--Luce model \citep{Plackett1975Permutations,Luce1959IndividualChoice} by minorization--maximization \citep{Hunter2004MM}; a donor prior, the mean of the coalition's donor orderings with one vote per source (so two sources with one ordering carry it twice), entered as an anchor with trust weight $\lambda$, fixed or chosen after every event by its prequential log-score \citep{Dawid1984Prequential} (Section~S1).

\paragraph{Contribution measures.} Each game is a transferable-utility attribution game with $v(\emptyset)\neq 0$. Its players are knowledge sources or members without preferences or strategies, so no bargaining, core or equilibrium claim is made, and each value is an allocation rule chosen for its axioms: the Shapley value is the one rule that is efficient, symmetric and additive and gives a member that changes no outcome zero \citep{Shapley1953Value}.

Harsanyi dividends are $d(S)=\sum_{T\subseteq S}(-1)^{|S|-|T|}v(T)$, so that $v(S)=\sum_{T\subseteq S}d(T)$ and $\phi_i=\sum_{S\ni i}d(S)/|S|$ \citep{Harsanyi1963Bargaining,Rota1964Moebius}. Pairs are compared by their dividend rather than by the Shapley interaction index \citep{Grabisch1999Interaction}, which sums the dividends of every coalition that holds the pair, each divided by its size less one, because the dividend isolates what the pair adds beyond its two members alone. A communication structure restricts the game: with meetings $H$ (sets of members), only meetings contained in $S$ convene, $S$ splits into groups linked by convened meetings, and the restricted value of $S$ is the sum (sum rule \citep{Myerson1977Graphs,Myerson1980Conference}; a meeting is Myerson's conference structure \citep{Myerson1980Conference,VandenNouweland1992Hypergraph}) or the maximum (best-group rule) of the zero-normalized group values $v(C)-v(\emptyset)$, so the empty team is worth zero and under the best-group rule the values sum to $\max_C [v(C)-v(\emptyset)]$; the contribution is the Shapley value of the restricted game.

Teams split into fixed groups use the Owen value \citep{Owen1977APrioriUnions}, and groups that change during a campaign the level structure value \citep{Winter1989Levels}, averaged over the structures in force; this average has no axiomatic characterization and is used as a robustness check.

With 8 to 10 seeds or campaigns a percentile interval tends to under-cover, so the seven-source intervals, used for the meetings, and the agent-team intervals of sets S2 and S3 are read as descriptive.

\clearpage
\section{The eight-donor campaign in full}
\label{si:databases}
The eight donors of the follow-up game in main-text Section 2.2 advised its measured-gap campaign; the split and the lower entry threshold were chosen after the seven-source game had been run, so this game is post hoc. They are the seven databases of the archive with the Materials Project GGA records split at the Materials Project GGA+U rule: the records whose formula holds Co, Cr, Fe, Mn, Mo, Ni, V, or W together with O or F (31{,}828 of the 95{,}920 MP GGA records) form the donor MP GGA+U (TM-O/F) subset, and the other 64{,}092 form MP GGA (no +U); the ICSD subset was not split. A donor enters the prior when it holds at least two anion families of at least 30 records. The seven-source game used four, which the TM-O/F subset, confined to the oxide and halide families, does not meet; on the unsplit archive both thresholds keep the same seven sources. Every one of the 256 donor subsets was run with 10 matched seeds (pairwise arm, early window), so the value table is a lookup into campaign logs. The campaign without donors reached $v(\emptyset)=0.6949$ and the full pool $v(N)=0.7509$. Intervals in this section and the next are $t$ intervals over the 10 seeds (9 degrees of freedom), and the $p$ values for a zero Shapley value are Holm-adjusted over the eight donors. With seven anion families one exchange of two families in the fitted ranking moves $\tau$ by $2/21=0.095$; smaller differences are averages of such steps over checkpoints and seeds.

\textbf{The split leaves the rest of the game in place.} The Shapley values of the two halves of MP GGA sum to $-0.0292$, against $-0.0294$ for MP GGA unsplit in the seven-source game, and the other donors keep their signs except JARVIS OptB88vdW, which moves from $-0.005$ to $+0.002$ (Table~\ref{tab:sevensource}). The TM-O/F subset holds the campaign back wherever other donors can carry it: its mean marginal change is $-0.116$ alone, between $-0.101$ and $-0.054$ with one to five other donors, $-0.022$ with six, and $+0.002$ when it joins the other seven. At the full pool the TM-O/F subset and the ICSD subset, which still holds the GGA records of the TM-O/F chemistries with ICSD provenance, mask each other: removing either alone changes the value little (leave-one-out $+0.002$ and $+0.006$), but with the ICSD subset absent the leave-one-out value of the TM-O/F subset is $-0.025$, with the TM-O/F subset absent that of the ICSD subset is $-0.021$, and removing both raises the full pool by $0.019$.

\begin{table}[h]
\centering
\caption{Shapley values in the seven-source game (MP GGA unsplit, 128 subsets) and in the eight-donor game (256 subsets), with the 95\% $t$ interval over the 10 seeds for the eight-donor values. Full pool: $0.7368$ and $0.7509$; $v(\emptyset)=0.6949$ in both.}
\label{tab:sevensource}
\small
\begin{tabular}{@{}lrrr@{}}
\toprule
Database & Seven sources & Eight donors & 95\% interval \\
\midrule
JARVIS TB-mBJ & $-0.0134$ & $-0.0086$ & $[-0.020, +0.002]$ \\
JARVIS OptB88vdW & $-0.0051$ & $+0.0019$ & $[-0.008, +0.012]$ \\
MP GGA (unsplit) & $-0.0294$ & & \\
\quad MP GGA (no +U) & & $+0.0376$ & $[+0.023, +0.053]$ \\
\quad MP GGA+U (TM-O/F) & & $-0.0668$ & $[-0.091, -0.042]$ \\
MP GGA (ICSD) & $-0.0234$ & $-0.0194$ & $[-0.040, +0.001]$ \\
MP $r^{2}$SCAN & $+0.0054$ & $+0.0104$ & $[-0.001, +0.022]$ \\
SNUMAT GGA & $+0.0539$ & $+0.0505$ & $[+0.027, +0.074]$ \\
SNUMAT HSE06 & $+0.0539$ & $+0.0505$ & $[+0.027, +0.074]$ \\
\bottomrule
\end{tabular}
\end{table}

\paragraph{Selection check, seven-source game.}\label{sec:selcheck7} The best pool of three in the seven-source game (MP $r^{2}$SCAN, SNUMAT GGA and SNUMAT HSE06; $\tau=0.851$ against $0.737$ for all seven, early window) was chosen again on five of the ten seeds and scored on the other five, in both directions. Both halves chose the same pool, so the held-out mean equals the ten-seed mean and the interval below is the in-sample interval, not adjusted for the selection, and the held-out pool exceeded the full pool by $+0.115$ (95\% $t$ interval $[+0.037,+0.193]$, 9 degrees of freedom). In 2000 bootstrap resamples the best pool had three sources in 71\%, two in 22\% and four in 7\%. In the final window the two halves chose different pools (MP $r^{2}$SCAN or JARVIS OptB88vdW, each with both SNUMAT sources), and the held-out difference, $+0.057$ ($[-0.023,+0.137]$), is not resolved. Result file: \texttt{ateki/results/stats\_reanalysis/seven\_source.json} (keys \texttt{early.crossfit} and \texttt{final.crossfit}, size 3), written by \texttt{ateki/experiments/22\_stats\_reanalysis.py}.

\begin{table}[h]
\centering
\caption{Best attainable early ranking quality by pool size (eight-donor campaign, Kendall $\tau$ against the measured gaps). Pools of two to four sources reach 0.853 to 0.879; pools of five or more reach at most 0.816 (0.749 to 0.781 for six to eight).}
\label{tab:bestsize}
\small
\begin{tabular}{@{}rrp{11cm}@{}}
\toprule
Size & Best $v(S)$ & Pool \\
\midrule
0 & 0.6949 & (no donors) \\
1 & 0.8019 & SNUMAT GGA \\
2 & 0.8686 & MP GGA (no +U), SNUMAT GGA \\
3 & 0.8533 & MP GGA (no +U), SNUMAT GGA, SNUMAT HSE06 \\
4 & 0.8789 & JARVIS TB-mBJ, MP GGA (no +U), SNUMAT GGA, SNUMAT HSE06 \\
5 & 0.8156 & JARVIS TB-mBJ, MP GGA (no +U), MP $r^{2}$SCAN, SNUMAT GGA, SNUMAT HSE06 \\
6 & 0.7813 & MP GGA (no +U), MP GGA (ICSD), MP GGA+U (TM-O/F), MP $r^{2}$SCAN, SNUMAT GGA, SNUMAT HSE06 \\
7 & 0.7490 & all except MP GGA+U (TM-O/F) \\
8 & 0.7509 & all eight \\
\bottomrule
\end{tabular}
\end{table}

\paragraph{Control for the split.} A control asks whether the value of the TM-O/F subset comes from its +U records or from the chemistries they cover. The records in the +U chemistries (Co, Cr, Fe, Mn, Mo, Ni, V, or W with O or F) were removed from one donor not computed with +U and the game was rerun. Removing the 1{,}471 such records of JARVIS TB-mBJ left every value unchanged; removing the 1{,}903 of SNUMAT HSE06 moved the TM-O/F subset from $-0.067$ to $-0.058$ and SNUMAT HSE06 from $+0.051$ to $+0.060$. The loss stays with the +U records when a beyond-GGA source no longer covers the same chemistries, but the shift ($-0.067$ to $-0.058$) lies inside both intervals, so the control does not separate the +U records from the chemistries. The runs are in \texttt{ateki/results/split\_controls/} (\texttt{ateki/experiments/08d\_split\_controls.py}).

\paragraph{Harsanyi dividends.} The dividends $d(S)=\sum_{T\subseteq S}(-1)^{|S|-|T|}v(T)$ were obtained by M{\"o}bius inversion of the 256 values; they reproduce every $v(S)$ and the Shapley values to $10^{-12}$. Table~\ref{tab:dividendorder} gives the dividends summed by order and Table~\ref{tab:pairs} every pairwise dividend. The singleton dividends are SNUMAT GGA and HSE06 $+0.1070$ each, MP GGA (no +U) $+0.0968$, MP $r^{2}$SCAN $+0.0267$, JARVIS OptB88vdW $+0.0103$, JARVIS TB-mBJ $+0.0065$, MP GGA (ICSD) $-0.0190$, and MP GGA+U (TM-O/F) subset $-0.1162$. Of the 28 pairs, 17 have a positive dividend, none has an interval above zero, and only the two SNUMAT databases have an interval below zero. The high-order dividends are large, noisy, and of alternating sign, so no single high-order coalition should be read on its own; their sums at orders four to seven have intervals that exclude zero.

\begin{table}[h]
\centering
\caption{Harsanyi dividends by coalition order (eight-donor campaign, early window). Interval: 95\% $t$ interval of the order sum over the 10 seeds.}
\label{tab:dividendorder}
\small
\begin{tabular}{@{}rrrrrr@{}}
\toprule
Order & Coalitions & Sum $d$ & Positive / negative & Sum of $|d|$ & 95\% interval of sum \\
\midrule
1 & 8 & $+0.2190$ & 6 / 2 & 0.4895 & $[-0.232, +0.670]$ \\
2 & 28 & $-0.0312$ & 17 / 11 & 0.7360 & $[-1.787, +1.725]$ \\
3 & 56 & $-2.1672$ & 11 / 44 & 2.6305 & $[-6.047, +1.712]$ \\
4 & 70 & $+6.0430$ & 65 / 5 & 6.3158 & $[+0.242, +11.844]$ \\
5 & 56 & $-7.6979$ & 3 / 53 & 7.8396 & $[-13.500, -1.896]$ \\
6 & 28 & $+5.1714$ & 28 / 0 & 5.1714 & $[+1.470, +8.873]$ \\
7 & 8 & $-1.6476$ & 0 / 8 & 1.6476 & $[-3.119, -0.177]$ \\
8 & 1 & $+0.1665$ & 1 / 0 & 0.1665 & $[-0.182, +0.515]$ \\
\midrule
all & 255 & $+0.0560$ & & 24.9970 & \\
\bottomrule
\end{tabular}
\end{table}

\begin{table}[h]
\centering
\caption{All pairwise dividends of the eight-donor campaign, most negative first, with 95\% $t$ intervals over the 10 seeds.}
\label{tab:pairs}
\small
\begin{tabular}{@{}lrr@{}}
\toprule
Pair & $d$ & 95\% interval \\
\midrule
SNUMAT GGA + SNUMAT HSE06 & $-0.1070$ & $[-0.200, -0.014]$ \\
JARVIS TB-mBJ + MP GGA (no +U) & $-0.0510$ & $[-0.119, +0.017]$ \\
MP GGA (no +U) + MP $r^{2}$SCAN & $-0.0438$ & $[-0.108, +0.020]$ \\
JARVIS OptB88vdW + MP GGA (no +U) & $-0.0350$ & $[-0.109, +0.039]$ \\
MP GGA (no +U) + SNUMAT GGA & $-0.0301$ & $[-0.133, +0.073]$ \\
MP GGA (no +U) + SNUMAT HSE06 & $-0.0301$ & $[-0.133, +0.073]$ \\
MP GGA (no +U) + MP GGA (ICSD) & $-0.0293$ & $[-0.090, +0.031]$ \\
JARVIS OptB88vdW + SNUMAT GGA & $-0.0244$ & $[-0.106, +0.057]$ \\
JARVIS OptB88vdW + SNUMAT HSE06 & $-0.0244$ & $[-0.106, +0.057]$ \\
JARVIS TB-mBJ + MP $r^{2}$SCAN & $-0.0065$ & $[-0.072, +0.059]$ \\
MP GGA (ICSD) + MP $r^{2}$SCAN & $-0.0019$ & $[-0.066, +0.062]$ \\
MP GGA (no +U) + MP GGA+U (TM-O/F) & $+0.0057$ & $[-0.084, +0.096]$ \\
JARVIS TB-mBJ + MP GGA (ICSD) & $+0.0107$ & $[-0.044, +0.065]$ \\
MP GGA+U (TM-O/F) + SNUMAT GGA & $+0.0114$ & $[-0.053, +0.076]$ \\
MP GGA+U (TM-O/F) + SNUMAT HSE06 & $+0.0114$ & $[-0.053, +0.076]$ \\
JARVIS TB-mBJ + SNUMAT GGA & $+0.0130$ & $[-0.075, +0.101]$ \\
JARVIS TB-mBJ + SNUMAT HSE06 & $+0.0130$ & $[-0.075, +0.101]$ \\
MP GGA (ICSD) + SNUMAT GGA & $+0.0156$ & $[-0.048, +0.079]$ \\
MP GGA (ICSD) + SNUMAT HSE06 & $+0.0156$ & $[-0.048, +0.079]$ \\
MP $r^{2}$SCAN + SNUMAT GGA & $+0.0171$ & $[-0.056, +0.091]$ \\
MP $r^{2}$SCAN + SNUMAT HSE06 & $+0.0171$ & $[-0.056, +0.091]$ \\
JARVIS OptB88vdW + MP $r^{2}$SCAN & $+0.0210$ & $[-0.048, +0.090]$ \\
JARVIS TB-mBJ + JARVIS OptB88vdW & $+0.0248$ & $[-0.064, +0.114]$ \\
MP GGA+U (TM-O/F) + MP $r^{2}$SCAN & $+0.0274$ & $[-0.058, +0.113]$ \\
JARVIS OptB88vdW + MP GGA (ICSD) & $+0.0286$ & $[-0.049, +0.107]$ \\
JARVIS TB-mBJ + MP GGA+U (TM-O/F) & $+0.0324$ & $[-0.055, +0.120]$ \\
MP GGA (ICSD) + MP GGA+U (TM-O/F) & $+0.0381$ & $[-0.031, +0.107]$ \\
JARVIS OptB88vdW + MP GGA+U (TM-O/F) & $+0.0495$ & $[-0.012, +0.111]$ \\
\bottomrule
\end{tabular}
\end{table}

\subsection{Further detail}

With seven families one exchange of two neighboring families moves $\tau$ by $0.095$, so the shortfall amounts to between one and two such exchanges in the early ranking, averaged over seeds and checkpoints. The two SNUMAT databases carried a third of the positive contribution each ($+0.051$), MP GGA (no +U) contributed $+0.038$, MP $r^{2}$SCAN $+0.010$, and JARVIS TB-mBJ about zero ($-0.009$), and the TM-O/F subset held the campaign back most ($-0.067$, interval $[-0.091,-0.042]$), followed by the ICSD subset ($-0.019$), 5{,}433 of whose records fall under the +U rule.

\clearpage
\section{Contribution measures compared}
\label{si:measures}
Table~\ref{tab:semivalues} gives every measure on the seed-averaged eight-donor game. Three arguments from data valuation were tested. \textbf{Duplicates.} Leave-one-out is expected to give exact duplicates zero \citep{Ghorbani2019DataShapley}. The TM-O/F subset and the ICSD subset, which mask each other at the full pool, received two of the three smallest leave-one-out values ($+0.002$ and $+0.006$; JARVIS TB-mBJ ties the TM-O/F subset), and every leave-one-out value is positive, so a removal screen would keep all eight donors. The identical SNUMAT pair received the largest leave-one-out value ($+0.044$ each): the second copy adds nothing alone, but a mean $+0.040$ (range $-0.019$ to $+0.093$, positive in 59 of the 63 pools) to pools holding another donor. \textbf{Small-pool weighting.} Beta$(16,1)$ gave JARVIS TB-mBJ a positive value ($+0.005$, against $-0.009$ under Shapley) and the TM-O/F subset the most negative value ($-0.109$), because the harm of the TM-O/F subset is largest in small pools. \textbf{Consistency across runs.} Banzhaf had the highest mean seed-to-seed Spearman correlation ($0.89$, against $0.85$ for Shapley and $0.28$ for leave-one-out; Table~\ref{tab:consistency}), in line with the safety-margin argument of \citet{Wang2023DataBanzhaf}, but its margin over Shapley lies within the bootstrap interval of the difference, $[-0.011, +0.107]$; leave-one-out named the Shapley-lowest donor in 2 of the 10 seeds. \textbf{Truncated sampling.} On the seven-source game, at the $3n=21$ permutations often used for Monte-Carlo Data Shapley, 16 of 50 untruncated repeats reversed the sign of at least one database, and truncation at the seed noise of the full-pool gain (tolerance $0.0216$) biased values by up to $0.019$ (Table~\ref{tab:truncation}); games of a few members should be enumerated exactly.

\begin{table}[h]
\centering
\caption{Semivalues of the eight donors (seed-averaged game, early window, Kendall-$\tau$ gain over the campaign without donors, $w(N)=+0.0560$).}
\label{tab:semivalues}
\small
\begin{tabular}{@{}lrrrrrrr@{}}
\toprule
Database & Shapley & Banzhaf & LOO & B$(4,1)$ & B$(16,1)$ & B$(1,4)$ & B$(1,16)$ \\
\midrule
JARVIS TB-mBJ & $-0.0086$ & $-0.0146$ & $+0.0019$ & $-0.0018$ & $+0.0053$ & $-0.0101$ & $-0.0044$ \\
JARVIS OptB88vdW & $+0.0019$ & $-0.0020$ & $+0.0084$ & $+0.0055$ & $+0.0099$ & $+0.0015$ & $+0.0051$ \\
MP GGA (no +U) & $+0.0376$ & $+0.0232$ & $+0.0343$ & $+0.0589$ & $+0.0832$ & $+0.0280$ & $+0.0318$ \\
MP GGA (ICSD) & $-0.0194$ & $-0.0258$ & $+0.0057$ & $-0.0208$ & $-0.0180$ & $-0.0127$ & $-0.0025$ \\
MP GGA+U (TM-O/F) & $-0.0668$ & $-0.0743$ & $+0.0019$ & $-0.0975$ & $-0.1089$ & $-0.0309$ & $-0.0083$ \\
MP $r^{2}$SCAN & $+0.0104$ & $+0.0001$ & $+0.0248$ & $+0.0154$ & $+0.0246$ & $+0.0129$ & $+0.0206$ \\
SNUMAT GGA & $+0.0505$ & $+0.0358$ & $+0.0438$ & $+0.0748$ & $+0.0976$ & $+0.0376$ & $+0.0409$ \\
SNUMAT HSE06 & $+0.0505$ & $+0.0358$ & $+0.0438$ & $+0.0748$ & $+0.0976$ & $+0.0376$ & $+0.0409$ \\
\bottomrule
\end{tabular}
\end{table}

Table~\ref{tab:bestsubset} gives the decision that the measured values support directly. The best pool of each size and the removal path, which drops from the full pool at each step the source whose removal raises the value most or lowers it least, are read from the 256 seed-averaged values; no contribution measure is involved. The removal path reaches the best of all 256 pools (JARVIS TB-mBJ, MP GGA (no +U), SNUMAT GGA, SNUMAT HSE06, $\tau=0.879$) after four steps. Its first step drops the TM-O/F subset and lowers the value slightly ($0.751$ to $0.749$; dropping JARVIS TB-mBJ instead gives exactly the same mean, and the script breaks such ties by a fixed donor order), because the ICSD subset masks them; the second step, dropping the ICSD subset, raises it to $0.770$. From four to seven sources the best attainable value falls with each added source ($0.749$ at seven), and all eight reach $0.751$. The best pools of two to four sources lie within $0.03$ of each other ($0.869$, $0.853$, $0.879$): over 2000 bootstrap resamples of the seeds the best size was four in 1757, two in 214, and three in 29, and the pools chosen on seeds 1--5 and on seeds 6--10 coincide at sizes 1, 2, and 4. The full pool fell short of the best pool by $0.128$ (95\% $t$ interval $[+0.061, +0.195]$).

\begin{table}[h]
\centering
\caption{Best pool of each size and the greedy removal path in the eight-donor game (seed-averaged early-window Kendall $\tau$ over the 10 seeds). Removal path: at each step the source whose removal raises the value most, or lowers it least, is dropped.}
\label{tab:bestsubset}
\small
\begin{tabular}{@{}rp{6.2cm}rlr@{}}
\toprule
Size & Best pool of this size & $\tau$ & Removal path: source dropped & $\tau$ after \\
\midrule
8 & all eight & $0.751$ & -- & $0.751$ \\
7 & JARVIS TB-mBJ, JARVIS OptB88vdW, MP GGA (no +U), MP GGA (ICSD), MP $r^{2}$SCAN, SNUMAT GGA, SNUMAT HSE06 & $0.749$ & MP GGA+U (TM-O/F) & $0.749$ \\
6 & MP GGA (no +U), MP GGA (ICSD), MP GGA+U (TM-O/F), MP $r^{2}$SCAN, SNUMAT GGA, SNUMAT HSE06 & $0.781$ & MP GGA (ICSD) & $0.770$ \\
5 & JARVIS TB-mBJ, MP GGA (no +U), MP $r^{2}$SCAN, SNUMAT GGA, SNUMAT HSE06 & $0.816$ & JARVIS OptB88vdW & $0.816$ \\
4 & JARVIS TB-mBJ, MP GGA (no +U), SNUMAT GGA, SNUMAT HSE06 & $0.879$ & MP $r^{2}$SCAN & $0.879$ \\
3 & MP GGA (no +U), SNUMAT GGA, SNUMAT HSE06 & $0.853$ & JARVIS TB-mBJ & $0.853$ \\
2 & MP GGA (no +U), SNUMAT GGA & $0.869$ & SNUMAT HSE06 & $0.869$ \\
1 & SNUMAT GGA & $0.802$ & MP GGA (no +U) & $0.802$ \\
0 & none & $0.695$ & SNUMAT GGA & $0.695$ \\
\bottomrule
\end{tabular}
\end{table}
\textbf{Pooling per distinct ordering.} The donor prior of a coalition is the mean of its donors' orderings with one vote per source, so the two SNUMAT sources, whose orderings of the receiver's families are identical, carry the SNUMAT ordering twice in every pool that holds both; the other six donors have distinct orderings. Under the alternative rule with one vote per distinct ordering, a pool holding both SNUMAT sources has the same prior as the pool without SNUMAT HSE06, and with matched seeds the same campaign, so the 256 values of this game follow from the measured ones without a new run (Table~\ref{tab:dedup}). The second copy then adds exactly zero to all 64 pools that hold the first, against a mean of $+0.040$ under per-source pooling; the full pool falls from $0.751$ to $0.707$, the two copies still receive equal values by symmetry, and their sum falls from $+0.101$ to $+0.063$.

\begin{table}[h]
\centering
\caption{Shapley values of the eight donors under per-source pooling (as in the main text) and per-distinct-ordering pooling (seed-averaged early-window game; 95\% $t$ interval over the 10 seeds for the per-ordering values). Full pool: $0.751$ per source, $0.707$ per ordering; $v(\emptyset)=0.695$.}
\label{tab:dedup}
\small
\begin{tabular}{@{}lrr@{}}
\toprule
Database & Per source & Per distinct ordering \\
\midrule
JARVIS TB-mBJ & $-0.009$ & $-0.009$ $[-0.019, -0.000]$ \\
JARVIS OptB88vdW & $+0.002$ & $+0.005$ $[-0.006, +0.017]$ \\
MP GGA (no +U) & $+0.038$ & $+0.040$ $[+0.024, +0.056]$ \\
MP GGA (ICSD) & $-0.019$ & $-0.023$ $[-0.043, -0.003]$ \\
MP GGA+U (TM-O/F) & $-0.067$ & $-0.072$ $[-0.101, -0.043]$ \\
MP $r^{2}$SCAN & $+0.010$ & $+0.008$ $[-0.004, +0.021]$ \\
SNUMAT GGA & $+0.051$ & $+0.032$ $[+0.017, +0.047]$ \\
SNUMAT HSE06 & $+0.051$ & $+0.032$ $[+0.017, +0.047]$ \\
\bottomrule
\end{tabular}
\end{table}

\begin{table}[h]
\centering
\caption{Consistency of each measure across the 10 seeds of the eight-donor game: mean Spearman correlation between the values of two seeds (45 pairs) with its 95\% bootstrap interval over seeds, the minimum over the 45 pairs, mean correlation with the seed-averaged Shapley values, seeds in which the measure names the Shapley-last donor last, and seeds with a sign different from that seed's own Shapley value for at least one donor.}
\label{tab:consistency}
\small
\resizebox{\textwidth}{!}{%
\begin{tabular}{@{}lrrrrrr@{}}
\toprule
Measure & Mean pairwise & 95\% interval & Minimum & To averaged Shapley & Shapley-last named & Sign flips \\
\midrule
Shapley & 0.850 & $[0.79, 0.96]$ & 0.542 & 0.923 & 10 & 0 \\
Banzhaf & 0.890 & $[0.80, 0.98]$ & 0.542 & 0.947 & 10 & 4 \\
Leave-one-out & 0.280 & $[0.11, 0.66]$ & $-0.655$ & 0.547 & 2 & 10 \\
Beta$(4,1)$ & 0.694 & -- & 0.036 & 0.831 & 9 & 6 \\
Beta$(16,1)$ & 0.619 & -- & $-0.133$ & 0.788 & 8 & 7 \\
Beta$(1,4)$ & 0.718 & -- & 0.229 & 0.843 & 6 & 7 \\
Beta$(1,16)$ & 0.518 & -- & $-0.205$ & 0.718 & 5 & 8 \\
\bottomrule
\end{tabular}}
\end{table}

\begin{table}[h]
\centering
\caption{Truncated Monte-Carlo Shapley on the seven-source game (50 repeats per setting): tolerance as a fraction of $w(N)$, permutations, median and 90th-percentile maximum error, repeats with a sign reversal, and value lookups.}
\label{tab:truncation}
\small
\begin{tabular}{@{}rrrrrr@{}}
\toprule
Tolerance & Permutations & Median max error & 90th pct & Sign reversals & Lookups \\
\midrule
0.00 & 21 & 0.0099 & 0.0139 & 16/50 & 147 \\
0.00 & 1000 & 0.0014 & 0.0025 & 0/50 & 7000 \\
0.01 & 21 & 0.0126 & 0.0183 & 23/50 & 135 \\
0.01 & 1000 & 0.0039 & 0.0052 & 0/50 & 6425 \\
0.05 & 21 & 0.0120 & 0.0190 & 17/50 & 131 \\
0.05 & 1000 & 0.0048 & 0.0066 & 0/50 & 6283 \\
0.25 & 21 & 0.0156 & 0.0236 & 18/50 & 113 \\
0.25 & 1000 & 0.0089 & 0.0101 & 0/50 & 5323 \\
\bottomrule
\end{tabular}
\end{table}

\paragraph{The same checks on the computed Al--Mg--Si games.} These games are unaffected by the archive rerun (every query is a calculation); Tables~\ref{tab:dvalmgsi} and~\ref{tab:dvhot} give the measures for the PBE and HSE06 receivers and for the higher-temperature MD receivers.
\begin{table}[h]
\centering
\caption{Semivalues of the donors in the computed Al--Mg--Si games (receivers QE PBE and QE HSE06, every coalition run; value: early share of the ten compounds with the lowest formation energy per atom, 40 runs). Fixed: constant weight on the pooled donor ordering; learned: weight re-estimated after every batch. In each run every MD formation energy is drawn from its measured sampling error (0~K, DFT and archive values fixed), so the 40 per-run games ($k = 1$ and $4$, 20 runs each) differ by MD sampling noise. The last four rows give the mean Spearman correlation between the values of two runs over the 780 run pairs of each game.}
\label{tab:dvalmgsi}
\footnotesize
\setlength{\tabcolsep}{4pt}
\begin{tabular}{@{}llrrrrr@{}}
\toprule
Game & Donor & Shapley & Banzhaf & Beta(16,1) & Beta(1,16) & LOO \\
\midrule
PBE, fixed & MACE-MP-0 relaxation & $+0.226$ & $+0.193$ & $+0.326$ & $+0.216$ & $+0.224$ \\
 & MP archive & $+0.148$ & $+0.147$ & $+0.149$ & $+0.150$ & $+0.146$ \\
 & MEAM MD & $-0.109$ & $-0.109$ & $-0.242$ & $+0.024$ & $+0.037$ \\
 & NEP89 MD & $-0.197$ & $-0.214$ & $-0.179$ & $-0.166$ & $-0.159$ \\
PBE, learned & MP archive & $+0.114$ & $+0.091$ & $+0.265$ & $+0.024$ & $+0.016$ \\
 & MACE-MP-0 relaxation & $+0.105$ & $+0.071$ & $+0.234$ & $+0.068$ & $+0.072$ \\
 & NEP89 MD & $+0.043$ & $+0.017$ & $+0.140$ & $+0.019$ & $+0.022$ \\
 & MEAM MD & $+0.024$ & $-0.004$ & $+0.084$ & $+0.039$ & $+0.047$ \\
HSE06, fixed & QE PBE & $+0.171$ & $+0.132$ & $+0.344$ & $+0.116$ & $+0.119$ \\
 & MACE-MP-0 relaxation & $+0.167$ & $+0.130$ & $+0.343$ & $+0.103$ & $+0.105$ \\
 & MP archive & $+0.120$ & $+0.107$ & $+0.167$ & $+0.118$ & $+0.122$ \\
 & MEAM MD & $-0.084$ & $-0.088$ & $-0.133$ & $-0.016$ & $-0.008$ \\
 & NEP89 MD & $-0.209$ & $-0.237$ & $-0.145$ & $-0.187$ & $-0.178$ \\
HSE06, learned & QE PBE & $+0.145$ & $+0.118$ & $+0.345$ & $+0.022$ & $+0.009$ \\
 & MACE-MP-0 relaxation & $+0.101$ & $+0.069$ & $+0.267$ & $+0.027$ & $+0.025$ \\
 & MP archive & $+0.064$ & $+0.034$ & $+0.215$ & $+0.000$ & $-0.006$ \\
 & MEAM MD & $-0.002$ & $+0.001$ & $-0.049$ & $+0.038$ & $+0.040$ \\
 & NEP89 MD & $-0.051$ & $-0.084$ & $+0.084$ & $-0.091$ & $-0.091$ \\
\midrule
\multicolumn{2}{@{}l}{Run-pair Spearman, PBE fixed} & $1.000$ & $1.000$ & $1.000$ & $1.000$ & $1.000$ \\
\multicolumn{2}{@{}l}{Run-pair Spearman, PBE learned} & $0.162$ & $0.195$ & $0.685$ & $0.169$ & $0.373$ \\
\multicolumn{2}{@{}l}{Run-pair Spearman, HSE06 fixed} & $0.982$ & $0.962$ & $0.919$ & $0.914$ & $0.902$ \\
\multicolumn{2}{@{}l}{Run-pair Spearman, HSE06 learned} & $0.840$ & $0.871$ & $0.897$ & $0.276$ & $0.281$ \\
\bottomrule
\end{tabular}
\end{table}

Table~\ref{tab:dvhot} gives the consistency checks for the six higher-temperature MD games, with every MD value redrawn from its sampling error in each of the 20 runs.

\begin{table}[h]
\centering
\caption{Data-valuation checks on the higher-temperature MD games (receiver, trust mode; six donors, all 64 coalitions; 20 runs, each drawing every MD formation energy from its measured sampling error). Spearman: mean correlation between the values of two runs over 190 run pairs. Top: runs in which the same potential at 300~K received the largest value. Other potentials: combined value of the two other potentials at the target temperature in the run-averaged game, and the runs in which its Shapley value was negative. The NEP89 spreads may be underestimated.}
\label{tab:dvhot}
\footnotesize
\setlength{\tabcolsep}{3pt}
\begin{tabular}{@{}lrrrrrrrrr@{}}
\toprule
 & \multicolumn{4}{c}{Run-pair Spearman} & \multicolumn{2}{c}{Top (runs)} & \multicolumn{3}{c}{Other potentials} \\
\cmidrule(lr){2-5}\cmidrule(lr){6-7}\cmidrule(l){8-10}
Receiver, trust & Shapley & Banzhaf & Beta(16,1) & LOO & Shapley & LOO & Shapley & LOO & runs $<0$ \\
\midrule
MEAM 453~K, learned & 0.846 & 0.834 & 0.875 & 0.302 & 20 & 11 & $-0.075$ & $-0.064$ & 20 \\
MEAM 453~K, fixed & 0.905 & 0.880 & 0.970 & 0.701 & 20 & 20 & $-0.103$ & $+0.013$ & 20 \\
MEAM 803~K, learned & 0.852 & 0.885 & 0.620 & 0.637 & 20 & 19 & $-0.106$ & $-0.128$ & 20 \\
MEAM 803~K, fixed & 0.971 & 0.953 & 0.989 & 0.607 & 20 & 19 & $-0.061$ & $+0.018$ & 20 \\
NEP89 453~K, learned & 0.932 & 0.945 & 0.855 & 0.419 & 20 & 18 & $-0.043$ & $-0.026$ & 18 \\
NEP89 453~K, fixed & 0.910 & 0.868 & 0.935 & 0.769 & 20 & 20 & $-0.154$ & $-0.006$ & 20 \\
NEP89 803~K, learned & 0.760 & 0.820 & 0.706 & 0.040 & 20 & 4 & $-0.036$ & $-0.012$ & 18 \\
NEP89 803~K, fixed & 0.977 & 0.934 & 0.973 & 0.628 & 20 & 20 & $-0.164$ & $-0.016$ & 20 \\
MACE-MP-0 453~K, learned & 0.897 & 0.820 & 0.936 & 0.534 & 20 & 8 & $-0.010$ & $-0.084$ & 11 \\
MACE-MP-0 453~K, fixed & 0.884 & 0.899 & 0.970 & 0.882 & 20 & 0 & $-0.004$ & $+0.094$ & 10 \\
MACE-MP-0 803~K, learned & 0.902 & 0.888 & 0.936 & 0.609 & 17 & 5 & $-0.005$ & $+0.057$ & 9 \\
MACE-MP-0 803~K, fixed & 0.973 & 0.927 & 0.973 & 0.915 & 20 & 9 & $-0.095$ & $+0.030$ & 19 \\
\bottomrule
\end{tabular}
\end{table}

\subsection{Further detail}

This resembles the redundancy of overlapping ties \citep{Granovetter1973WeakTies,Burt1992StructuralHoles}, here as substitutes in a database portfolio.

The contribution measures serve a different purpose, dividing the gain among the sources, where their axioms matter.

Whether redundant knowledge is worthless therefore depends on how the campaign integrates its sources, not on the redundancy alone.

The TM-O/F subset does most harm as one of a few advisers, and a TB-mBJ ordering advances a small pool and holds back a large one.

\clearpage
\section{Communication structure}
\label{si:structure}
The meeting games of this section restrict the seven-source band-gap game (MP GGA unsplit, 128 subsets; Table~\ref{tab:sevensource}); they were not rerun with the TM-O/F subset as a separate donor.

\paragraph{Weighting by team structure.} A post-hoc option reweights unrestricted Shapley values by a centrality weight $w_i(G)$ of the team graph,
\begin{equation}
\phii^{G} \;=\; n\, w_i(G)\, \phii \;+\; \frac{1}{n}\Bigl(\vN - \textstyle\sum_j n\, w_j(G)\, \phi_j\Bigr),
\label{eq:topoweight}
\end{equation}
where the second term restores efficiency additively. The principled alternative restricts the game itself (Myerson value; main-text Section~2.5).

\subsection{The degree-centrality weight}
\label{app:math-degree}

The communication weight in Eq.~\ref{eq:topoweight} is the normalized degree centrality, $w_i(G) = \deg(i) / \sum_j \deg(j)$. On a star graph with $n$ bodies the hub has degree $n-1$ and each leaf degree $1$, so the hub receives $(n-1)/(2(n-1)) = 0.5$ and each leaf $1/(2(n-1))$; for the $n = 7$ databases of Section~\ref{si:structure} this is $6/12 = 0.5$ for the hub and $1/12$ for each leaf. After the mean normalization of Eq.~\ref{eq:topoweight}, the hub's Shapley value is multiplied by $n/2 = 3.5$ and each leaf's by $n/(2(n-1)) = 7/12$ before the additive shift, so the leaf ordering is preserved. This is the graded correction quoted in Section~\ref{si:structure}. In contrast, betweenness centrality is exactly $1$ for the hub and $0$ for every leaf.

\subsection{Two failure modes of centrality weighting}
\label{app:math-topology}

Both failure modes of centrality weighting return plausible values
without any error signal:

\begin{enumerate}
\item \textbf{Star-graph betweenness fails silently, by construction.}
The natural choice of betweenness centrality \citep{Freeman1977Betweenness} assigns exactly $1$ to the hub and exactly $0$ to every leaf on a star graph (one supervisor, $n-1$ leaves), so betweenness weighting assigns the hub the entire (shifted) outcome regardless of any member's contribution: winner-takes-all by construction, not by evidence. Every leaf receives the same weight (zero), so after the efficiency shift all leaves receive one identical value and the ranking carries no information about any leaf. We therefore weight by \emph{degree centrality} (Section~\ref{app:math-degree}).
\item \textbf{Efficiency must be restored additively.} Rescaling
  $\phii^{G}$ multiplicatively to match $\vN$ inverts every sign
  whenever the scale factor, the ratio of $\vN$ to the sum of the $\phii^{G}$, is negative, as when $\vN < 0$
  (which happens routinely when the value is a negated cost) while the weighted values sum to a positive number. The additive shift in Eq.~\ref{eq:topoweight} is
  sign-safe.
\end{enumerate}

The post-hoc weighting is transparent but ad hoc; the principled
alternative, the Myerson value, makes the structure part of the game itself.

\subsection{The pairwise reduction check}
\label{app:math-reduction}

When every hyperedge has exactly two members, containment of a hyperedge in $S$ is containment of a graph edge in the induced subgraph. In this case, $v_H$ and every resulting attribution coincide exactly with the graph restriction. We verify this executably on a synthetic four-player superadditive game over all 16 coalitions and on the measured seven-source game: feeding the clique expansion of the size-three conferences through the hypergraph code as pure pairs reproduces the graph-restricted value on all 128 coalitions and the graph Myerson value with maximum absolute deviation $0.0$ at tolerance $10^{-9}$.

\paragraph{One database as hub of a star.} Each database was placed in turn at the hub of a star (Table~\ref{tab:star}). Betweenness weighting gave every leaf the same value; with MP GGA as hub it gave the hub $-0.1707$ and every leaf $+0.0354$. Degree weighting kept the leaf ordering for every hub. The Myerson value placed every hub below its unrestricted value, because on a star every interaction between members arises only in subsets that contain the hub and on this game interactions of three or more sources sum to a loss in point estimate; the best-group rule roughly halved that charge.

\begin{table}[h]
\centering
\caption{Each database as the hub of a star (seven-source campaign, early window): unrestricted Shapley value, degree-weighted value, Myerson value under the sum rule, and graph value under the best-group rule; hub rank after restriction.}
\label{tab:star}
\small
\resizebox{\textwidth}{!}{%
\begin{tabular}{@{}lrrrrr@{}}
\toprule
Hub & Unrestricted & Degree-weighted & Myerson, sum & Graph value, best-group rule & Rank (sum, best) \\
\midrule
JARVIS TB-mBJ & $-0.0134$ & $-0.0389$ & $-0.0633$ & $-0.0349$ & 7, 7 \\
JARVIS OptB88vdW & $-0.0051$ & $-0.0131$ & $-0.0580$ & $-0.0314$ & 7, 7 \\
MP GGA & $-0.0294$ & $-0.0883$ & $-0.0933$ & $-0.0479$ & 7, 7 \\
MP GGA (ICSD) & $-0.0234$ & $-0.0696$ & $-0.0835$ & $-0.0432$ & 7, 7 \\
MP $r^{2}$SCAN & $+0.0054$ & $+0.0192$ & $-0.0442$ & $-0.0244$ & 7, 7 \\
SNUMAT GGA & $+0.0539$ & $+0.1687$ & $+0.0243$ & $+0.0174$ & 2, 2 \\
SNUMAT HSE06 & $+0.0539$ & $+0.1687$ & $+0.0243$ & $+0.0174$ & 2, 2 \\
\bottomrule
\end{tabular}}
\end{table}

\paragraph{Declared meetings.} Structure A holds one meeting per organization (MP GGA, MP GGA (ICSD), MP $r^{2}$SCAN; JARVIS TB-mBJ, OptB88vdW; SNUMAT GGA, HSE06) and a beyond-GGA meeting (JARVIS TB-mBJ, MP $r^{2}$SCAN, SNUMAT HSE06); Structure B adds a GGA-level meeting (JARVIS OptB88vdW, MP GGA, SNUMAT GGA). Each restricted value was compared with the Myerson value of the clique expansion of the same meetings (any two members of a meeting may combine without the rest). Both structures connect all seven databases, so every column sums to $w(N)=0.0419$ (Table~\ref{tab:meetings}).

\begin{table}[h]
\centering
\caption{Contributions of the seven databases under declared meetings (early window; rank in parentheses). Hyper: all-members-present meetings; Clique: pairwise links between meeting members. Sum and Best: rule for valuing a separated team.}
\label{tab:meetings}
\scriptsize
\resizebox{\textwidth}{!}{%
\begin{tabular}{@{}lrrrrrrrrr@{}}
\toprule
& & \multicolumn{4}{c}{Structure A} & \multicolumn{4}{c}{Structure B} \\
\cmidrule(lr){3-6}\cmidrule(l){7-10}
Database & None & Hyper sum & Hyper best & Clique sum & Clique best & Hyper sum & Hyper best & Clique sum & Clique best \\
\midrule
SNUMAT GGA & $+.0539$ (1) & $+.0633$ (1) & $+.0579$ (1) & $+.0682$ (1) & $+.0658$ (1) & $+.0545$ (1) & $+.0482$ (2) & $+.0521$ (2) & $+.0501$ (2) \\
SNUMAT HSE06 & $+.0539$ (1) & $+.0277$ (2) & $+.0322$ (2) & $+.0288$ (2) & $+.0314$ (2) & $+.0482$ (2) & $+.0491$ (1) & $+.0568$ (1) & $+.0589$ (1) \\
MP $r^{2}$SCAN & $+.0054$ (3) & $+.0006$ (4) & $-.0081$ (4) & $-.0112$ (5) & $-.0221$ (7) & $+.0045$ (3) & $-.0065$ (4) & $+.0032$ (3) & $-.0023$ (3) \\
JARVIS OptB88vdW & $-.0051$ (4) & $+.0110$ (3) & $+.0032$ (3) & $+.0040$ (3) & $+.0012$ (3) & $+.0023$ (4) & $-.0065$ (4) & $-.0054$ (4) & $-.0064$ (4) \\
JARVIS TB-mBJ & $-.0134$ (5) & $-.0070$ (5) & $-.0108$ (5) & $-.0103$ (4) & $-.0069$ (4) & $-.0061$ (5) & $-.0058$ (3) & $-.0119$ (5) & $-.0104$ (5) \\
MP GGA (ICSD) & $-.0234$ (6) & $-.0215$ (6) & $-.0153$ (6) & $-.0167$ (6) & $-.0132$ (5) & $-.0206$ (6) & $-.0098$ (6) & $-.0227$ (6) & $-.0186$ (6) \\
MP GGA & $-.0294$ (7) & $-.0322$ (7) & $-.0171$ (7) & $-.0208$ (7) & $-.0143$ (6) & $-.0409$ (7) & $-.0268$ (7) & $-.0302$ (7) & $-.0294$ (7) \\
\bottomrule
\end{tabular}}
\end{table}

\subsection{Further detail}

Knowledge moves more readily between units that are tied to each other, and weak ties help a unit search for knowledge but hinder the transfer of complex knowledge \citep{Hansen1999SearchTransfer}.

\clearpage
\section{Language-model agent teams}
\label{si:teams}
\paragraph{Protocol.} Each agent held the family ordering of one database as private knowledge and, in each of 12 rounds, loaded one four-family tray, for which the archive returned measured orders (each record drawn at most once per campaign). In the flat team all agents acted at once on a shared notebook and could message any agent; in the coordinator-led team agent 1 (an MP GGA holder) assigned the trays. Meetings were read from the logs (a message from $i$ to $T$ is the meeting $\{i\}\cup T$; a tray that cites other agents' notebook entries is the meeting of its author and the cited authors; meetings recurring at least twice were kept), and $v(S)$ was measured by replaying the trays of the members of $S$ as a Plackett--Luce ranking over all $2^{n}$ subsets ($n=10$: 1{,}024 subsets).

\paragraph{Prompts, parsing, and fallback.} Every agent received the same system prompt with its own knowledge filled in: it is one of the five (or ten) agents of a materials screening team that must rank the families by measured optical band gap, widest first; each query names four different families, the laboratory measures one archived compound of each and returns their order, and one measurement may not represent its family; its private knowledge is its database's family ordering by computed gap, which can disagree with measured gaps, and other agents hold other databases; it may write one notebook entry (observation, fact, failure, claim, or summary), send one message to chosen agents, and must list the notebook entries and query results its query builds on; it replies in JSON only. Each turn's user message gave the round, the 30 most recent query results, the 20 most recent notebook entries, and the eight most recent messages addressed to the agent. In the coordinator-led team the coordinator was asked, each round, to assign one four-family query to every agent including itself, with a message of at most 40 words; the other agents saw their assignment and could follow or change it, and could message only the coordinator. Replies were sampled at temperature 0.7 (at most 900 tokens for agents and 1000 for the coordinator). A reply that did not parse or did not name four valid families was requested again with a new seed, up to three attempts; after three failures the agent's tray was its assignment, or else four random families, and the failure was counted. Sets S1, S2, and S3 used 2{,}642, 1{,}340, and 2{,}017 model calls with no fallback; the finer task used 8{,}499 calls with 26 fallbacks. The templates are in \texttt{ateki/experiments/11\_recorded\_meetings.py} and \texttt{13\_recorded\_meetings\_4b.py}.

\paragraph{Team ranking quality and tray statistics.} Table~\ref{tab:teams} gives the values plotted in the main-text agent-team figure.

\begin{table}[h]
\centering
\caption{Coordination in language-model agent teams ranking seven anion families by measured gap (12 rounds of four-family trays). Team $\tau$: Kendall $\tau$ against the measured ranking, mean over campaigns. Distinct trays: share of distinct trays among all trays a team loaded. Random trays at the same budget, a search baseline, reach $0.876$, sd $0.096$ (5 agents, 240 evaluations) and $0.929$, sd $0.073$ (10 agents, 480 evaluations), each over 20 random searches; the five-agent sd of $0.096$ is the spread of the random trays and is not the coordinator effect of $+0.096$ below. Coordinator effect: coordinator-led minus flat, paired by seed, with 95\% bootstrap interval.}
\label{tab:teams}
\small
\resizebox{\textwidth}{!}{%
\begin{tabular}{@{}llrrrrr@{}}
\toprule
Team & Campaigns & \multicolumn{2}{c}{Team $\tau$} & \multicolumn{2}{c}{Distinct trays} & Coordinator effect \\
\cmidrule(lr){3-4}\cmidrule(lr){5-6}
& & flat & coordinated & flat & coordinated & \\
\midrule
Qwen3-4B, 5 agents & 20 + 20 & 0.756 & 0.852 & 9.6\% & 11.3\% & $+0.096$ $[+0.029, +0.165]$ \\
Nemotron 30B-A3B, 5 agents & 10 + 10 & 0.670 & 0.790 & 5.3\% & 17.0\% & $+0.121$ $[+0.002, +0.229]$ \\
Qwen3-4B, 10 agents & 8 + 8 & 0.812 & 0.857 & 5.6\% & 8.4\% & $+0.045$ $[-0.019, +0.105]$ \\
\bottomrule
\end{tabular}}
\end{table}

\paragraph{Sequential turns.} To test whether the flat team duplicates trays because its agents choose at the same time from the same snapshot, a third S1 arm let the five Qwen3-4B agents choose in a fixed order that rotated each round, each agent seeing the trays already claimed that round but not their results; results, notes and messages were applied at the end of the round as in the flat arm, with the same seeds (0--19), prompts, budget and archive draws (1200 calls, no fallback replies). The sequential team reached team $\tau$ $0.771$ and a distinct-tray share of 11.9\%, against $0.756$ and 9.6\% flat and $0.852$ and 11.3\% coordinated. Paired by seed (95\% percentile bootstrap, 5000 resamples), sequential minus flat was $+0.015$ $[-0.027,+0.062]$ in team $\tau$ and $+0.023$ $[+0.014,+0.033]$ in distinct share; sequential minus coordinated was $-0.081$ $[-0.152,-0.010]$ in team $\tau$ and $+0.006$ $[-0.007,+0.018]$ in distinct share. Taking turns removed about as much duplication as the coordinator did but not the coordinator's gain in ranking, and the sequential team stayed below the random-tray search baseline ($0.876$, sd $0.096$).

\paragraph{Shares by database.} Table~\ref{tab:teamshares} gives each database's share of team $\tau$ as the unrestricted Shapley value of the replayed team, summed over the agents holding the database; in a coordinated team the MP GGA column is the coordinator's unrestricted share. The last two columns come from a different game, the team restricted to the recorded meetings: the coordinator's Myerson value (sum rule) and its value under the best-group rule. The columns therefore have no common total and are not compared with each other. Because team $\tau$ is bounded, the sum rule adds the values of disconnected groups as if each could raise $\tau$ on its own, so it overstates the value of a fragmented team; the best-group rule does not.

\begin{table}[h]
\centering
\caption{Share of team $\tau$ by the database an agent holds (unrestricted Shapley value of the replayed team, mean over campaigns), and the coordinator's own share in the game restricted to the recorded meetings, under the sum rule (Myerson value) and the best-group rule. The two groups of columns come from different games and are not compared.}
\label{tab:teamshares}
\small
\resizebox{\textwidth}{!}{\begin{tabular}{@{}llrrrrrrr@{}}
\toprule
Set & Team & MP GGA & MP $r^{2}$SCAN & OptB88vdW & TB-mBJ & SNUMAT HSE06 & Coord.\ sum & Coord.\ best \\
\midrule
S1 Qwen 5 & flat & 0.099 & 0.201 & 0.219 & 0.149 & 0.332 & & \\
S1 Qwen 5 & coordinated & 0.129 & 0.255 & 0.138 & 0.213 & 0.265 & $-0.825$ & $+0.061$ \\
S2 Nemotron 5 & flat & 0.104 & 0.193 & 0.224 & 0.077 & 0.403 & & \\
S2 Nemotron 5 & coordinated & 0.185 & 0.167 & 0.245 & 0.199 & 0.203 & $-0.792$ & $+0.144$ \\
S3 Qwen 10 & flat & 0.077 & 0.216 & 0.160 & 0.212 & 0.335 & & \\
S3 Qwen 10 & coordinated & 0.167 & 0.207 & 0.139 & 0.197 & 0.290 & $-2.963$ & $-0.064$ \\
\bottomrule
\end{tabular}}
\end{table}

\paragraph{Registered predictions (sets S1--S3).} Recorded on 4 Oct 2026, before any campaign of the extension, in time-stamped commits of the project repository (files PREDICTION.md in the code release). P1 (S1: coordinator-led ranks better, interval excluding zero; flat distinct trays below 20\%): pass ($+0.096$, $[+0.029,+0.165]$; 9.6\%). P2 (S2: flat distinct trays below 30\%; coordinator-led at least as good): pass (5.3\%; 0.790 against 0.670). P3 (S3: flat distinct trays at 10 agents no higher than at 5; coordinator-led at least as good): pass (5.6\% against 9.6\%; 0.857 against 0.812). P4 (the MP $r^{2}$SCAN and SNUMAT HSE06 holders take the two largest shares in every set and team): fail (S1 flat: JARVIS OptB88vdW second). P5 (coordinator share negative under the sum rule and positive under the best-group rule in every coordinated set): fail (S3 best-group share $-0.064$, positive in 2 of 8 campaigns). P6 (no hub in the 10-agent flat team: no agent cited more than twice the median): pass (maximum 1.24 times the median).

\paragraph{The finer task with a rotated coordinator.} Five Qwen3-4B agents ranked the 13 anion-element groups of the measured compilation (groups with at least 30 compounds), where the databases' standalone orderings all differ (Kendall $\tau$ against the measured order: SNUMAT HSE06 0.90, MP $r^{2}$SCAN 0.85, JARVIS TB-mBJ 0.82, JARVIS OptB88vdW 0.79, MP GGA 0.69, computed from each database's full group order; the same TB-mBJ $>$ OptB88vdW $>$ GGA order against measured gaps that \citet{Borlido2019BandGapBenchmark} report). There were 20 flat campaigns and 100 coordinator-led campaigns, 20 with each database's holder as coordinator (8{,}499 model calls, 26 unparseable replies replaced by the fallback rule). Team $\tau$: flat $0.773$ $[0.724,0.819]$; coordinated by MP GGA $0.828$, MP $r^{2}$SCAN $0.808$, JARVIS OptB88vdW $0.836$, JARVIS TB-mBJ $0.822$, SNUMAT HSE06 $0.835$. Registered predictions: Q1 (role premium, pooled, interval excluding zero) fail, $-0.004$ $[-0.030,+0.021]$ (seed-level cluster bootstrap as for Q4); Q2 (premium positive in at least 4 of 5 rotations) fail, 3 of 5; Q3 (SNUMAT HSE06 in the top two and MP GGA last with probability at least 0.9 in both team types) fail (coordinated 0.73 and 0.995; flat 0.36 and 0.83); Q4 (coordinator raises team $\tau$, interval excluding zero) pass, $+0.052$ $[+0.006,+0.102]$, positive on average but not robust across seeds: the lower bound is just above zero and the effect was uneven across seeds (better in 62 of 100 pairs and in 10 of 20 seeds). Because each seed's flat run serves all five coordinator rotations, this interval is a seed-level cluster bootstrap ($B=20{,}000$) that resamples the 20 seeds with all five rotations of each; a $t$ interval on the 20 seed means gives $[-0.001,+0.105]$; Q5 (SNUMAT HSE06 coordinator outranks MP GGA coordinator with probability at least 0.9) fail, difference $+0.007$, probability 0.62; Q6 (flat team issues fewer distinct queries) pass, 0.125 against 0.146, difference $-0.021$ $[-0.038,-0.003]$. Shares by database in the flat team: MP $r^{2}$SCAN 0.253, JARVIS TB-mBJ 0.220, SNUMAT HSE06 0.210, JARVIS OptB88vdW 0.184, MP GGA 0.133 (Spearman correlation with standalone accuracy 0.7; coordinated teams 0.9).

\subsection{Further detail}

The meetings were read from the logs as recurring message and notebook-citation links, not scheduled meetings, and $v(S)$ was measured by replaying the trays of the members of $S$ (campaign counts and tray statistics in Section~S5).

\textbf{Flat teams loaded few distinct trays and trailed the random-tray search baseline.} Only 5--10\% of the trays of a flat team were distinct, and the five-agent flat team trailed the random-tray search baseline at the same budget ($0.756$ against $0.876$, sd $0.096$ over 20 random searches of 240 evaluations each, a search baseline whose seeds are not paired with those of the team runs); its members often measured the same families, although tray overlap is not shown to cause the shortfall: a coordination loss against the random-tray search baseline rather than against a nominal group of the same members, akin to the process loss by which a group falls short of what its members could reach with their efforts combined well \citep{Steiner1972GroupProcess}. Coordination, the management of dependencies between activities \citep{Malone1994Coordination}, is meant to reduce such loss, and misaligned or duplicated work between agents is a recurring failure of multi-agent language-model systems \citep{Cemri2025MASfail}.

As a process count, coordinated minus flat, paired by seed, the coordinator raised the share of distinct trays in every set, by $+1.75$ ($[+0.42,+3.25]$), $+11.67$ ($[+8.17,+15.33]$), and $+2.81$ ($[+1.56,+3.85]$) percentage points in sets S1, S2, and S3, and coordinated teams still repeated most of their trays. In set S1 only, sequential minus coordinated (five Qwen agents choosing in turn without a coordinator), the share differed by $+0.58$ percentage points ($[-0.67,+1.75]$). Two confounds stay open: the coordinator chose in an assigned order while flat agents chose at once, and in sets S1--S3 it always held the MP GGA ordering.

Under replay valuation, which values what each member did rather than an intervention on it \citep{Janzing2020FeatureRelevance,Heskes2020CausalShapley}, the coordinator's effect is instead assigned to the agents it directed: its role premium, the coordinator's share of team $\tau$ minus the share of the agent holding the same database in the flat team of the same seed, was $-0.004$ $[-0.030,+0.021]$ on the finer task (seed-level cluster bootstrap, since each flat run again serves all five rotations; 10 of 20 seeds positive).

In the flat teams the agent holding SNUMAT HSE06 carried the largest share (0.33 of team $\tau$ for five Qwen agents), the database that, with SNUMAT GGA, also contributed most in the database campaign.

\paragraph{Language-model teams.} Agents ran Qwen3-4B-Instruct-2507 or Nemotron 3.5 Lightning 30B-A3B (NVFP4) (no thinking mode), served with vLLM 0.29.0 \citep{Kwon2023vLLM} on an NVIDIA DGX Spark at temperature 0.7.

Twelve predictions for the extended runs were recorded on 4 October 2026, before the campaigns, in time-stamped commits of the project's version-controlled repository (files PREDICTION.md, included in the code release and listed in Section~S12), and six of the twelve held, including the three on the coordinator and on duplicated trays; prompts, the retry rule, and parse failures are given with them (Section~S5).

\clearpage
\section{The emergent hierarchical swarm and several bodies of one kind}
\label{app:swarm}

\subsection{Setting and results}
\label{sec:swarm}

Here a \emph{swarm} is a team of computational members whose coordinator re-ranks them into tiers by their record after every batch (the adaptive team of the main text), not a self-organizing swarm without central control. In the games above the structure of the team is declared before the campaign: a pool, a star, a set of meetings, or a coordinator. A materials group that runs many advisers at once can instead let the team organize itself during the campaign, as a swarm \citep{Bonabeau1999Swarm} of heterogeneous members that follow local rules and share part of what they learn. The question here is whether such a team outperforms a structure fixed by its designer, and how the contribution of each member changes when roles are not assigned in advance. Seven decision bodies run the Al--Mg--Si campaign of Section~2.4 of the main text together (168 compounds, batches of four, PBE as the receiver): five hold a low-cost ordering (MACE-MP-0 relaxation, NEP89 MD and relaxation, MEAM MD, and the Materials Project archive, which covers 36 of the compounds), one is an active learner with no prior ordering, and one is a random scout. Every member scores the remaining compounds with its own rule, mixing its ordering with a composition surrogate fitted on the receiver results it has seen and trusting the ordering by its agreement with those results, and every member sees only the results its position in the team gives it access to.

\begin{figure}[!ht]
\centering
\includegraphics[width=\textwidth]{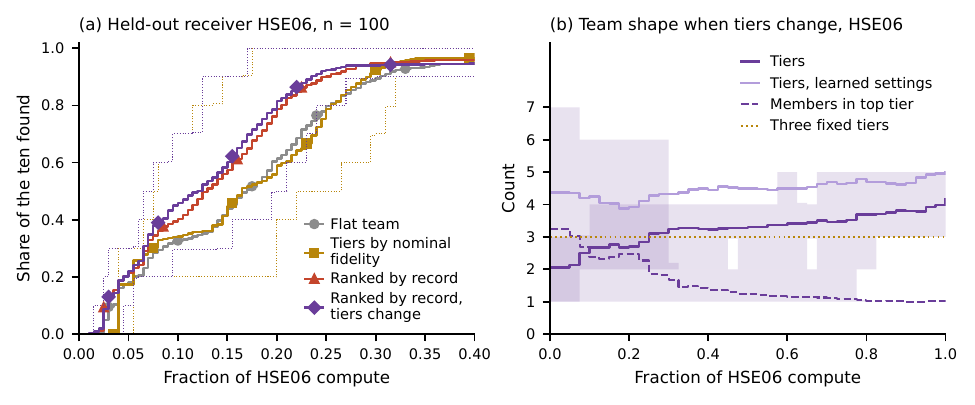}
\caption{On the held-out HSE06 receiver the number of tiers of the team whose tiers change varied between one and six during a campaign without lowering its recall below that of the three fixed tiers. (a) Share of the ten compounds with the lowest formation energy per atom found against the fraction of HSE06 compute; line, mean over n = 100 campaigns; dotted lines, 5--95\% range over campaigns for the fixed tiers and for tiers that change. (b) Number of tiers and number of members in the top tier of the team whose tiers change, mean and 5--95\% band over campaigns (learned settings: mean number of tiers only); dotted line, the three fixed tiers.}
\label{fig:swarmdyn}
\end{figure}

Eight structures were compared over 100 campaigns each. Four are fixed: no communication, a flat team with equal batch slots and full sharing, a fixed supervisor (the active learner) that fills each batch from the members' scores, and fixed tiers assigned by nominal fidelity (MACE-MP-0 and the archive above the potentials, the potentials above the learner and the scout, with batch slots in the ratio 4:2:1 and results passed down from higher tiers). The other four are emergent, in the sense of a non-egalitarian emergent hierarchical swarm: all members start in one tier; after every batch each member's prequential score (the agreement between the scores it gave the batch before the results arrived and the results it can see) updates a running average and the members are re-ranked; a member with a negative score falls to a floor weight; and members whose current orderings agree merge into a temporary cell that proposes once, with the weight of its strongest member, and dissolves when the agreement drops. Every position in an emergent team is therefore temporary. The emergent structures differ in whether the shape of the hierarchy is also allowed to change. With a \emph{three-tier} shape the members are re-ranked into one leader, a second tier of two and a third tier, so the number and size of the tiers are fixed. With a \emph{changing shape} the tiers are re-derived after every batch from the gaps between the running scores: a new tier begins wherever the drop from one member to the next exceeds a threshold, so the team stays flat while the records are close, deepens as they separate, and flattens again where they converge, and the advantage an incumbent receives for its current tier expires after a fixed number of batches, so a position must then be held on the record alone. Each protocol is either implemented with fixed cold-start parameters or learned: its parameters were trained with the cross-entropy method \citep{Rubinstein1999CrossEntropy} on the PBE receiver only and then applied unchanged to two held-out receivers, HSE06 \citep{Heyd2003HSE,Heyd2006HSEErratum} and $r^{2}$SCAN \citep{Kingsbury2022r2SCAN} computed for the same compounds. All 128 sub-teams were run for each structure, so each member's Shapley value is measured within that structure (protocol and role statistics in the subsections below).

\textbf{The emergent team outperforms every fixed structure on all three receivers} (Figure~9a of the main text). With the three-tier shape, on PBE its early share is $0.70$, against $0.64$ for the fixed tiers, $0.61$ for the flat team and $0.58$ for the fixed supervisor; on the held-out receivers it is $0.49$ against at most $0.40$ (HSE06) and $0.46$ against at most $0.44$ ($r^{2}$SCAN), gaps of $+0.088$ (90\% bootstrap interval $+0.071$ to $+0.105$) and $+0.021$ ($+0.006$ to $+0.034$) over 100 campaigns per structure, whose seeds differ between structures, so the intervals are unpaired; and the compute needed to find eight of the ten falls from $0.23$ to $0.19$ of the HSE06 total. The fixed tiers, which rank the members by their nominal fidelity, improve on the flat team by at most $0.03$ and fall below it on HSE06. \textbf{Merging is the part of the protocol that matters most.} The cell that forms most often (about half of the early rounds on every receiver) contains every member except MACE-MP-0: once the potentials and the archive have lost the trust of their own orderings, they all propose from the same surrogate fitted on the same results, and merging them returns their duplicated batch slots to the leader. Without cells the protocol falls below the flat team on HSE06 ($0.34$); without tiers it keeps most of the gain ($0.46$; both ablations with their 90\% bootstrap intervals in Figure~9a of the main text). MACE-MP-0 holds the leader position in 80--88\% of the early rounds and the archive in the rest. \textbf{Learning the promotion protocol does not add to the implemented one}: the learned parameters change the leader in most campaigns, against at most one campaign in five for the implemented parameters, but the learned and implemented teams reach the same value on the held-out receivers ($0.49$ and $0.49$ on HSE06, $0.46$ and $0.46$ on $r^{2}$SCAN), so the implemented parameters are an adequate cold start for this team.

\textbf{Tiers that are temporary and change shape keep the gain} (Figure~9a of the main text). Under the three-tier protocol the hierarchy is nearly static: 96\% of the members keep their tier from one batch to the next, and in most campaigns the leader holds its position for the whole campaign. With a changing shape the number of tiers changes about 13 times in a campaign of 42 batches and ranges from one to six. After the first batch the team has two tiers on average (it is still flat in about one campaign in eight), and it deepens to about four tiers by the end, as the records of the members separate. The top tier first holds about three members and narrows towards one, MACE-MP-0, as the campaign proceeds (fewer than 1.4 members on average after 40\% of the HSE06 compute); a member's stay in the top tier lasts a median of three to four batches, and the NEP89 potentials, MEAM MD and the active learner each lead in 6--13\% of the early batches, when a few PBE results cannot yet separate their orderings from that of MACE-MP-0. The changing shape reaches the value of the three-tier shape on PBE ($0.69$ against $0.70$, difference $-0.008$, 90\% interval $-0.023$ to $+0.006$) and $r^{2}$SCAN ($0.47$ against $0.46$) and exceeds it on HSE06 ($0.51$ against $0.49$, difference $+0.024$, interval $+0.002$ to $+0.046$). The learned version sets a narrower gap, builds four to five tiers on average, and reaches the same values. The contribution follows the positions: the active learner receives $+0.028$ under the changing shape against $+0.014$ under three tiers (PBE), and on HSE06 the NEP89 relaxation receives $+0.021$ against $0.000$, while MACE-MP-0 keeps the largest contribution ($0.22$ on PBE, $0.20$ on HSE06). For a group running many agents, the organization chart therefore does not need to be fixed in advance, either in who holds a tier or in how many tiers there are: a hierarchy that forms, deepens and dissolves with the record performs as well as a fixed three-tier one, and members contribute most in the stages of the campaign in which they led.

\textbf{The structure moves contribution to the members it promotes} (Figure~9b of the main text). Relative to the flat team, the emergent team gives MACE-MP-0 $+0.06$ and the archive $+0.04$ (PBE) and $+0.09$ (HSE06) more contribution, and takes it from MEAM MD ($-0.008$ PBE, $-0.045$ HSE06) and the NEP89 relaxation ($-0.005$, $-0.017$); NEP89 MD, the third member whose potential wrongly predicts Mg$_2$Si to be unstable, is not down-weighted ($+0.001$, $+0.021$). The archive gains most on HSE06, where it holds the leader position in a fifth of the early rounds (Table~\ref{tab:swarmroles}); a fixed tier cannot give a member a position that changes with the campaign. \textbf{No team of seven reaches the best single member}: MACE-MP-0 alone has an early share of $0.81$ on PBE and $0.73$ on HSE06. The emergent structure closes 43\% of the gap between the flat team and this member on PBE and 30\% on HSE06, without being told which member is the best. For a group that runs many agents, a team that promotes members by their record and merges members that have become redundant recovers part of the value lost to pooling; the attribution computed within the emergent structure separates the contribution a member receives for its knowledge from the contribution it receives for the position it has been promoted to.

\subsection{Several bodies of one kind}
\label{sec:replicated}

\begin{figure}[t]
\centering
\includegraphics[width=\textwidth]{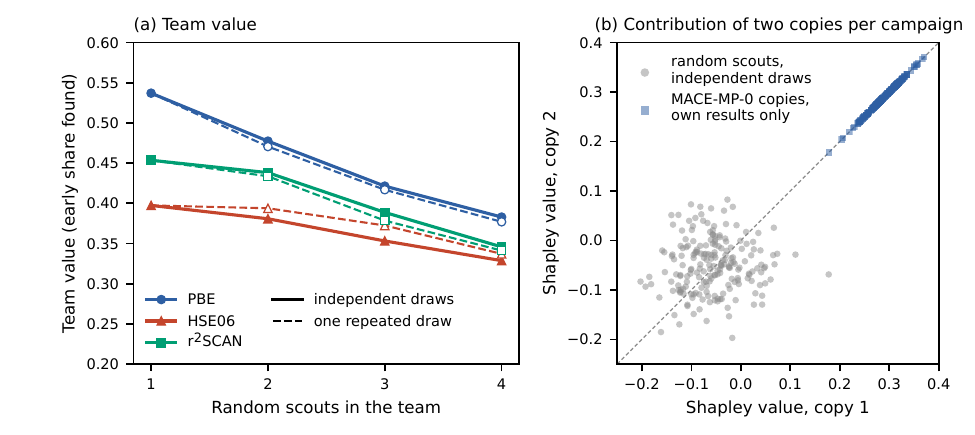}
\caption{Several decision bodies of one kind as separate players (computed Al--Mg--Si campaign, 200 campaigns per coalition). (a) Early share of the ten compounds with the lowest formation energy per atom found by a team of MACE-MP-0, the active learner and one to four random scouts, per receiver; solid: each scout draws its own random order; dashed: the scouts repeat one draw. (b) Shapley value of copy 1 against copy 2 in each campaign (PBE receiver): two of three random scouts with independent draws (gray), and two copies of MACE-MP-0 that see only the results they proposed (blue); diagonal: equal contribution.}
\label{fig:replicated}
\end{figure}

The swarm above holds one body of each kind. A materials group often runs several of one kind: three students each choosing compositions at random as scouts, a random-search script started with three seeds, two group members who each relax the pool with MACE-MP-0, or several agents of one language model. Whether such copies form one player or several decides how a contribution statement is written, so the question was tested on the Al--Mg--Si campaign of Section~\ref{sec:swarm} (flat team, batches of four, PBE, HSE06 and $r^{2}$SCAN receivers). In the first game MACE-MP-0 and the active learner are joined by one to four random scouts, each drawing its own random order of the pool, or, as a control, all repeating one draw. In the second game the active learner and one scout are joined by two copies of MACE-MP-0, which either share all results or see only the results they proposed. Every coalition of every game was run in the same 200 campaigns, so the copies of a campaign meet identical conditions, and the Shapley values are exact and computed per campaign (Figure~\ref{fig:replicated}).

\textbf{Random scouts with their own draws are separate players.} Two such scouts receive different contribution in 86--93\% of the campaigns, by about $0.06$ on average, because each draw places different compounds in the early batches (Figure~\ref{fig:replicated}b). Their expected contributions are equal ($-0.056$, $-0.050$ and $-0.053$ for three scouts on PBE, with overlapping intervals), and a given scout receives positive contribution in 16--17\% of the PBE campaigns and in up to 39\% of the $r^{2}$SCAN campaigns. An audit of one campaign that merged the scouts into one player would hide which scout found the compound; merging them also changes the total, because the scouts then enter every coalition together: with four scouts their joint contribution is $-0.204$ as one player against $-0.178$ as the sum of four (PBE). Scouts that repeat one draw receive identical contribution in every campaign, as the symmetry axiom (members that add the same to every coalition receive the same value) requires. \textbf{The number of copies is a design variable}: each added scout takes batch slots from MACE-MP-0, and the early share falls from $0.54$ with one scout to $0.38$ with four (PBE; $0.40$ to $0.33$ for HSE06, $0.45$ to $0.35$ for $r^{2}$SCAN), whether the scouts draw independently or not. The independent scouts lower the outcome additively: the Harsanyi dividend of a pair of scouts is between $-0.003$ and $+0.012$.

\textbf{Copies of one potential are redundant even when their outputs differ.} When the two MACE-MP-0 copies see only their own results, their surrogate parts are fitted on different compounds, and in a team of the two copies alone their orderings differ in 89--94\% of the batches (0\% when the results are shared). Their contribution is nevertheless equal in every campaign ($+0.289$ each on PBE), because every sub-team containing copy 1 performs as the same sub-team containing copy 2: the copies are interchangeable players, and interchangeability, not identical outputs within the team, is what the symmetry axiom tests. The pair dividend is $-0.68$ on PBE ($-0.60$ HSE06, $-0.34$ $r^{2}$SCAN), equal in size and opposite in sign to the value of one copy over an empty team, so the second copy adds nothing to the first. For a group, two members running one potential on one pool are one adviser whose contribution is split between them, whereas scouts with their own random draws are several advisers, each of which can gain or lose contribution for the compounds its own draw reached. The same distinction applies to agents of one language model: sampled at nonzero temperature, or holding different knowledge as in Section~\ref{si:teams}, they are separate players; queried deterministically with one prompt, they are one.

\subsection{Protocols and role statistics}

This subsection specifies the structures of Section~\ref{sec:swarm} and gives the role and shape statistics of the emergent teams.

\paragraph{Members and local rules.} A member holding an ordering scores each remaining compound by $w\,r_{\mathrm{own}} + (1-w)\,r_{\mathrm{sur}}$, where $r_{\mathrm{own}}$ is its ordering as an evenly spaced rank score (0.5 for compounds it does not hold), $r_{\mathrm{sur}}$ is the rank score of a composition-kernel lower confidence bound (bandwidth 0.12 in the Al, Mg, Si fractions) fitted on the receiver results the member has seen, and $w = \max(0,\tau)\,n/(n+8)$ with $\tau$ the Kendall $\tau$ between its ordering and those $n$ results ($w=0.5$ for $n<3$). The active learner uses $r_{\mathrm{sur}}$ alone; the scout uses a fixed random order.

\paragraph{Batch slots.} Each unit (a member, or a cell of merged members) carries a persistent slot contribution that grows by $k\,w_u/\sum w$ per batch ($k=4$) and falls by one for every compound the unit places; the unit with the largest contribution places its best-scored compound not yet in the batch. A cell scores compounds by the mean of its members' scores. The fixed supervisor places all four compounds by the mean score of all members.

\paragraph{Sharing.} In the flat team and under the fixed supervisor every result is shared. Without communication a member sees only the results of the compounds it placed. In the tiered structures a member sees the results of compounds placed at its own tier or above, the leader sees all results, and in the emergent structures the members of one cell share within the cell.

\paragraph{Promotion protocol.} After each batch, member $i$ receives the prequential score $q_i = \tau_i\,n_i/(n_i+8)$, where $\tau_i$ is the Kendall $\tau$ between the scores $i$ gave the evaluated compounds just before their results arrived and those results, over the $n_i$ results $i$ can see. The running average is $m_i \leftarrow (1-\alpha)\,m_i + \alpha\,q_i$. Members are ranked by $m_i$ plus an incumbency margin $\delta$ per tier above the third; the first $n_\ell$ become leaders, the next two the second tier, the rest the third. Slot weights are $\rho^{2}$, $\rho$ and 1 for the three tiers, and $f$ for a member with $m_i<0$. Two members merge when the Kendall $\tau$ between their current scores (on at most 60 remaining compounds) reaches $\mu$; cells are the connected components and are recomputed every batch. The implemented parameters are $\alpha=0.4$, $\rho=2$, $\delta=0.05$, $\mu=0.85$, $f=0.25$, $n_\ell=1$; the learned parameters (cross-entropy method, population 16, six iterations, 20 training campaigns on PBE) are $\alpha=0.42$, $\rho=3.8$, $\delta=0.007$, $\mu=0.78$, $f=0.29$, $n_\ell=1$. The ablations set $\mu>1$ (no cells) or $\rho=1$, $f=1$ (no tiers).

\paragraph{Changing shape.} The changing-shape protocol uses the same scores, running averages, floor weight and cells, but no fixed number of tiers. After each batch the members are sorted by $m_i + \delta\,(T-t_i)\,[s_i \le L]$, where $T$ is the current number of tiers, $t_i$ the current tier of member $i$ (1 at the top), $s_i$ the number of batches since its tier last changed, and $L$ the tenure of the incumbency margin. A new tier begins between consecutive members whenever their sorted scores differ by more than the gap $g$, so $T$ ranges from one (a flat team) to the number of members. Slot weights are $\rho^{\,T-t_i}$, and sharing follows the tiered rule above. The team starts flat. The implemented parameters are those of the three-tier protocol with $g=0.08$ and $L=4$; the learned parameters (same training) are $\alpha=0.41$, $\rho=2.8$, $\delta=0.005$, $\mu=0.83$, $f=0.11$, $g=0.034$, $L=4.9$.

\paragraph{Value and attribution.} The team value is the early share of the ten compounds with the lowest formation energy per atom found over the first quarter of the receiver compute. Team values use 100 campaigns per structure and receiver. The Shapley values use all 128 sub-teams per structure, each the mean of 16 campaigns, so they sum to the 100-campaign team value only approximately; the empty team selects at random.

\begin{table}[!ht]
\centering
\small
\caption{Roles in the emergent teams with the three-tier shape over the first twelve batches (48 compounds), 100 campaigns per receiver.}
\label{tab:swarmroles}
\begin{tabular}{lcccccc}
\toprule
 & \multicolumn{3}{c}{Implemented protocol} & \multicolumn{3}{c}{Learned protocol} \\
\cmidrule(lr){2-4}\cmidrule(lr){5-7}
Receiver & PBE & HSE06 & $r^{2}$SCAN & PBE & HSE06 & $r^{2}$SCAN \\
\midrule
Leader share, MACE-MP-0 & 0.88 & 0.80 & 0.85 & 0.82 & 0.74 & 0.78 \\
Leader share, MP archive & 0.11 & 0.20 & 0.14 & 0.18 & 0.25 & 0.22 \\
Leader changes per campaign & 0.08 & 0.21 & 0.12 & 0.85 & 0.96 & 0.76 \\
Tier persistence between batches & 0.96 & 0.96 & 0.96 & 0.92 & 0.92 & 0.92 \\
Cells per batch & 1.18 & 1.14 & 1.16 & 1.08 & 1.05 & 1.06 \\
Share of batches with the five-member cell$^{a}$ & 0.52 & 0.55 & 0.57 & 0.52 & 0.48 & 0.54 \\
\bottomrule
\end{tabular}

\smallskip
{\footnotesize $^{a}$ MP archive, NEP89 MD, NEP89 relaxation, MEAM MD and the active learner.}
\end{table}

\begin{table}[!ht]
\centering
\small
\caption{Shape of the emergent teams over the whole campaign (42 batches), 100 campaigns per receiver. Each entry is the range over the PBE, HSE06 and $r^{2}$SCAN receivers; the persistence and the leader changes refer to the first twelve batches, as in Table~\ref{tab:swarmroles}.}
\label{tab:swarmshape}
\setlength{\tabcolsep}{3pt}
\begin{tabular}{@{}p{0.43\textwidth}ccc@{}}
\toprule
 & Three tiers & Changing shape & \shortstack{Changing shape,\\ learned} \\
\midrule
Number of tiers, range over all batches & 3 & 1--6 & 1--7 \\
Mean number of tiers after the first batch & 3.00 & 2.05--2.08 & 4.37--4.40 \\
Mean number of tiers at the end & 3.00 & 4.04--4.22 & 5.00--5.24 \\
Share of campaigns flat after the first batch & 0 & 0.12--0.13 & 0 \\
Changes in the number of tiers per campaign & 0 & 12.8--13.4 & 12.8--13.7 \\
Members in the top tier after the first batch & 1 & 3.14--3.24 & 1.51--1.53 \\
Stay in the top tier, median (batches) & 42 & 3--4 & 6--14 \\
Tier persistence between batches & 0.96 & 0.73--0.75 & 0.72--0.73 \\
Leader changes per campaign & 0.08--0.21 & 3.22--3.32 & 1.43--1.88 \\
Early share of the ten found (PBE / HSE06 / $r^{2}$SCAN) & 0.70 / 0.49 / 0.46 & 0.69 / 0.51 / 0.47 & 0.70 / 0.51 / 0.47 \\
\bottomrule
\end{tabular}
\end{table}

\subsection{Further detail}

The team closed 43\% of the gap (computed from unrounded means) between the flat team ($0.61$) and MACE-MP-0 alone ($0.81$) without being told which member was best, a simple analogue of the credibility part of transactive memory, by which a group learns whom to trust for what \citep{Wegner1987TransactiveMemory,Lewis_2003}.

These organization results generate hypotheses; they do not confirm them.

A member allowed to sit in a second group, like a student shared between two laboratories and akin, for computational members such as these, to a boundary spanner \citep{Tushman1977BoundaryRoles}, raised it only with the learned, near-zero agreement threshold for sharing.

\clearpage
\section{Several swarms in one campaign}
\label{app:multiswarm}

\subsection{Setting and results}
\label{sec:multiswarm}

A large campaign is often run by several groups at once, each organized as its own swarm, that share one furnace, one beamline allocation, or one compute allocation. Three questions then arise that a single swarm does not pose: whether the same members do better as one swarm or as several, how the shared batch should be divided between swarms of the same or of different size, and how the contribution is divided first between the swarms and then between their members. The seven bodies of Section~\ref{sec:swarm} were grouped into two, three, or seven swarms (4\,+\,3; 2\,+\,5 with MACE-MP-0 and the active learner as the small swarm, or with NEP89 MD and the scout; 1\,+\,6 with MACE-MP-0 alone; 2\,+\,2\,+\,3 by nominal fidelity; seven swarms of one), and in a second set the seven-body swarm was joined by copies of itself (7\,+\,7, 7\,+\,7\,+\,7, and 7\,+\,3 with a small copy of MACE-MP-0, the learner and a scout). Every swarm runs the emergent organization of Section~\ref{sec:swarm} among its own members, and all swarms share one batch of four compounds, so every configuration spends the same receiver compute. The batch is divided between swarms in one of three ways: an equal share per swarm, a share proportional to swarm size, or a share by record, in which the promotion rule of the emergent team is applied to the swarms themselves (the swarm whose averaged scores have best anticipated the results it has seen receives the largest share). Results of one swarm are either published to the others after every batch or kept within the swarm. Contribution is computed at two levels: the Shapley value of each swarm, and for the members the Owen value \citep{Owen1977APrioriUnions}, which treats the swarms as fixed unions: it first divides the outcome between the swarms by their Shapley values in the game played by whole swarms, and then divides each swarm's share between its members. Equivalently, each Harsanyi dividend is split equally among the swarms it touches, then equally among the members of that coalition inside each swarm (not among all of the swarm's members). The difference between a member's Owen and Shapley values is therefore the contribution it gains or loses through the swarm it belongs to. Team values use 100 matched campaigns per configuration on the three receivers; the contribution games run every sub-team in 100 campaigns on the PBE and HSE06 receivers (configurations and values in the subsections below).

\begin{figure}[t]
\centering
\includegraphics[width=\textwidth]{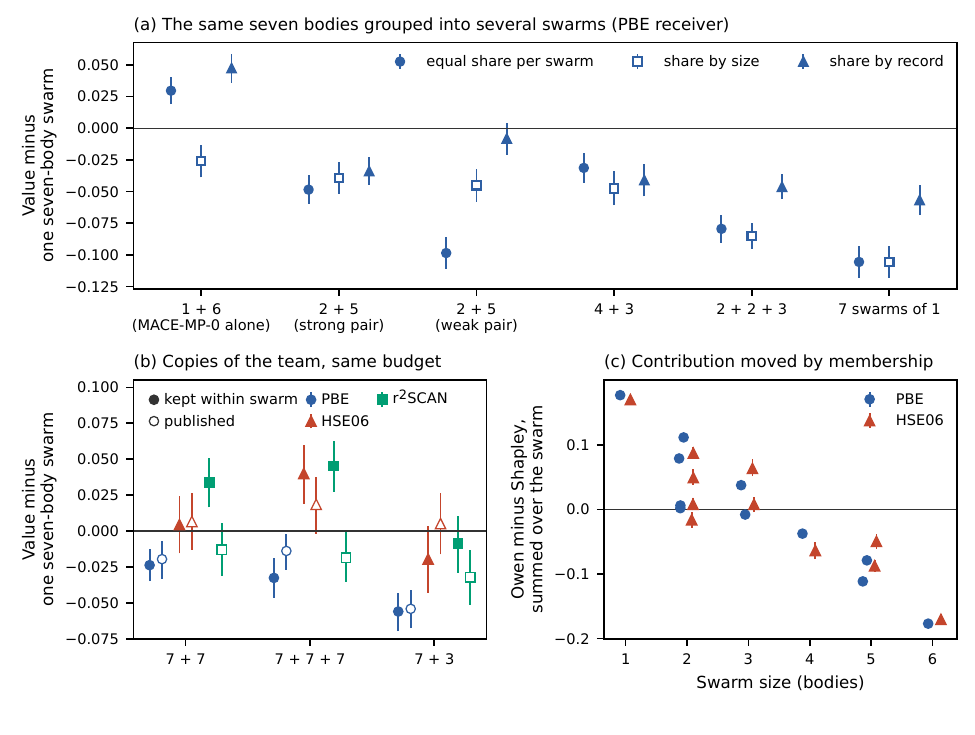}
\caption{Several swarms sharing one batch of four compounds in the Al--Mg--Si campaign; every swarm uses the emergent organization of Section~\ref{sec:swarm} among its members. Color marks the receiver as in Figure~9a of the main text (PBE blue, HSE06 red, r$^2$SCAN green). (a) The seven bodies grouped into several swarms, results published between swarms, PBE receiver: early share of the ten compounds with the lowest formation energy per atom found minus the value of one seven-body swarm, per rule for dividing the batch between swarms (marker shape); mean and 90\% bootstrap interval over 100 matched campaigns. ``Strong pair'': MACE-MP-0 and the active learner; ``weak pair'': NEP89 MD and the random scout. (b) The seven-body swarm joined by copies of itself, equal share per swarm, per receiver, with results kept within each swarm (filled) or published (open). (c) Owen value (the Shapley value computed with the swarms as fixed blocs: swarms first divide the team value among themselves, then each swarm divides its share among its members) minus Shapley value summed over the members of each swarm, against the size of the swarm, for the two- and three-swarm groupings of (a), equal share, published results; mean and 90\% interval over 100 campaigns.}
\label{fig:multiswarm}
\end{figure}

\textbf{Splitting the same members into several swarms lowers the outcome unless the split gives the strongest member more of the batch} (Figure~\ref{fig:multiswarm}a). With an equal share per swarm, seven swarms of one lose $0.106$ against one swarm on PBE, and three swarms grouped by nominal fidelity lose $0.08$; members in different swarms cannot merge into one cell, so the members that have lost trust in their own orderings keep their duplicated slots. The one grouping that gains places MACE-MP-0 in a swarm of its own: with a share by record it gains $+0.047$ (90\% interval $+0.036$ to $+0.059$) on PBE, $+0.097$ on HSE06 and $+0.025$ on $r^{2}$SCAN, because a separate swarm receives a fixed share of every batch from the first batch on, whereas inside one swarm MACE-MP-0 starts level with the others and acquires weight only through its record. The gain depends on knowing in advance which member to separate: with an equal share, a small swarm of the two weakest members takes half of every batch and loses $0.098$ on PBE and HSE06. \textbf{Dividing the batch by record protects the campaign when the strong swarm is not known}: it brings the weak pair back to the value of one swarm ($-0.008$, interval $-0.021$ to $+0.004$, on PBE; $-0.009$ on HSE06; $+0.013$ on $r^{2}$SCAN) and gives the highest value of the three rules in five of the six groupings on every receiver. A share by size is never the best rule here: it gives each member the same slot weight wherever it sits, so it removes the advantage of a separate swarm without restoring the cells of a single one.

\textbf{Copies of the team help where no member's ordering is reliable, and only when they do not share results} (Figure~\ref{fig:multiswarm}b). Copies of equal size receive equal contribution (7\,+\,7: $+0.28$ and $+0.27$ on PBE, $+0.20$ each on HSE06), as the symmetry axiom expects of swarms that are interchangeable up to run-to-run noise. On PBE, which MACE-MP-0 orders closely, a second or third copy lowers the outcome (three copies: $-0.033$ with results kept within each swarm, $-0.014$ published), because the leading member of each copy now receives a smaller part of the batch. On the held-out receivers three copies that keep their results to themselves gain $+0.040$ (HSE06) and $+0.045$ ($r^{2}$SCAN) over one swarm, and on $r^{2}$SCAN publishing the results removes the gain (isolated minus published, $+0.064$, interval $+0.046$ to $+0.081$). Copies that do not share results fit their surrogates on different compounds, so their orderings separate (Section~\ref{sec:replicated}) and their batches cover more of the pool early; copies that share results converge on one ordering. A small copy joined to the full team (7\,+\,3) takes half of every batch under an equal share and lowers the outcome on PBE by $0.056$.

\textbf{The contribution of a member depends on the size of its swarm} (Figure~\ref{fig:multiswarm}c). The Shapley value of a swarm follows what the swarm achieves alone, not how many members it has: MACE-MP-0 as a swarm of one receives $+0.38$ and the six-body swarm $+0.18$ (PBE, equal share, results kept within each swarm). Because the Owen value divides each swarm's share only among its own members, a member of the smaller swarm receives more contribution than its Shapley value in all 16 two-swarm settings (four groupings, two receivers, equal and size shares), and members of the larger swarm receive less, by up to $0.19$ summed over the swarm. The shift does not require the small swarm to be strong: NEP89 MD, with the scout as its only partner, gains $+0.10$ over its Shapley value on PBE and HSE06. For swarms of nearly equal size (2\,+\,2\,+\,3) the two values differ by at most $0.02$ per swarm. For a consortium of computational members such as these, the two values answer different questions. The Owen value is the division when the swarms are organizations with their own budgets whose split between groups is set first, by the game played between the swarms as wholes, such as two laboratories sharing one facility; the Shapley value over members is the division of contribution for knowledge, independent of how the members were grouped. A contribution statement for several swarms should name which of the two it reports.

\subsection{Definitions}

\paragraph{Configurations.} The seven bodies are MACE-MP-0 relaxation, NEP89 MD, NEP89 relaxation, MEAM MD, the Materials Project archive, the active learner and the random scout. The groupings are: 4\,+\,3, \{MACE-MP-0, NEP89 MD, MEAM MD, scout\} and \{archive, NEP89 relaxation, learner\}; 2\,+\,5 strong, \{MACE-MP-0, learner\} and the other five; 2\,+\,5 weak, \{NEP89 MD, scout\} and the other five; 1\,+\,6, \{MACE-MP-0\} and the other six; 2\,+\,2\,+\,3, \{MACE-MP-0, archive\}, \{NEP89 MD, NEP89 relaxation\} and \{MEAM MD, learner, scout\}; and seven swarms of one. The copies are the seven-body swarm joined by one or two copies of itself, or by a copy of \{MACE-MP-0, learner, scout\}. A copied ordering differs from its original only through the results it sees; a copied scout draws its own random order.

\paragraph{Dividing the batch.} Each swarm $j$ carries a slot contribution that grows by $k\,W_j/\sum W$ per batch ($k=4$) and falls by one for each compound it places, as for the units inside a swarm (Section~\ref{app:swarm}). The swarm with the largest contribution fills the next slot, and the slot is passed to one of its units by the swarm's own slot contributions. The weights are $W_j=1$ (equal share), $W_j=|j|$ (share by size), or, for the share by record, the tier weights of the three-tier protocol applied to swarms: the swarm prediction for a compound is the mean score its members gave it before the result arrived, its prequential score is $\tau_j\,n_j/(n_j+8)$ over the $n_j$ results the swarm can see, a running average with rate 0.4 ranks the swarms with an incumbency margin of 0.05, and the leading swarm receives weight 4, the next two weight 2, the others weight 1, and a swarm with a negative average weight 0.25.

\paragraph{Sharing and value.} Inside a swarm, results are shared by the rule of the emergent team (Section~\ref{app:swarm}). Published results of other swarms are visible to every member after the batch. The team value is the early share of the ten compounds with the lowest formation energy per atom found over the first quarter of the receiver compute, averaged over 100 campaigns; the campaign seed does not depend on the configuration or the coalition, so differences are paired.

\paragraph{Owen value.} For unions $B_1,\dots,B_u$ with $i\in B_k$,
\begin{equation*}
\mathrm{Ow}_i=\sum_{T\subseteq U\setminus\{k\}}\ \sum_{R\subseteq B_k\setminus\{i\}} \frac{|T|!\,(u-|T|-1)!}{u!}\,\frac{|R|!\,(|B_k|-|R|-1)!}{|B_k|!}\Big[v(Q_T\cup R\cup\{i\})-v(Q_T\cup R)\Big],
\end{equation*}
where $Q_T$ is the union of the swarms in $T$. The values of each union sum to its Shapley value in the game between whole swarms, and with one union, or with every member in its own union, the Owen value equals the Shapley value; both properties are checked in the self-test of the experiment script. Each contribution game runs all 128 sub-teams of the seven bodies in 100 campaigns; a sub-team keeps its members' swarms, and a swarm with no members drops out.

\begin{table}[!ht]
\centering
\small
\caption{Early share of the ten compounds with the lowest formation energy per atom found by several swarms sharing one batch (emergent organization inside each swarm), mean over 100 campaigns. Columns: share of the batch per swarm (equal, by size, by record) with published results, and equal share with results kept within each swarm. The one-swarm row ($0.702$, $0.503$, $0.481$ before rounding) comes from runs separate from those of Section~\ref{sec:swarm} and Figure~9 of the main text ($0.696$, $0.487$, $0.465$).}
\label{tab:multiswarm}
\resizebox{\textwidth}{!}{%
\begin{tabular}{lcccccccccccc}
\toprule
 & \multicolumn{4}{c}{PBE} & \multicolumn{4}{c}{HSE06} & \multicolumn{4}{c}{$r^{2}$SCAN} \\
\cmidrule(lr){2-5}\cmidrule(lr){6-9}\cmidrule(lr){10-13}
Swarms & eq. & size & rec. & isol. & eq. & size & rec. & isol. & eq. & size & rec. & isol. \\
\midrule
One swarm of 7 & 0.70 & 0.70 & 0.70 & 0.70 & 0.50 & 0.50 & 0.50 & 0.50 & 0.48 & 0.48 & 0.48 & 0.48 \\
4 + 3 & 0.67 & 0.65 & 0.66 & 0.68 & 0.49 & 0.47 & 0.46 & 0.49 & 0.46 & 0.45 & 0.45 & 0.50 \\
2 + 5, strong pair & 0.65 & 0.66 & 0.67 & 0.64 & 0.48 & 0.45 & 0.51 & 0.47 & 0.47 & 0.45 & 0.47 & 0.49 \\
2 + 5, weak pair & 0.60 & 0.66 & 0.69 & 0.57 & 0.41 & 0.47 & 0.49 & 0.38 & 0.45 & 0.46 & 0.49 & 0.44 \\
1 + 6, MACE-MP-0 alone & 0.73 & 0.68 & 0.75 & 0.69 & 0.55 & 0.47 & 0.60 & 0.53 & 0.50 & 0.46 & 0.51 & 0.48 \\
2 + 2 + 3 & 0.62 & 0.62 & 0.66 & 0.62 & 0.40 & 0.39 & 0.44 & 0.46 & 0.44 & 0.42 & 0.46 & 0.50 \\
7 swarms of 1 & 0.60 & 0.60 & 0.65 & 0.54 & 0.37 & 0.37 & 0.39 & 0.40 & 0.42 & 0.42 & 0.43 & 0.44 \\
7 + 7 (copies) & 0.68 & 0.68 & 0.69 & 0.68 & 0.51 & 0.51 & 0.51 & 0.51 & 0.47 & 0.47 & 0.46 & 0.51 \\
7 + 7 + 7 (copies) & 0.69 & 0.69 & 0.69 & 0.67 & 0.52 & 0.52 & 0.51 & 0.54 & 0.46 & 0.46 & 0.47 & 0.53 \\
7 + 3 (copies) & 0.65 & 0.67 & 0.64 & 0.65 & 0.51 & 0.49 & 0.48 & 0.48 & 0.45 & 0.47 & 0.45 & 0.47 \\
\bottomrule
\end{tabular}}
\end{table}

With a flat organization inside each swarm instead of the emergent one, the same groupings reach lower values throughout (one flat swarm of seven: 0.61, 0.38 and 0.44 on the three receivers), and MACE-MP-0 in its own swarm with a share by record gains $+0.13$ on PBE and $+0.18$ on HSE06 over the flat swarm of seven.

\clearpage
\section{Swarms that combine, nest, and separate}
\label{app:morph}

\subsection{Setting and results}
\label{sec:morphswarm}

In Section~\ref{sec:multiswarm} the grouping of the members into swarms is fixed before the first batch. A consortium does not stay fixed: two groups whose recommendations agree hold joint meetings, a group whose recommendations keep failing against the measurements is placed under the group whose recommendations hold, and a group whose members disagree divides. Experiment~17 lets the structure of the Al--Mg--Si campaign change in this way while it runs. The structure has two levels above the members: independent groups, each made of a head swarm and zero or more nested swarms. Every second batch a coordinator, who sees every result, may apply five operations, each triggered by the orderings the members have just given the open pool and by the groups' records: \emph{fuse} two single swarms whose orderings agree (Kendall $\tau\ge0.6$) and whose records differ by at most $0.1$; \emph{nest} the weaker of two agreeing groups under the stronger; \emph{adopt} a group with a negative record into the group with the best record; \emph{split} a swarm into the components of its agreement graph ($\tau\ge0.2$); and \emph{detach} a nested swarm that no longer agrees with its head. Nesting changes who sees which results: the head swarm's results reach its nested swarms, and their results reach the head's leader, while separate groups see only their own. Inside every swarm the emergent organization of Section~\ref{sec:swarm} runs unchanged. Four rule sets were compared with the fixed structure from five starting structures (seven single members; 2\,+\,2\,+\,3 by level of theory; 4\,+\,3; 2\,+\,5 with the weak pair; one swarm of seven): combine only (fuse, nest, adopt), separate only (split, detach), both kinds with the thresholds above as cold-start defaults, and both with thresholds learned by the cross-entropy method on training seeds disjoint from the evaluation seeds. When the structure changes, no fixed set of unions exists, so contribution uses the level structure value \citep{Winter1989Levels}, the generalization of the Owen value to nested unions, evaluated for each structure that was in force and averaged with weights equal to the compute spent under it in the early window (definitions, learned thresholds and all values in the subsections below).

\begin{figure}[t]
\centering
\includegraphics[width=\textwidth]{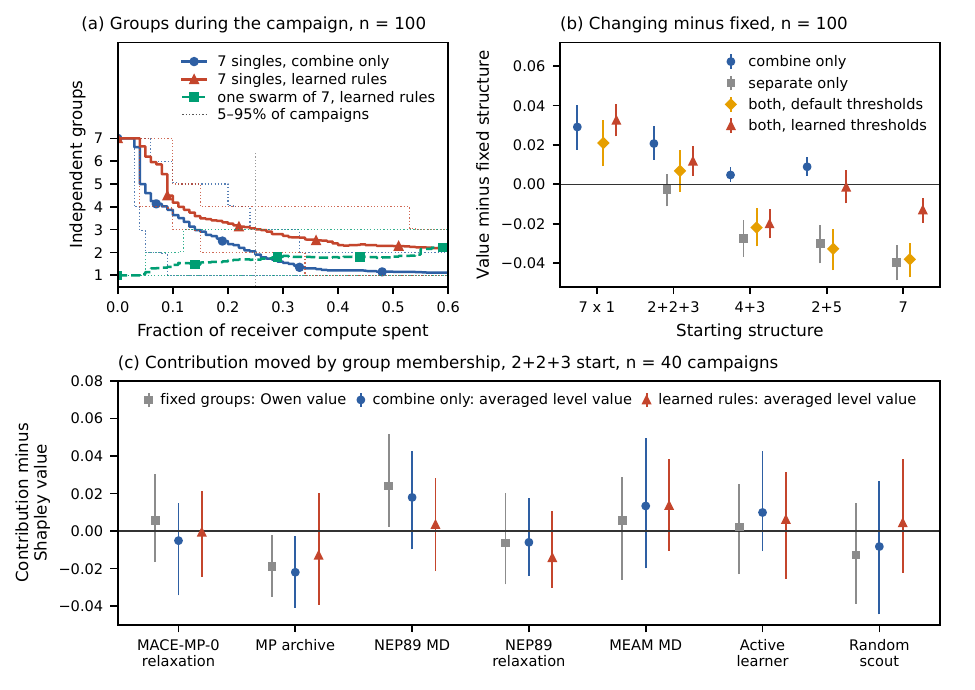}
\caption{Swarms that combine, nest and separate during the Al--Mg--Si campaign (results kept within each group, batch divided between groups by record). (a) Number of independent groups against the receiver compute spent, PBE receiver; mean (solid or dashed line) and 5th--95th percentile range over 100 campaigns (thin dotted outlines of the same color); the gray vertical line marks the end of the early window. (b) Early share of the ten compounds with the lowest formation energy per atom found with a changing structure minus the value of the fixed structure with the same start, PBE receiver; starts are seven single members, 2\,+\,2\,+\,3 by level of theory, 4\,+\,3, 2\,+\,5 with the weak pair, and one swarm of seven; mean and 90\% bootstrap interval over 100 matched campaigns. (c) Contribution minus Shapley value per member for the 2\,+\,2\,+\,3 start, PBE receiver: Owen value of the fixed groups, and the level structure value averaged over the structures in force for the two changing rule sets; mean and 90\% interval over 40 campaigns.}
\label{fig:morphswarm}
\end{figure}

\textbf{Combining never lowered the outcome; separating lowered it in most settings} (Figure~\ref{fig:morphswarm}b). With combination only, the changing structure matched or exceeded the fixed one from all four multi-group starts on all three receivers, with intervals above zero in 10 of the 12 settings; the largest gains start from seven single members ($+0.029$ on PBE, $+0.027$ on HSE06, $+0.018$ on $r^{2}$SCAN). The operation behind the gain is adoption: in the early window a campaign started from seven single members adopts 2.2 to 2.4 groups, nests 0.8 to 1.3 and fuses at most 0.2, and the adopted members then see the results of the group whose orderings hold. The number of independent groups falls from seven to between 1.9 and 2.6 at a quarter of the compute and to between 1.2 and 1.6 at half of it (Figure~\ref{fig:morphswarm}a). Separation alone lowered the mean outcome in all 12 settings in which a swarm can divide (every start except seven single members; 90\% interval below zero in at least 9), by up to $0.060$ (2\,+\,5 with the weak pair, HSE06), and never raised it, because a separated swarm loses the results of its former partners. With results published between groups, the 4\,+\,3 start does not lose with both operations ($+0.002$, $+0.010$, $+0.013$ on the three receivers); seven single members that already share every result gain nothing from regrouping, which then changes only the division of the batch ($-0.027$, $-0.014$, $-0.010$). The learned thresholds keep both operations in use (from seven single members on PBE, 3.8 fusions, 4.3 nestings, 2.7 splits and 3.9 detachments per campaign in the early window), so the number of groups keeps moving, and they remove most of the loss from separation (one swarm of seven at the start: $-0.013$, $-0.004$ and $-0.001$ against $-0.038$, $-0.030$ and $-0.014$ with the default thresholds), but stay below combination only in 10 of the 12 multi-group settings.

\textbf{Small groups by level of theory that combine as records accumulate outperform one large swarm on the held-out receivers.} Started as 2\,+\,2\,+\,3 (MACE-MP-0 with the archive; the two NEP89 members; MEAM, the learner and the scout) with combination only, the campaign exceeds the fixed seven-member swarm by $+0.044$ (90\% interval $+0.023$ to $+0.065$) on HSE06 and $+0.083$ ($+0.067$ to $+0.100$) on $r^{2}$SCAN, and falls short by $0.018$ on PBE, the receiver that MACE-MP-0 orders most closely. In these campaigns the two weaker groups are adopted under the MACE-MP-0 group (1.2 to 1.4 adoptions per campaign) but keep their own swarms inside it, so the cells that merge agreeing members stay within one level of theory, while the adopted swarms receive the leading group's results. For a consortium of computational members such as these this suggests, as a conjecture not tested here, beginning with groups that share a level of theory or a technique and placing a failing group under the supervision of the group with the best record, rather than dividing a large group.

\textbf{When the structure changes, each member's contribution stays close to its Shapley value} (Figure~\ref{fig:morphswarm}c). With the fixed 2\,+\,2\,+\,3 groups the Owen value moves contribution away from the archive ($-0.019$) and toward NEP89 MD ($+0.024$), both intervals excluding zero on PBE. Over the twelve games with a changing structure the averaged level structure value differs from the Shapley value by at most $0.043$ for any member and by $0.004$ to $0.028$ on average, and only one member in one game (the archive, combination only from 2\,+\,2\,+\,3 on PBE, where the structure stays close to its start) has an interval excluding zero. A grouping moves contribution between members only as long as it persists. For a consortium of computational members such as these that regroups during a campaign, the Shapley value over members is therefore an adequate contribution statement, whereas a persistent organization of groups, such as two laboratories with separate budgets, needs the union-structured value of Section~\ref{sec:multiswarm}.

\subsection{Definitions}

\paragraph{Structure and sharing.} A structure is a list of independent groups; a group is a head swarm followed by zero or more nested swarms; a swarm is a set of members. The batch of $k=4$ compounds is divided by stride slot contributions held per member at three levels: between groups, between the swarms of a group, and between the cells of a swarm (Section~\ref{app:swarm}); because the contributions are held per member, they carry over when the structure changes. Groups and the swarms within a group receive tier weights from their records (leader $r^{2}$, second $r$, others 1, a negative record the floor weight, with $r=2$ and floor $0.25$ as in Section~\ref{app:swarm}). Inside a swarm, results are shared by the rule of the emergent team. Within a group, results proposed by the head swarm reach every member of the group, and results proposed by a nested swarm reach its own members and the leader of the head swarm. Results of another group are visible only when results are published between groups.

\paragraph{Records.} The record of a set of members $B$ is $q(B)=\tau_B\,n/(n+8)$, where $\tau_B$ is the Kendall $\tau$ between the mean score the members of $B$ gave each evaluated compound before its result arrived and the receiver value, over all $n$ results so far. The coordinator that divides the batch between groups and changes the structure therefore uses every result, whereas each member uses only the results it can see. The fixed structures of this annex use the same record to divide the batch, so their values differ from the share by record of Section~\ref{app:multiswarm}, which uses only the results a swarm can see.

\paragraph{Operations.} Every $p$ batches, while more than $k$ compounds remain, the agreement $a(B,C)$ between two sets of members is the Kendall $\tau$ between their mean scores on up to 60 open compounds. The operations are applied in this order:
\begin{enumerate}
\item \emph{split}: a swarm with two or more members is divided into the connected components of the graph with an edge between two members when their agreement is at least $\theta_{s}$; when the swarm contains a member with a positive running score, members with a negative running score lose their edges. The component with the best running score keeps the position of the swarm; another component stays in the group as a nested swarm if it agrees with the head swarm at least $\theta_{n}$ or its record is negative, and otherwise forms a new group;
\item \emph{detach}: a nested swarm whose agreement with its head swarm is below $\theta_{s}$ and whose record is non-negative forms a new group;
\item \emph{promote}: a nested swarm whose record exceeds the head swarm's by more than $\delta$ becomes the head swarm;
\item \emph{fuse} and \emph{nest}: pairs of groups with agreement at least $\theta_{j}$ are taken in order of decreasing agreement, each group at most once, provided the stronger record is non-negative. Two single-swarm groups with non-negative records that differ by at most $\delta$ fuse into one swarm; otherwise the swarms of the weaker group are appended to the stronger group as nested swarms;
\item \emph{adopt}: when the best group record exceeds $\delta$, every other group with a negative record is appended to the best group as nested swarms.
\end{enumerate}
Combination only applies operations 3 to 5, separation only operations 1 to 3 (a component that leaves a swarm may stay nested in the same group), and both kinds apply all five. The cold-start defaults are $\theta_{j}=0.6$, $\theta_{s}=0.2$, $\theta_{n}=0.4$, $\delta=0.1$ and $p=2$. The learned thresholds were fitted by the cross-entropy method (population 16, 6 iterations, elite quarter) to the mean early value over the five starts on 12 training campaigns per start with the PBE receiver, with seeds disjoint from the evaluation seeds: $\theta_{j}=0.05$, $\theta_{s}=0.80$, $\theta_{n}=0.21$, $\delta=0.33$, $p=3$ (training value 0.672 against 0.656 with the defaults). The learned set fuses and nests almost any two groups and divides almost any swarm, so the structure changes at nearly every opportunity.

\paragraph{Contribution with a changing structure.} For a structure with groups $\mathcal{G}$ and swarms $\mathcal{S}$, the level structure value \citep{Winter1989Levels} of member $i$ is the mean marginal contribution $v(P_i^{\pi}\cup\{i\})-v(P_i^{\pi})$ over the orderings $\pi$ of the seven members in which the members of every swarm, and the members of every group, appear consecutively; $P_i^{\pi}$ is the set of members before $i$ in $\pi$. With no unions it is the Shapley value and with one level of unions it is the Owen value; both reductions, efficiency, and the property that the values of a group sum to the group's Shapley value in the game between whole groups are checked in the self-test of the experiment script. For each campaign seed the structure trajectory of the full team is logged, and the contribution is the level structure value of each structure in force weighted by the compute spent under it in the first quarter of the receiver compute. This compute-weighted average is descriptive, not a new value: it is efficient because each structure's value is efficient and the weights sum to one, but the other axioms are not established for the average. The coalition values $v(S)$ run every sub-team with every swarm of the starting structure restricted to its members in $S$, with the same rule set, in 40 campaigns per sub-team.

\begin{table}[!ht]
\centering
\small
\caption{Early share of the ten compounds with the lowest formation energy per atom found with a changing structure (results kept within each group), mean over 100 matched campaigns, and the mean number of independent groups at a quarter of the receiver compute. Separation only is omitted for seven single members, which cannot divide.}
\label{tab:morph}
\begin{tabular}{llcccccc}
\toprule
 & & \multicolumn{3}{c}{Value} & \multicolumn{3}{c}{Groups at a quarter} \\
\cmidrule(lr){3-5}\cmidrule(lr){6-8}
Start & Rule set & PBE & HSE06 & $r^{2}$SCAN & PBE & HSE06 & $r^{2}$SCAN \\
\midrule
Seven single members & fixed & 0.592 & 0.424 & 0.459 & 7.0 & 7.0 & 7.0 \\
 & combine only & 0.621 & 0.451 & 0.477 & 1.9 & 2.3 & 2.6 \\
 & both, default & 0.613 & 0.440 & 0.478 & 2.4 & 3.2 & 3.4 \\
 & both, learned & 0.624 & 0.435 & 0.462 & 3.0 & 3.5 & 3.6 \\
\addlinespace
2\,+\,2\,+\,3 & fixed & 0.661 & 0.534 & 0.534 & 3.0 & 3.0 & 3.0 \\
 & combine only & 0.681 & 0.542 & 0.551 & 1.1 & 1.5 & 1.5 \\
 & separate only & 0.658 & 0.506 & 0.527 & 3.8 & 3.7 & 3.7 \\
 & both, default & 0.667 & 0.513 & 0.540 & 1.5 & 1.7 & 1.9 \\
 & both, learned & 0.672 & 0.538 & 0.541 & 2.3 & 2.4 & 2.6 \\
\addlinespace
4\,+\,3 & fixed & 0.645 & 0.446 & 0.464 & 2.0 & 2.0 & 2.0 \\
 & combine only & 0.650 & 0.448 & 0.472 & 1.2 & 1.4 & 1.4 \\
 & separate only & 0.618 & 0.391 & 0.452 & 3.3 & 2.7 & 3.0 \\
 & both, default & 0.623 & 0.394 & 0.454 & 1.4 & 1.8 & 2.0 \\
 & both, learned & 0.625 & 0.429 & 0.455 & 2.1 & 2.1 & 2.1 \\
\addlinespace
2\,+\,5, weak pair & fixed & 0.633 & 0.475 & 0.487 & 2.0 & 2.0 & 2.0 \\
 & combine only & 0.642 & 0.493 & 0.491 & 1.3 & 1.2 & 1.4 \\
 & separate only & 0.603 & 0.415 & 0.465 & 2.7 & 2.6 & 2.5 \\
 & both, default & 0.600 & 0.426 & 0.462 & 1.4 & 1.7 & 1.8 \\
 & both, learned & 0.631 & 0.474 & 0.491 & 2.2 & 2.3 & 2.4 \\
\addlinespace
One swarm of 7 & fixed & 0.699 & 0.498 & 0.468 & 1.0 & 1.0 & 1.0 \\
 & separate only & 0.659 & 0.470 & 0.454 & 1.6 & 1.6 & 1.5 \\
 & both, default & 0.661 & 0.468 & 0.454 & 1.1 & 1.1 & 1.2 \\
 & both, learned & 0.686 & 0.494 & 0.467 & 1.7 & 1.6 & 1.3 \\
\bottomrule
\end{tabular}
\end{table}

\clearpage
\section{Members that move between swarms or belong to two}
\label{app:mobile}

\subsection{Setting and results}
\label{sec:mobile}

In Sections~\ref{sec:multiswarm} and~\ref{sec:morphswarm} a member belongs to one swarm and changes swarm only when its whole swarm fuses, nests or splits. In a consortium a single member often changes group on its own, for example a researcher whose calculations keep agreeing with another laboratory's, and some members belong to two groups at once, such as a student shared between two laboratories who attends both groups' meetings. Experiment~18 adds two member rules to the campaign of Section~\ref{sec:morphswarm}, applied by the same coordinator every second batch. Under \emph{move}, a member leaves its swarm for the other swarm whose ordering it agrees with most, when that agreement exceeds its agreement with the rest of its own swarm by at least $0.2$ (Kendall $\tau$ on the open pool); the receiving swarm accepts it when its record is non-negative, and a member with a negative record is not accepted into a swarm with a positive one. The mover enters at the lowest tier and waits two restructuring steps before it may move again. Under \emph{share}, a member keeps its own swarm and also sits in the one other swarm it agrees with most, at agreement $0.6$ or more and on the same acceptance rule: it sees that swarm's results and that swarm sees its proposals, while its batch slots still come from its own swarm; the second membership is reassessed at every step. These thresholds are cold-start defaults; a learned set was fitted with the cross-entropy method on training seeds disjoint from the evaluation seeds (a $\tau$ margin of $0.25$ to move, agreement $0.07$ to share, one step between moves). The campaigns use the seeds of Section~\ref{sec:morphswarm}, so every comparison with a fixed grouping, and with publishing every result to all swarms, is matched campaign by campaign. With shared members the structure is not a partition, so the level structure value of Section~\ref{sec:morphswarm} is averaged over the partitions obtained by placing each shared member in one or the other of its swarms; this splits the member's place between its two groups and keeps the contributions summing to the team value. The construction is related to the configuration value of \citet{Albizuri2006Configuration}, which extends the Owen value to overlapping unions, and to cooperative games with overlapping coalitions \citep{Chalkiadakis2010Overlapping} (definitions and all values in the subsections below).

\begin{figure}[t]
\centering
\includegraphics[width=\textwidth]{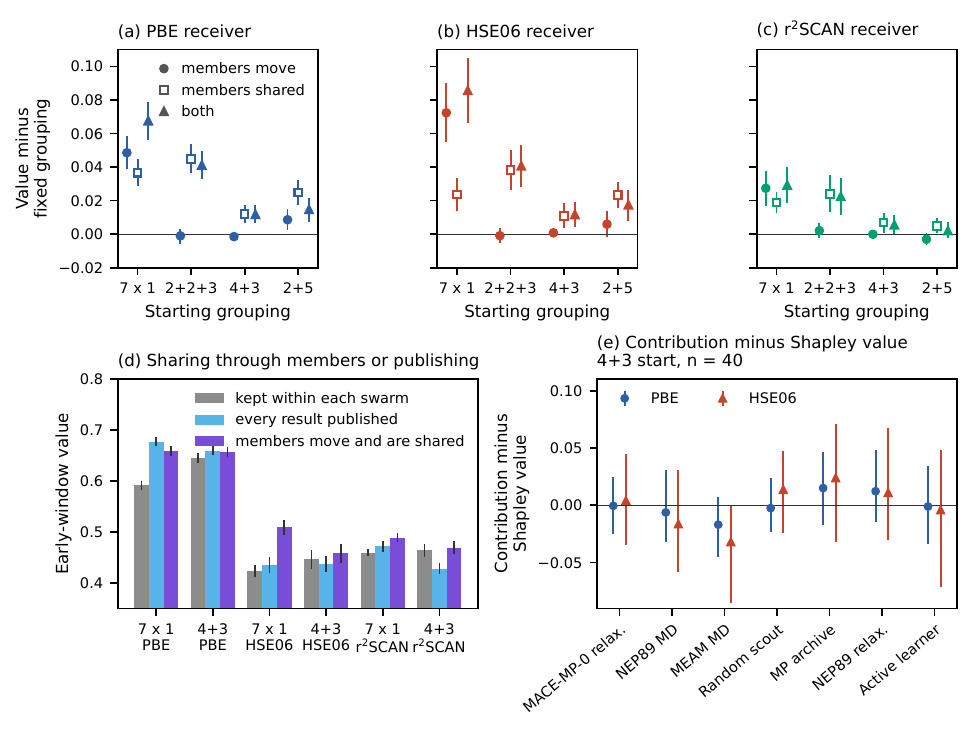}
\caption{Members that move between swarms or belong to two swarms during the Al--Mg--Si campaign (learned thresholds; results kept within each swarm except through shared members). (a--c) Early share of the ten compounds with the lowest formation energy per atom found minus that of the fixed grouping with the same start, for moving members, shared members and both (marker shape), per receiver (color, as in Figure~\ref{fig:multiswarm}); mean and 90\% bootstrap interval over 100 matched campaigns. (d) Early-window value of the fixed grouping with results kept within each swarm, with every result published to all swarms, and with moving and shared members, for seven single members and the 4\,+\,3 start. (e) Contribution with moving and shared members minus the Shapley value for the 4\,+\,3 start, PBE and HSE06; mean and 90\% interval over 40 campaigns.}
\label{fig:mobile}
\end{figure}

\textbf{A member that moves on its own reassembles a fragmented team; within designed groups it rarely moves} (Figure~\ref{fig:mobile}a--c). Started from seven single members with a fixed grouping, the move rule with default thresholds raises the early value by $+0.048$ on PBE, $+0.071$ on HSE06 and $+0.026$ on $r^{2}$SCAN, with intervals above zero, against $+0.029$, $+0.027$ and $+0.018$ when whole swarms combine (Section~\ref{sec:morphswarm}). Each member goes to the member whose ordering it shares: on PBE, NEP89 MD joins NEP89 relaxation in 0.78 of the campaigns in the early window, MACE-MP-0 joins MEAM MD in 0.76, and the active learner joins the archive in 0.43. From the 2\,+\,2\,+\,3 and 4\,+\,3 groupings a member moves 0.1 to 1.3 times per campaign and the value does not change (all intervals include zero on the three receivers), because these groupings already place members with agreeing orderings together. From 2\,+\,5 with the weak pair, NEP89 MD leaves the scout for the five-member swarm in 0.86 of the campaigns, with a gain of $+0.011$ on PBE.

\textbf{A shared member carries most of the benefit of publishing every result, and on the hybrid and meta-GGA receivers more} (Figure~\ref{fig:mobile}d). At the default agreement of $0.6$ fewer than one member is shared on average for most starts, and the gains stay below $0.008$. The learned threshold ($0.07$) shares 1.5 to 2.8 members at a quarter of the compute in all starts except 2\,+\,5, and sharing then raises the value from every multi-swarm start on all three receivers, by $+0.024$ to $+0.045$ from 2\,+\,2\,+\,3 and $+0.007$ to $+0.012$ from 4\,+\,3. Publishing every result to every swarm raises seven single members on PBE further ($0.678$ against $0.659$ with moving and shared members), but lowers 4\,+\,3 on HSE06 and $r^{2}$SCAN below the isolated groups ($0.438$ and $0.428$ against $0.446$ and $0.464$), whereas moving and shared members raise both ($0.458$ and $0.469$). A shared member links two swarms only when their orderings agree and the receiving swarm's record holds, so results proposed by members whose orderings disagree with the receiver, chiefly NEP89 MD, do not reach the swarms they would mislead. With both rules and learned thresholds, the 2\,+\,2\,+\,3 grouping by level of theory exceeds one swarm of seven by $+0.077$ (90\% interval $+0.056$ to $+0.098$) on HSE06 and $+0.089$ ($+0.074$ to $+0.105$) on $r^{2}$SCAN, matches it on PBE ($+0.003$, $-0.009$ to $+0.013$), and exceeds the same grouping with combination of whole swarms by $+0.020$ on PBE and $+0.033$ on HSE06. For a consortium of computational members such as these, a few members who attend a second group's meetings, chosen by agreement and record, pass on knowledge more selectively than a common results server.

\textbf{Moving and shared members leave contribution close to the Shapley value} (Figure~\ref{fig:mobile}e). Over the sixteen games (PBE and HSE06; 2\,+\,2\,+\,3 and 4\,+\,3; move, share and both from a fixed grouping, and both with combination of swarms) the averaged level structure value differs from the Shapley value by at most $0.043$ for any member and by $0.008$ to $0.028$ on average, and the contributions sum to the team value to within $4\times10^{-15}$. Five games have one member with an interval excluding zero; in three of them it is MEAM MD in the 4\,+\,3 grouping on HSE06 ($-0.032$ to $-0.035$), which starts in the swarm of MACE-MP-0 and NEP89 MD. As for swarms that regroup, a membership that changes during the campaign, or is split between two groups, moves contribution between members only to the extent that it persists.

\subsection{Definitions}

\paragraph{Member rules.} The structure and the swarm-level operations are those of Section~\ref{app:morph}; the member rules are applied after them at every restructuring step, while more than $k$ compounds remain. The agreement $a(i,C)$ between member $i$ and swarm $C$ is the Kendall $\tau$ between the scores of $i$ and the mean scores of the members of $C$ on up to 60 open compounds. \emph{Move}: in order of decreasing running score, a member $i$ in swarm $B$ whose last move is at least $c$ steps ago moves to the swarm $C\neq B$ with the largest $a(i,C)$ when $a(i,C)-a(i,B\setminus\{i\})\ge\theta_{m}$ (for a single-member swarm the second term is zero), $q(C)\ge0$, and not both $\bar{s}_i<0$ and $q(C)>0$, where $q$ is the record of Section~\ref{app:morph} and $\bar{s}_i$ the running score of Section~\ref{sec:swarm}; a swarm left empty is removed, and a group whose head swarm empties takes its next swarm as head. \emph{Share}: each member $i$ is attached to the swarm $C\neq B$ with the largest $a(i,C)$ when $a(i,C)\ge\theta_{b}$, on the same acceptance rule; it then sees every result proposed by a member of $C$, and the members of $C$ see every result it proposes. The cold-start defaults are $\theta_{m}=0.2$, $\theta_{b}=0.6$ and $c=2$. The learned thresholds, fitted for the fixed grouping with both rules on 12 PBE training seeds per start, five starts, population 16 and six iterations, are $\theta_{m}=0.251$, $\theta_{b}=0.073$ and $c=1.13$ (moves are counted in whole batches, so a member may move again two batches after its last move; training value $0.685$ against $0.667$ at the defaults).

\paragraph{Contribution with overlapping membership.} When the members in a set $H$ also sit in a second swarm, each assignment of every member of $H$ to its own or to its second swarm defines a partition of the members into swarms and groups. The contribution of a structure in force is the mean, over the $2^{|H|}$ assignments, of the level structure value of Section~\ref{app:morph}. It is efficient because every term is, and it equals the level structure value when $H$ is empty; both properties are checked in the self-test of the experiment script. As in Section~\ref{app:morph}, the contribution of a campaign weights the structures in force by the compute spent under each in the first quarter of the receiver compute, and the coalition values run every sub-team with the same rules in 40 campaigns.

\begin{table}[!ht]
\centering
\small
\caption{Early share of the ten compounds with the lowest formation energy per atom found with moving and shared members from a fixed grouping, mean over 100 matched campaigns, and the mean number of shared members at a quarter of the receiver compute.}
\label{tab:mobile}
\begin{tabular}{llcccccc}
\toprule
 & & \multicolumn{3}{c}{Value} & \multicolumn{3}{c}{Shared at a quarter} \\
\cmidrule(lr){3-5}\cmidrule(lr){6-8}
Start & Member rule & PBE & HSE06 & $r^{2}$SCAN & PBE & HSE06 & $r^{2}$SCAN \\
\midrule
Seven single members & none & 0.592 & 0.424 & 0.459 & 0.0 & 0.0 & 0.0 \\
 & move, default & 0.640 & 0.494 & 0.485 & 0.0 & 0.0 & 0.0 \\
 & share, default & 0.596 & 0.431 & 0.462 & 1.4 & 0.8 & 0.9 \\
 & both, default & 0.642 & 0.494 & 0.487 & 1.2 & 0.8 & 0.4 \\
 & move, learned & 0.640 & 0.496 & 0.487 & 0.0 & 0.0 & 0.0 \\
 & share, learned & 0.628 & 0.448 & 0.478 & 2.1 & 2.1 & 2.4 \\
 & both, learned & 0.659 & 0.509 & 0.489 & 2.7 & 2.8 & 2.5 \\
\addlinespace
2\,+\,2\,+\,3 & none & 0.661 & 0.534 & 0.534 & 0.0 & 0.0 & 0.0 \\
 & move, default & 0.659 & 0.536 & 0.537 & 0.0 & 0.0 & 0.0 \\
 & share, default & 0.664 & 0.539 & 0.535 & 1.0 & 0.4 & 0.3 \\
 & both, default & 0.660 & 0.537 & 0.536 & 0.5 & 0.2 & 0.1 \\
 & move, learned & 0.660 & 0.533 & 0.536 & 0.0 & 0.0 & 0.0 \\
 & share, learned & 0.705 & 0.573 & 0.558 & 2.1 & 2.3 & 2.4 \\
 & both, learned & 0.702 & 0.575 & 0.557 & 1.5 & 1.9 & 2.0 \\
\addlinespace
4\,+\,3 & none & 0.645 & 0.446 & 0.464 & 0.0 & 0.0 & 0.0 \\
 & move, default & 0.645 & 0.447 & 0.464 & 0.0 & 0.0 & 0.0 \\
 & share, default & 0.648 & 0.449 & 0.464 & 1.0 & 1.0 & 0.5 \\
 & both, default & 0.647 & 0.449 & 0.464 & 0.6 & 0.7 & 0.3 \\
 & move, learned & 0.644 & 0.447 & 0.464 & 0.0 & 0.0 & 0.0 \\
 & share, learned & 0.658 & 0.457 & 0.471 & 2.4 & 2.5 & 1.7 \\
 & both, learned & 0.657 & 0.458 & 0.469 & 1.7 & 1.9 & 1.6 \\
\addlinespace
2\,+\,5, weak pair & none & 0.633 & 0.475 & 0.487 & 0.0 & 0.0 & 0.0 \\
 & move, default & 0.644 & 0.482 & 0.486 & 0.0 & 0.0 & 0.0 \\
 & share, default & 0.635 & 0.476 & 0.487 & 0.3 & 0.1 & 0.2 \\
 & both, default & 0.644 & 0.482 & 0.486 & 0.0 & 0.0 & 0.0 \\
 & move, learned & 0.641 & 0.481 & 0.484 & 0.0 & 0.0 & 0.0 \\
 & share, learned & 0.658 & 0.499 & 0.492 & 1.6 & 2.0 & 1.9 \\
 & both, learned & 0.648 & 0.493 & 0.489 & 0.1 & 0.1 & 0.4 \\
\bottomrule
\end{tabular}
\end{table}

\clearpage
\section{Contribution that is not shared equally}
\label{app:unequal}

\subsection{Setting and results}
\label{sec:unequal}

Every value in the preceding sections divides a joint gain equally: the Shapley value gives each member of a coalition the same share of the gain that the coalition creates beyond its sub-coalitions (its Harsanyi dividend, Section~2.1 of the main text), and Section~\ref{sec:mobile} places a member shared by two swarms in each with weight one half. A consortium often agrees on an unequal division instead: contribution in proportion to the compute each partner spent, to what each partner achieves alone, or to seniority. The weighted Shapley value of \citet{Kalai1987Weighted} covers these agreements. Each member keeps its own solo gain, and the dividend of a coalition $T$ is divided in proportion to weights $w_i$, so member $i$ receives $w_i/\sum_{j\in T}w_j$ of it; with an order of seniority instead of weights, the dividend goes to the most senior members present in $T$, the ordered partition that \citet{Kalai1987Weighted} allow in their weight systems. Every such rule keeps the contributions summing to the team gain. Experiment~19 reruns all 128 sub-teams of the seven computed Al--Mg--Si advisers in one swarm (100 matched campaigns per sub-team on each of the PBE, HSE06 and $r^{2}$SCAN receivers) and divides the outcome by six rules: equal shares; shares in proportion to the compute each body spends producing its advice over the pool (NEP89 MD 1{,}027~s, NEP89 relaxation 282~s, MEAM MD 200~s, MACE-MP-0 relaxation 166~s, 1~s for the archive and for the two proposal rules); in proportion to the inverse of that compute; in proportion to the solo gain; by the three tiers of Section~\ref{sec:swarm} as seniority; and, outside the Shapley family, the team gain divided in proportion to the solo gains. It then repeats the overlapping-membership games of Section~\ref{sec:mobile} with the shared member placed in its own swarm with weight $h$ from 0 to 1 (definitions and all values in the subsections below).

\begin{figure}[!ht]
\centering
\includegraphics[width=\textwidth]{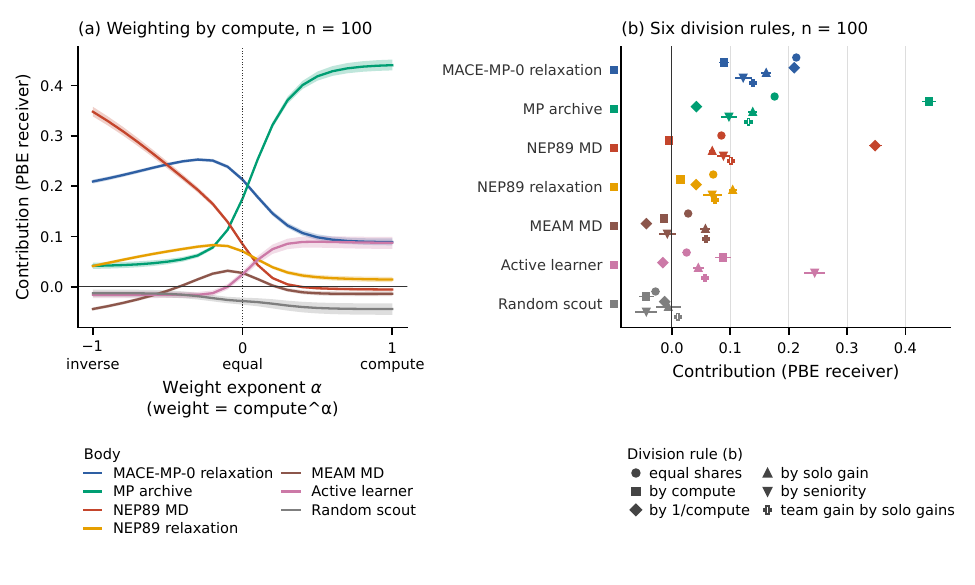}
\caption{Dividing joint gains by compute moved the contribution of the MP archive from $+0.18$ to $+0.44$ and that of NEP89 MD from $+0.09$ to about zero, more than any team structure moved any contribution (at most $0.043$). Computed Al--Mg--Si campaign, one swarm of seven members, PBE receiver, n = 100 matched campaigns. (a) Contribution when each joint gain or loss is divided in proportion to compute$^{\alpha}$, from the inverse of the compute ($\alpha=-1$) through equal shares ($\alpha=0$) to the compute ($\alpha=1$); line, mean; band, 90\% bootstrap interval. (b) Contribution under six division rules (equal shares; in proportion to compute, to its inverse, or to the solo gain; seniority by tier; team gain in proportion to solo gains): one row per member in the order and color of the legend, marker for the rule, horizontal line for the 90\% interval. The weight of a member shared by two swarms is examined below (Figure~\ref{fig:sharedmember}).}
\label{fig:unequal}
\end{figure}

\textbf{The division rule moves contribution more than any structure tested} (Figure~\ref{fig:unequal}a,b). Against the equal division, the other rules change a member's contribution by up to $0.27$ of the early-window value (NEP89 MD on HSE06 under inverse-compute weights; $0.26$ on PBE, where the team gain is $0.57$), whereas changing or overlapping structures moved it by at most $0.043$ (Sections~\ref{sec:morphswarm} and~\ref{sec:mobile}). The cause is the size of the joint dividends: pairs of advisers that point to the same low-formation-energy compounds lose value together (on PBE, MACE-MP-0 relaxation with the archive $-0.725$ and the archive with NEP89 MD $-0.644$), while larger coalitions gain (the archive, NEP89 MD and MACE-MP-0 together $+0.667$), so a weight decides how much of each overlap a member is charged as well as how much of each joint gain it receives. Dividing by compute charges the overlaps to the potentials that spent the most: NEP89 MD falls from $+0.085$ to $-0.005$ on PBE and the archive, which spent no compute, rises from $+0.176$ to $+0.441$. Dividing by the inverse of compute reverses this (NEP89 MD $+0.348$, archive $+0.042$). Under seniority the senior tiers carry the losses of every overlap with the junior members, so the active learner is charged none of them and rises from $+0.025$ to $+0.244$ (its solo gain is $+0.285$). The intermediate weights of panel (a) move every contribution smoothly between these limits.

\textbf{The members a group would keep are the same under every rule, up to ties for third; the member it would drop is not.} MACE-MP-0 relaxation and the MP archive are among the three highest-contributing members, or tied for third, under all six rules on all three receivers, so the advisers a group would keep do not depend on the rule. MEAM MD, the member with negative contribution on HSE06 and $r^{2}$SCAN under equal shares, stays negative when contribution follows the compute or its inverse, but becomes positive on HSE06 when it follows the solo gain ($+0.029$) and indistinguishable from zero under seniority ($+0.006$, interval $-0.008$ to $+0.021$), and the proportional rule gives every member non-negative contribution on every receiver. A rule weighted by solo merit therefore protects a member whose solo gain is high but whose ordering adds nothing to the team, and the proportional rule, which ignores interactions altogether, cannot identify a member to drop. For a consortium of computational members such as these the practical sequence is to identify members to drop with the equal division, which charges overlaps without regard to cost or rank, and to apply an agreed weighting only to divide the outcome among the members that remain.

\begin{figure}[!ht]
\centering
\includegraphics[width=0.75\textwidth]{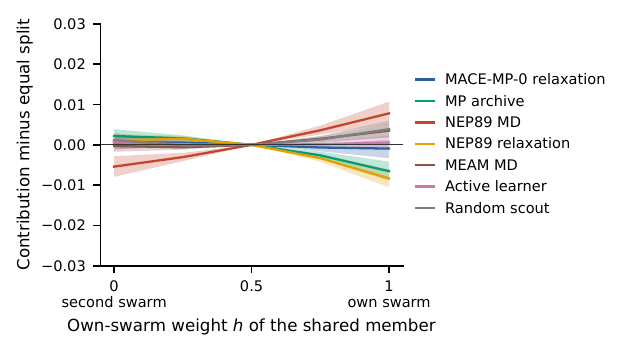}
\caption{Moving the weight of a shared member between its two swarms changed any contribution by at most $0.013$. Two swarms of two and one of three, members that move and members shared by two swarms, HSE06 receiver: contribution when the shared member is placed in its own swarm with weight $h$, minus the equal placement $h=1/2$; line, mean over n = 100 campaigns; band, 90\% bootstrap interval.}
\label{fig:sharedmember}
\end{figure}

\textbf{How a shared member is split between its swarms matters little} (Figure~\ref{fig:sharedmember}).

 With 1.4 to 1.7 shared members on average in the early window, moving the weight of the own swarm from $h=0$ (contribution follows the information the member passes to its second swarm) to $h=1$ (contribution follows its batch slots) changes any contribution by at most $0.013$ in the four games (PBE and HSE06; 4\,+\,3 and 2\,+\,2\,+\,3), below the structure effect itself (at most $0.028$ from the Shapley value) and an order of magnitude below the effect of the division rule. The equal placement of Section~\ref{sec:mobile} is therefore a default whose choice does not change the contribution statement, whereas the weights of the division rule do and have to be stated with it.

\subsection{Definitions}

\paragraph{Weighted values.} For a game $v$ on the bodies $N$, the Harsanyi dividends are $d(T)=\sum_{S\subseteq T}(-1)^{|T|-|S|}v(S)$ for non-empty $T$, so that $v(S)-v(\emptyset)=\sum_{\emptyset\neq T\subseteq S}d(T)$. For positive weights $w$, the weighted Shapley value of \citet{Kalai1987Weighted} is
\begin{equation}
\phi^{w}_i(v)=\sum_{T\ni i}\frac{w_i}{\sum_{j\in T}w_j}\,d(T),
\end{equation}
which equals the Shapley value for equal weights and gives each body its solo dividend $d(\{i\})=v(\{i\})-v(\emptyset)$ for any weights. For an ordered partition of $N$ into seniority classes, the dividend of $T$ is divided in equal parts among the members of $T$ in the most senior class present in $T$. Both rules satisfy $\sum_i\phi_i=v(N)-v(\emptyset)$. The proportional rule gives $\bigl(v(N)-v(\emptyset)\bigr)\,g_i/\sum_j g_j$ with $g_i=\max\bigl(v(\{i\})-v(\emptyset),0\bigr)$; it is efficient but is not additive over games. Weighted values give up symmetry by design: two bodies that change every coalition equally receive different contributions when their weights differ. The values are computed per campaign from the 128 coalition values of that campaign; the bootstrap is over the 100 matched campaigns.

\paragraph{Weights.} Compute weights are the seconds each body spends computing its advice for the whole pool (GPU seconds for MACE-MP-0 and NEP89, core seconds for MEAM): NEP89 MD 1{,}027, NEP89 relaxation 282, MEAM MD 200, MACE-MP-0 relaxation 166; the archive, the active learner and the random scout compute no advice table and receive the floor of 1~s. Solo weights are $\max\bigl(\bar v(\{i\})-\bar v(\emptyset),0.01\bigr)$ from the mean over campaigns. Seniority uses the tiers of Section~\ref{sec:swarm}: MACE-MP-0 relaxation and the archive first; NEP89 MD, NEP89 relaxation and MEAM MD second; the active learner and the random scout third. The family of panel (a) of Figure~\ref{fig:unequal} uses $w_i=c_i^{\alpha}$ for 21 values of $\alpha$ from $-1$ to $1$.

\paragraph{Shared members.} With a set $H$ of shared members, each assignment of the members of $H$ to their own or second swarm defines a partition; the assignment that places $m$ members in their own swarm receives weight $h^{m}(1-h)^{|H|-m}$, and the contribution of the structure in force is the weighted mean of the level structure values (Section~\ref{app:mobile}). The thresholds are the learned ones of Section~\ref{sec:mobile}; 100 campaigns per coalition.

\begin{table}[!ht]
\centering
\small
\caption{Contribution of the seven bodies in one swarm under six division rules, mean over 100 matched campaigns (90\% interval widths 0.001 to 0.044).}
\label{tab:unequal}
\begin{tabular}{lcccccc}
\toprule
Body & Equal & Compute & 1/compute & Solo gain & Seniority & Proportional \\
\midrule
\multicolumn{7}{l}{\emph{PBE receiver} ($v(N)=0.692$, $v(\emptyset)=0.123$)} \\
MACE-MP-0 relaxation & $+0.213$ & $+0.089$ & $+0.210$ & $+0.161$ & $+0.122$ & $+0.139$ \\
NEP89 MD & $+0.085$ & $-0.005$ & $+0.348$ & $+0.069$ & $+0.088$ & $+0.101$ \\
NEP89 relaxation & $+0.071$ & $+0.015$ & $+0.041$ & $+0.104$ & $+0.069$ & $+0.074$ \\
MEAM MD & $+0.028$ & $-0.014$ & $-0.044$ & $+0.057$ & $-0.008$ & $+0.058$ \\
MP archive & $+0.176$ & $+0.441$ & $+0.042$ & $+0.138$ & $+0.098$ & $+0.131$ \\
Active learner & $+0.025$ & $+0.088$ & $-0.016$ & $+0.045$ & $+0.244$ & $+0.057$ \\
Random scout & $-0.028$ & $-0.044$ & $-0.012$ & $-0.006$ & $-0.044$ & $+0.010$ \\
\addlinespace
\multicolumn{7}{l}{\emph{HSE06 receiver} ($v(N)=0.507$, $v(\emptyset)=0.123$)} \\
MACE-MP-0 relaxation & $+0.199$ & $+0.117$ & $+0.239$ & $+0.118$ & $+0.092$ & $+0.114$ \\
NEP89 MD & $+0.055$ & $-0.015$ & $+0.330$ & $-0.001$ & $-0.007$ & $+0.087$ \\
NEP89 relaxation & $+0.001$ & $+0.007$ & $-0.040$ & $+0.075$ & $+0.056$ & $+0.014$ \\
MEAM MD & $-0.033$ & $-0.073$ & $-0.238$ & $+0.029$ & $+0.006$ & $+0.036$ \\
MP archive & $+0.165$ & $+0.326$ & $+0.084$ & $+0.114$ & $+0.080$ & $+0.090$ \\
Active learner & $+0.004$ & $+0.053$ & $-0.001$ & $+0.041$ & $+0.157$ & $+0.034$ \\
Random scout & $-0.008$ & $-0.031$ & $+0.010$ & $+0.007$ & $-0.001$ & $+0.009$ \\
\addlinespace
\multicolumn{7}{l}{\emph{$r^{2}$SCAN receiver} ($v(N)=0.467$, $v(\emptyset)=0.111$)} \\
MACE-MP-0 relaxation & $+0.147$ & $+0.075$ & $+0.156$ & $+0.126$ & $+0.105$ & $+0.084$ \\
NEP89 MD & $+0.084$ & $-0.000$ & $+0.299$ & $+0.025$ & $+0.080$ & $+0.102$ \\
NEP89 relaxation & $+0.034$ & $-0.005$ & $-0.024$ & $+0.052$ & $+0.039$ & $+0.046$ \\
MEAM MD & $-0.055$ & $-0.064$ & $-0.115$ & $-0.019$ & $-0.032$ & $+0.001$ \\
MP archive & $+0.125$ & $+0.312$ & $+0.028$ & $+0.103$ & $+0.082$ & $+0.088$ \\
Active learner & $+0.021$ & $+0.045$ & $+0.004$ & $+0.032$ & $+0.082$ & $+0.025$ \\
Random scout & $-0.000$ & $-0.007$ & $+0.008$ & $+0.037$ & $+0.000$ & $+0.010$ \\
\bottomrule
\end{tabular}
\end{table}

\paragraph{Shared members.} Over the four overlapping-membership games, the largest change of any contribution between $h=0$ and $h=1$ is $0.012$ (HSE06, 4\,+\,3), $0.013$ (HSE06, 2\,+\,2\,+\,3), $0.005$ (PBE, 4\,+\,3) and $0.013$ (PBE, 2\,+\,2\,+\,3); the largest difference between the equal placement and the Shapley value in the same games is $0.025$, $0.028$, $0.011$ and $0.015$. The body with the largest change is NEP89 MD in the 2\,+\,2\,+\,3 grouping ($+0.013$ from $h=0$ to $h=1$ on both receivers). The contributions sum to the team gain to within $2\times10^{-14}$ for every $h$.

\clearpage
\section{Extended results of the Al--Mg--Si games}
\label{app:almgsi}

\paragraph{Mg$_2$Si in each source.} NEP89 and MEAM do not bind Mg$_2$Si; the other sources do (Table~\ref{tab:mg2si}). Mg$_2$Si was not a direct fitting target of the MEAM potential: its Si--Mg pair was fitted to the B1 reference structure \citep{Jelinek2012MEAM}.

\begin{table}[h]
\centering
\caption{Formation energy and volume of Mg$_2$Si by source.}
\label{tab:mg2si}
\resizebox{\textwidth}{!}{%
\begin{tabular}{lcc}
\hline
Source & Formation energy (eV/atom) & Volume (\AA$^{3}$/atom) \\
\hline
NEP89, relaxed at 0~K & $+0.088$ & 21.00 \\
MEAM, published (Table~IX of \citet{Jelinek2012MEAM}) & $+0.042$ & 23.05 \\
MACE-MP-0, relaxed at 0~K & $-0.081$ & 21.36 \\
Quantum ESPRESSO PBE relaxation (main text, Methods) & $-0.160$ & 21.43 \\
Materials Project mp-1367 (corrected entry) & $-0.109$ & -- \\
CALPHAD assessment \citep{Fries_1998}, as tabulated by \citet{Jelinek2012MEAM} & $-0.225$ (enthalpy) & -- \\
Measured (300~K, from the fitted lattice parameter) \citep{Kasai_2017} & -- & 21.3 \\
\hline
\end{tabular}}
\end{table}

\begin{figure}[!ht]
\centering
\includegraphics[width=\textwidth]{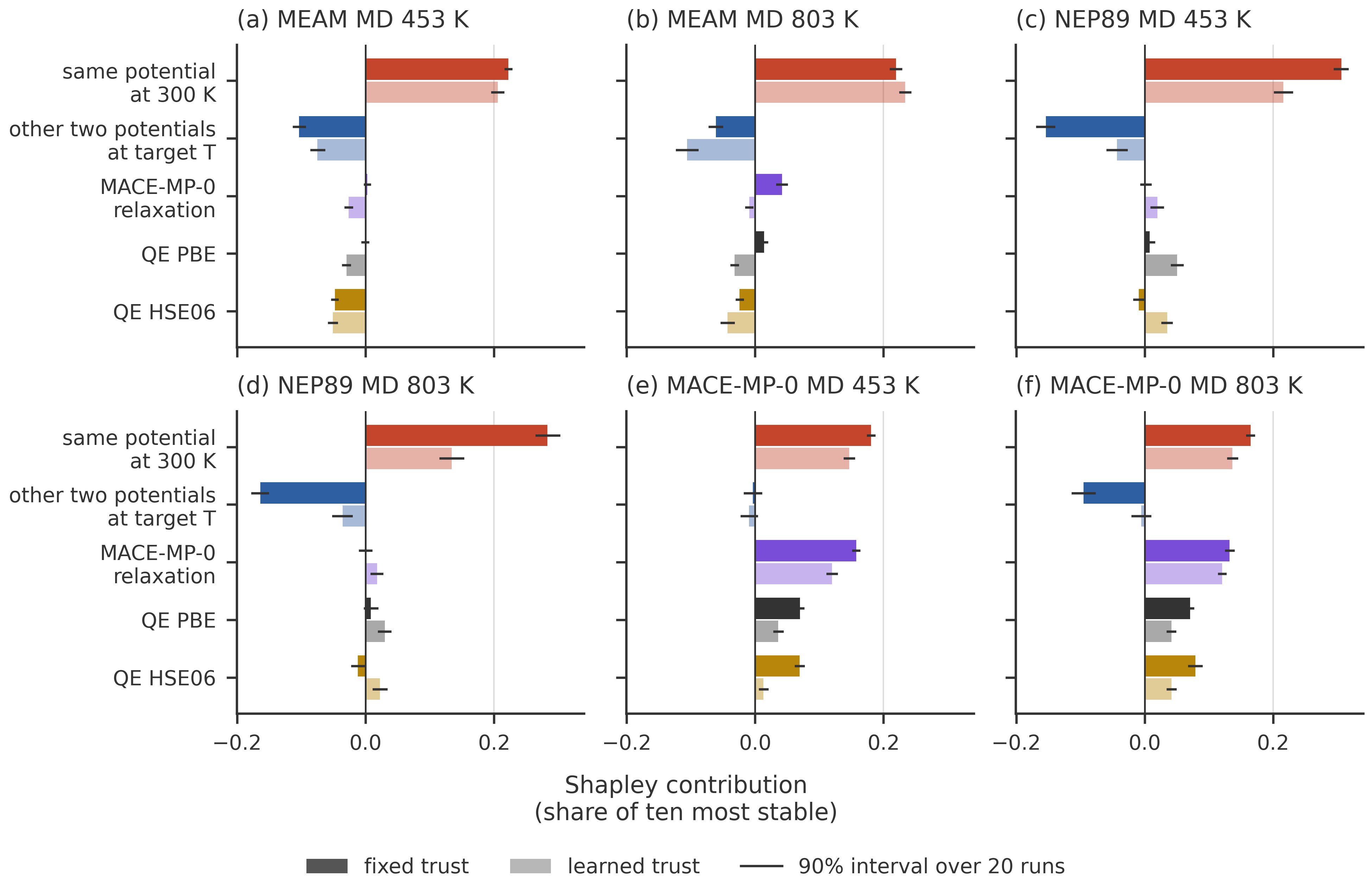}
\caption{In each of the six molecular-dynamics receivers at 453 and 803~K the same potential at 300~K received the largest contribution under both trust modes ($+0.13$ to $+0.23$ under learned trust), and the other two potentials at the target temperature a negative or near-zero combined contribution. (a--f) One receiver per panel (MD potential and temperature). Shapley contribution (early share of the ten compounds with the lowest formation energy per atom found) of the six donors, grouped by role (the two other potentials at the target temperature are summed); all 64 coalitions run per receiver. Dark, fixed trust; light, learned trust (the encoding of Figure~5 of the main text). Whiskers: 90\% percentile-bootstrap intervals of the mean over 20 runs, in each of which every MD value is drawn from its measured sampling error (Section~\ref{si:measures}, Table~\ref{tab:dvhot}).}
\label{fig:almgsi}
\end{figure}

This annex gives the extended games of the Al--Mg--Si case study of Section~2.4 of the main text: the HSE06 receiver, the molecular-dynamics receivers at heat-treatment temperatures, and two receivers closer to a heat-treatment decision. The HSE06 formation energies carry a k-point uncertainty of about 25~meV/atom; bracketing it by shifting only the elemental references, or the compound energies with them, left the ten HSE06 targets unchanged and the Kendall $\tau$ between the PBE and HSE06 orderings at $0.86$--$0.89$. The Shapley contributions of the PBE and HSE06 games, with their intervals over runs, are shown in Figure~5 of the main text; Figure~\ref{fig:almgsi} shows those of the temperature games.

The same game was played with the HSE06 hybrid functional \citep{Heyd2003HSE,Heyd2006HSEErratum,Krukau2006HSE06} as the receiver, computed at the PBE geometry; for these metallic compounds HSE06 is a different receiver, not a more accurate level \citep{Paier2007HybridMetals,Stroppa2008hybridmetals}, and PBE added as a fifth player (all 32 coalitions run; Figure~5c of the main text). Under fixed trust PBE ($+0.17$) and MACE-MP-0 ($+0.17$) receive nearly equal contribution, followed by the archive ($+0.12$), MEAM MD ($-0.08$), and NEP89 MD ($-0.21$). The best coalition is PBE with MACE-MP-0 ($v=0.77$), and the full pool is again lower ($v(N)=0.51$). The two leading players carry nearly the same ordering (Kendall $\tau$ with HSE06 of $+0.86$ and $+0.80$): each alone reaches an early share of about 0.75, and together they reach 0.77, so the symmetry and efficiency axioms split their overlapping contribution between them. Under learned trust the contributions are PBE $+0.14$, MACE-MP-0 $+0.10$, archive $+0.06$, MEAM $-0.002$, and NEP89 $-0.05$, with $v(N)=0.60$; NEP89 is the only player whose contribution remains clearly negative, because it orders the pool against both DFT levels, and the contribution of MEAM is indistinguishable from zero.

These values charge each query its HSE06 compute only, so they do not reflect that the PBE pool cost about 270 core-hours (61{,}285 wall-seconds on 16 cores) and the MACE-MP-0 pool under three GPU-minutes. Under the combined accounting each query is charged its HSE06 and PBE core-seconds (HSE06 on 32 cores), and a coalition that holds PBE is charged the PBE relaxation of all 168 compounds before its first query and then HSE06 per query; the compute of the potentials, under 0.01\% of the total, is not charged. The acquisition does not see cost, so both accountings choose the same compounds in the same order and only the early window, the first quarter of the compute charged, differs. Table~\ref{tab:costcredit} gives both accountings for the Al--Mg--Si game and for the band-gap game of Section~2.4 of the main text. Under the combined accounting the PBE ordering, which is the most accurate prior, receives negative contribution in both systems and under both trust modes, the contribution of the other donors changes by at most $0.04$ (Al--Mg--Si) and $0.02$ (band gaps), and the best coalition does not contain PBE.

\paragraph{Sharing the relaxation cost.} The combined accounting is one end of a range. When the PBE relaxations of the band-gap pool serve several campaigns, each can be charged a share $f$ of their cost, the division of a common cost among its users studied in airport games \citep{Littlechild1973Airport}. Because the acquisition does not see cost, every coalition chooses the same compounds at every $f$, and only the early window of a coalition that uses the PBE ordering changes; the early window was recomputed from the logged campaigns at 41 values of $f$ from 0 to 1. The contribution of the PBE ordering falls with its own charge, as monotonicity requires \citep{Young1985Monotonic}, from $+0.220$ at $f=0$ to $-0.155$ at $f=1$ under fixed trust and from $+0.135$ to $-0.261$ under learned trust, while the contributions of the four archives change by at most $0.02$ between $f=0$ and $f=1$ (Table~\ref{tab:costcredit}; only the contribution of the PBE ordering was recomputed at intermediate $f$). At $f=1$ the sweep reproduces the combined accounting; at $f=0$ it gives $+0.220$ and $+0.135$ rather than the $+0.203$ and $+0.128$ of the HSE06-only accounting, because every query is still charged its PBE core-seconds and only the upfront relaxation of the pool is shared. It is positive for $f<0.567$ (fixed) and $f<0.341$ (learned), so the relaxations pay once they serve $1/f\approx1.8$ or $2.9$ campaigns.

\begin{table}[!ht]
\centering
\small
\caption{Break-even charge share $f^{*}$ of the PBE relaxations (95\% bootstrap interval, 2000 resamples) and the reuse multiple $1/f^{*}$, by the window in which the campaign is judged (share of the receiver compute). Never: the PBE ordering stays positive when charged in full ($+0.029$, $[+0.025,+0.033]$).}
\label{tab:window}
{\setlength{\tabcolsep}{4pt}%
\begin{tabular}{@{}llcccc@{}}
\toprule
Game & Trust & $1/8$ & $1/4$ & $1/2$ & whole budget \\
\midrule
Band gap & fixed & $0.24$ $[0.21,0.27]$, $4.1$ & $0.57$ $[0.50,0.63]$, $1.8$ & $0.83$ $[0.73,0.94]$, $1.2$ & $0.84$ $[0.74,0.95]$, $1.2$ \\
Band gap & learned & $0.21$ $[0.18,0.23]$, $4.8$ & $0.34$ $[0.28,0.39]$, $2.9$ & $0.46$ $[0.36,0.56]$, $2.2$ & $0.46$ $[0.36,0.57]$, $2.2$ \\
Al--Mg--Si & fixed & $0.32$ $[0.30,0.33]$, $3.2$ & $0.62$ $[0.58,0.66]$, $1.6$ & $0.99$ $[0.95,1.00]$, $1.0$ & never \\
Al--Mg--Si & learned & $0.27$ $[0.25,0.29]$, $3.7$ & $0.43$ $[0.40,0.47]$, $2.3$ & $0.55$ $[0.51,0.59]$, $1.8$ & $0.61$ $[0.56,0.65]$, $1.7$ \\
\bottomrule
\end{tabular}}
\end{table}

Table~\ref{tab:window} gives the break-even by window. The rule of the main text is stated for the first quarter of the receiver compute; judged over the whole budget the relaxations pay after $1.2$ to $2.2$ campaigns, and the Al--Mg--Si fixed-trust PBE ordering stays positive even when charged in full, so the rule is a statement about the early window. In the same campaign neither MEAM nor NEP89 binds Mg$_2$Si; no run separates this from the full-budget result, and neither is offered as the cause of the other. With the hypothetical ordered alloys removed from the band-gap pool (118 compounds, all ten targets archived), the fixed-trust multiple at a quarter of the compute is $2.1$; with the alloys kept as candidates but barred as targets (142 compounds) it is $2.4$. Result files: \texttt{compute/live\_semiconductor/results/campaigns/window\_cost\_sweep.json}, \texttt{archived\_targets\_check.json} and \texttt{archived\_targets\_only.json}.

\paragraph{How much of the disagreement of the potentials with PBE is linear in composition.} A fit of the formation energy of each potential minus that of PBE to the atomic fractions of Al, Mg and Si (Table~\ref{tab:compfit}) explains about half of the NEP89 difference and less than a fifth of the MEAM difference. Without its linear part, NEP89 orders the pool with PBE ($\tau=+0.49$ in MD, $+0.50$ relaxed, against $-0.18$ and $-0.22$ raw); MEAM changes little. Two 0~K checks on Mg$_2$Si agree with the MD values: the NEP89 relaxation with an equation-of-state fit gives $+0.088$~eV/atom at $21.00$~\AA$^{3}$/atom, and the published MEAM value is $+0.042$~eV/atom at $23.05$~\AA$^{3}$/atom (Table~IX of \citet{Jelinek2012MEAM}). Result file: \texttt{compute/live\_almgsi/results/nep89\_mg2si\_check.json}, written by \texttt{nep89\_mg2si\_check.py}.

\begin{table}[!ht]
\centering
\small
\caption{Linear composition fit of the formation energy of each potential minus that of PBE (coefficients for the Al, Mg and Si fractions, eV/atom) and Kendall $\tau$ against PBE before and after removing the fitted part.}
\label{tab:compfit}
\begin{tabular}{@{}lrrrrrr@{}}
\toprule
Member & $n$ & Al, Mg, Si & $R^{2}$ & residual sd & $\tau$ raw & $\tau$ without linear part \\
\midrule
NEP89 MD & 188 & $-0.079$, $+0.014$, $-0.280$ & $0.47$ & $0.054$ & $-0.18$ & $+0.49$ \\
NEP89 relaxation & 179 & $-0.087$, $+0.019$, $-0.275$ & $0.48$ & $0.054$ & $-0.22$ & $+0.50$ \\
MEAM MD & 185 & $-0.150$, $+0.098$, $+0.253$ & $0.18$ & $0.143$ & $+0.46$ & $+0.36$ \\
MEAM relaxation & 179 & $-0.008$, $+0.014$, $+0.159$ & $0.08$ & $0.088$ & $+0.46$ & $+0.40$ \\
\bottomrule
\end{tabular}
\end{table}

\begin{table}[!ht]
\centering
\small
\caption{Shapley contribution (early share of the ten targets found) of the donors to the two HSE06 campaigns under the two cost accountings: HSE06 compute only, and HSE06 plus PBE compute with the PBE relaxation of the whole pool charged upfront to every coalition that uses the PBE ordering. Fixed / learned trust; all 32 coalitions run, 20 seeds and batches of one and four per coalition. Band gaps: 142 candidates without halogens, 170 of 171 HSE06 gaps computed. Best coalitions under HSE06 + PBE cost: MACE-MP-0 alone ($0.758$ / $0.676$) for Al--Mg--Si; the measured gaps, TB-mBJ, and MP GGA ($0.672$, fixed) and TB-mBJ alone ($0.909$, learned) for band gaps.}
\label{tab:costcredit}
\begin{tabular}{llcc}
\toprule
System & Donor & HSE06 only & HSE06 + PBE \\
\midrule
Al--Mg--Si & PBE & $+0.171$ / $+0.145$ & $-0.109$ / $-0.203$ \\
 & MACE-MP-0, 0~K & $+0.167$ / $+0.101$ & $+0.147$ / $+0.100$ \\
 & MP archive & $+0.120$ / $+0.064$ & $+0.094$ / $+0.067$ \\
 & MEAM MD & $-0.084$ / $-0.002$ & $-0.070$ / $-0.004$ \\
 & NEP89 MD & $-0.209$ / $-0.051$ & $-0.166$ / $-0.040$ \\
 & $v(\emptyset)$ ; $v(N)$ & $0.345$ ; $0.509$ / $0.602$ & $0.359$ ; $0.255$ / $0.279$ \\
\midrule
Band gaps & PBE (computed) & $+0.203$ / $+0.128$ & $-0.155$ / $-0.261$ \\
 & MP GGA & $+0.040$ / $+0.093$ & $+0.053$ / $+0.093$ \\
 & JARVIS TB-mBJ & $+0.040$ / $+0.121$ & $+0.055$ / $+0.120$ \\
 & SNUMAT HSE06 & $+0.018$ / $+0.026$ & $+0.024$ / $+0.030$ \\
 & measured gaps & $+0.020$ / $+0.088$ & $+0.034$ / $+0.089$ \\
 & $v(\emptyset)$ ; $v(N)$ & $0.377$ ; $0.697$ / $0.834$ & $0.384$ ; $0.395$ / $0.455$ \\
\bottomrule
\end{tabular}
\end{table}

\textbf{Accuracy of the computed band-gap receiver.} The band-gap campaign ranks candidates by their computed HSE06 gaps, so only the ordering of those gaps enters its games. For the 110 halogen-free compounds that are also in the measured compilation \citep{Zhuo2018BandGapML}, the HSE06 gaps (PBE geometry, self-consistent $k$ mesh, no spin--orbit coupling) lie below the measured gaps by $0.50$~eV on average (mean absolute deviation $0.82$~eV, Kendall $\tau=0.55$): by $0.79$~eV for 23 oxides, $0.65$~eV for 8 nitrides, $0.41$~eV for 55 chalcogenides, $0.23$~eV for 17 pnictides, and $0.72$~eV for 7 other compounds. The 28 halides, which the campaign leaves out, lie $1.26$~eV below, so the mean over all 138 compounds with a measured gap is $0.65$~eV.

A further family of games tests a change of processing environment. Each potential was run at 453~K (T6 artificial ageing) and 803~K (solution treatment; the 803~K receivers may be partly molten, so these games test transfer between potentials rather than solution-treatment physics), and six donors were offered to each of the six hot MD receivers: the same potential at 300~K, the two other potentials at the target temperature, the 0~K MACE-MP-0 relaxation, and PBE and HSE06 at 0~K. All 64 coalitions were run for each receiver (Figure~\ref{fig:almgsi}). The same potential at 300~K receives the largest contribution in every receiver ($+0.13$ to $+0.23$ under learned trust); the two other potentials at the target temperature have a negative combined contribution in all six receivers (for example $-0.106$ in the MEAM campaign at 803~K, of which NEP89 carries $-0.093$); with the MD sampling error included, this held in every run for the MEAM receivers and in about half of the runs for the MACE-MP-0 receivers (Section~\ref{si:measures}), so being at the right temperature does not compensate for a different potential; and for the MACE-MP-0 receivers the second-largest contribution goes to the 0~K relaxation with the same potential. The two DFT levels receive small contribution ($-0.05$ to $+0.08$; $-0.05$ to $+0.05$ under learned trust), and under learned trust HSE06 receives less than PBE in all six receivers, by a margin below $0.001$ for MACE-MP-0 MD at 803~K: the more expensive calculation is not the more useful prior to a potential fitted to PBE-level data. The best coalition for every receiver is the same potential at 300~K alone. For a materials group, when a campaign moves to a new heat-treatment step, the prior that contributes is the group's own model in the previous step, and the other models at the new step reduce the outcome.

Two further games use receivers closer to a heat-treatment decision. With MACE-MP-0 MD at 300~K as the receiver and six donors (all 64 coalitions run), the fixed-trust contributions are the 0~K relaxation of the same potential $+0.16$, PBE and the archive $+0.09$ each, HSE06 $+0.07$, MEAM MD $-0.03$, and NEP89 MD $-0.09$, and the full pool ($v(N)=0.74$) is again below the best single donor ($0.82$). With quasi-harmonic formation free energies at 803~K as the receiver, computed from MACE-MP-0 phonons \citep{Togo2023phonopy} for the 49 dynamically stable compounds, five donors (the 0~K relaxation, PBE, and the three potentials in MD at 803~K; 32 coalitions) receive fixed-trust contributions of $+0.07$, $+0.06$, $+0.06$ (MACE-MP-0 MD), $+0.05$ (NEP89 MD), and $-0.16$ (MEAM MD). Under learned trust MEAM's contribution rises to $-0.005$. Once entropy enters the receiver, the static and the MD donors share the contribution nearly equally, and the potential that orders the pool differently is the one whose removal improves the campaign.

In summary, with the HSE06 hybrid functional as the receiver and PBE as a fifth player, PBE and MACE-MP-0 share the leading contribution when only HSE06 compute is charged, PBE receives negative contribution once its relaxations are charged as well, the full pool again scores below the best pair of donors, and NEP89 is the only player whose contribution remains clearly negative under learned trust, because it orders the pool against both DFT levels. When each potential is run at the ageing and solution-treatment temperatures, the same potential at 300~K receives the largest contribution in each of the six receivers, while the two other potentials at the target temperature have a negative combined contribution: when a campaign moves to a new heat-treatment step, the prior that contributes is the group's own model in the previous step. Receivers closer to a heat-treatment decision, MD at 300~K and quasi-harmonic formation free energies at 803~K, give the same picture: the potential that orders the pool differently is the one whose removal improves the campaign.

\subsection{The observed precipitation sequence as an ordinal check}
\label{app:almgsi-seq}

Every value function of Section~2.4 of the main text is defined by a computed receiver, and against an ordering observed in the laboratory the contribution of a donor can change sign as well. The precipitation sequence of Al--Mg--Si alloys, established by electron microscopy and atom-probe studies \citep{Edwards1998Precipitation,Zandbergen1997BetaDoublePrime,Vissers2007BetaPrime,Andersen2007U1}, orders 17 solute states from the dispersed solid solution to equilibrium $\beta$-Mg$_2$Si. MEAM MD, which receives small positive contribution in the learned-trust PBE game, orders this sequence in reverse (Kendall agreement $-0.50$), while PBE ($+0.70$) and MACE-MP-0 ($+0.56$ at 0~K) reproduce it. The binding energies of nine Mg/Si solute clusters separate the donors in the same way (Kendall $\tau$ against PBE: MACE-MP-0 $+0.78$, MEAM MD $-0.67$).

\subsection{Further detail}

A subset without PBE pays the PBE relaxation of each compound it queries, which its HSE06 calculation needs as geometry, so no relaxation is charged twice; the borrowed ordering simply exists only once the whole pool has been relaxed.

The test target was the ten of 168 Al--Mg--Si compounds with the lowest formation energy per atom, 141 of them ordered fcc decorations (ordered arrangements of Al, Mg and Si on a face-centered cubic lattice) that no archive holds, with PBE as the receiver and four low-cost sources: the MACE-MP-0 foundation potential \citep{Batatia2024MACEMP} relaxed at 0~K, MD at 300~K with NEP89 \citep{Liang2025NEP89} and with the modified embedded-atom method (MEAM) potential of \citet{Jelinek2012MEAM}, and the Materials Project archive. Under fixed trust MACE-MP-0 contributed $+0.23$ and the archive $+0.15$, while NEP89 MD, which orders the pool against PBE ($\tau = -0.18$; about half of its offset from PBE is linear in composition, and without that part $\tau = +0.49$, Section~S11), and MEAM MD ($\tau = +0.46$) held the campaign back as misaligned members ($-0.20$, $-0.11$).

PBE supplies the prior closest to HSE06 ($\tau = +0.86$, against $+0.80$ for MACE-MP-0), but its relaxations spent the early budget that MACE-MP-0 leaves to the receiver.

\paragraph{Computed campaigns.} The band-gap pool held 171 compounds of at most six atoms per primitive cell, relaxed with PBE in Quantum ESPRESSO 7.4 \citep{Giannozzi2017QE} on 16 cores and with HSE06 single points at the PBE geometry; its ordered MgCaO and ZnBeO alloys come from strongly immiscible oxides and are hypothetical ranking candidates, not synthesizable phases. Three of the ten HSE06 targets are such alloys (CaMgO$_2$ in the CuPt-111 and CuAu-001 orderings, ranks 2 and 6, and layered-wurtzite BeZnO$_2$, rank 8), and NaNO$_2$ is also a target; with the alloys removed as targets, or from the pool, the break-even multiple stays near two (Section~S11). The Al--Mg--Si pool held 168 binary and ternary compounds (Materials Project structures near the hull and ordered fcc decorations).

MD formation energies (NEP89, release 2025-04-09, in GPUMD 5.8 \citep{Fan2022GPUMD}; MEAM in LAMMPS 2025.7.22 \citep{Thompson2022LAMMPS}; MACE-MP-0 with mace-torch 0.3.16) were time averages over 3~ps after 1~ps of equilibration, and in each of 20 runs every MD value was drawn from its measured standard error. $v(S)$ was the mean share of the receiver's ten best candidates found over the first quarter of its compute.

\clearpage
\section{Code, data, and result files}
\label{si:artifacts}
All quantitative results are produced by the scripts and stored in the result files below (archived release v1.3.0, \url{https://doi.org/10.5281/zenodo.23876457}). Exact computations are deterministic; sampled ones are seeded. The archive manifest hash is \texttt{f2158ecdc78e0229} (261{,}053 records, 8 sources); the campaigns use every record, and no held-out split of the archive enters any result. Campaigns on the archive draw each record at most once per campaign (\texttt{campaign\_protocols/\allowbreak{}code/\allowbreak{}experiments/\allowbreak{}\_draw.py}, called by \texttt{26\_paired\_campaign.py}).
{\raggedright
\begin{itemize}
\item Eight-donor campaign (MP GGA and its GGA+U (TM-O/F) subset as separate donors), 256 subsets with per-seed checkpoints (experiments 08a--08c): \texttt{ateki/\allowbreak{}results\_8donor/\allowbreak{}}.
\item Seven-source campaign, 128 subsets with per-seed checkpoints (experiment 08; meeting games and truncation checks): \texttt{ateki/\allowbreak{}results/\allowbreak{}database\_attribution/\allowbreak{}}.
\item Harsanyi dividends (experiment 09): \texttt{ateki/\allowbreak{}results/\allowbreak{}harsanyi\_dividends/\allowbreak{}expt\_zhuo/\allowbreak{}}.
\item Measures, star and declared meetings (experiment 10): \texttt{ateki/\allowbreak{}results/\allowbreak{}materials\_structure\_attribution/\allowbreak{}}.
\item Sum and best-group rules (experiment 12): \texttt{ateki/\allowbreak{}results/\allowbreak{}component\_rule/\allowbreak{}}.
\item Data-valuation measures (experiment 20): \texttt{ateki/\allowbreak{}results/\allowbreak{}data\_valuation\_estimators/\allowbreak{}}.
\item Language-model teams (experiments 11 and 13): \texttt{ateki/\allowbreak{}results/\allowbreak{}recorded\_meetings\_scale/\allowbreak{}} (sets S1--S3, with the sequential arm in \texttt{S1\_seq/\allowbreak{}} and the per-campaign coordinator shares in \texttt{coordinator\_per\_campaign.txt}) and \texttt{ateki/\allowbreak{}results/\allowbreak{}recorded\_meetings\_4b/\allowbreak{}}.
\item Organization experiments 14--19: \texttt{ateki/\allowbreak{}results/\allowbreak{}\{emergent\_swarm, replicated\_bodies, multiple\_swarms, morphing\_swarms, mobile\_members, unequal\_credit\}/\allowbreak{}}.
\item Sub-team coalition values of the emergent team, including MACE-MP-0 alone ($0.808$ on PBE in this swarm setting, the reference for the share of the gap closed; $0.88$ in the four-source game of Section~2.4 of the main text, a different game with other members, so the two values are not compared): \texttt{ateki/\allowbreak{}results/\allowbreak{}emergent\_swarm/\allowbreak{}shapley.json}; early shares in \texttt{team\_values.json}.
\item Computed campaigns: \texttt{compute/\allowbreak{}live\_semiconductor/\allowbreak{}} (band gaps) and \texttt{compute/\allowbreak{}live\_almgsi/\allowbreak{}} (Al--Mg--Si).
\item Statistical re-analyses (held-out pool choice, final-window split): \texttt{ateki/\allowbreak{}results/\allowbreak{}stats\_reanalysis/\allowbreak{}}.
\item Figures: \texttt{writings/\allowbreak{}papers\_2026/\allowbreak{}regen\_figs/\allowbreak{}} (\texttt{mgea\_*.py} for the main-text figures of this version).
\end{itemize}
\par}
\end{document}